\documentclass{edp-jp4}
\usepackage{graphicx}
\usepackage{amssymb}
\usepackage{amsmath}
\begin{document}

%
\title{Form Factors in Magnetic Scattering of Thermal Neutrons} 
\author{R. Ballou}
\address{Laboratoire Louis N\'eel - C.N.R.S., BP 166X, F-38042 Grenoble Cedex 9, France}
\maketitle

\begin{abstract}

This lecture addresses the concept of form factor in magnetic scattering of thermal neutrons, analyzing its meaning, discussing its measurement by polarized neutrons and detailing its computation for the ions by the spherical tensor operator formalism. 

\end{abstract}
%

\section{\label{sec:introduction}Introduction}

It is an experimental fact that the magnetic scattering of thermal neutrons from a magnetic material can be detected only over a fraction of the accessible range of the scattering vector and, unlike the nuclear scattering, vanishes out as this gets large. Ascribed to the spatial distribution of the magnetic densities, which is modulated at the atomic scale and thus at the scale of the wavelength of the thermal neutrons, the underlying variation in the scattering amplitudes is extracted out through form factors associated to ions. A concrete advantage is that these depend on radial integrals that can be computed at once and tabulated for each ion and valence state. Within the so-called dipole approximation the only unknowns  then are the proportions of the spin and orbital magnetic moments attached to the ions. A further interest is that the expression of the differential scattering cross section is considerably simplified and adapted for the quantitative analysis of the static and dynamic correlation functions of magnetic moments. A number of magnetic phenomena thus can be investigated with more ease, for instance the long range or  short range magnetic orders with the associated magnetic transitions or crossovers and the magnon excitations emerging from these orders, the configurations and dynamics of magnetic moments in spin glasses or in spin fluids, the excitation spectrum of low dimensional magnets or of arrays of molecular nanomagnets, et cetera. \medskip

With polarized neutrons, measurements can be performed at precisions which, interestingly, allow probing the actual spatial distribution of the magnetic densities. A wholly new set of features associated to these then become available to analysis, for instance the quantum states of the concerned ions, which requires the exact computation of the associated form factors, but also the degree of covalency with neighboring ions or of hybridization with conduction electrons or else the itinerant nature of the magnetic densities, which all might question about the use itself of form factors associated to single ions, the spin delocalization and polarization effects, which might inform about the exchange paths and mechanisms in magnets, et cetera. \medskip

We qualitatively examine in the following the concept of form factor, then describe the method used for its precise measurement by polarized neutron in collinear magnets and finally provide with a survey of the algebra involved for their computation by the spherical tensor operator formalism. The aim of the lecture was to refresh our mind with concepts and methodologies that are not new but unfamiliar because they are not of frequent use. Accordingly, there is nothing in this manuscript that cannot be found elsewhere. Nevertheless, no attempt will be made to exhaustively account for the literature, which instead will be limited to the minimum.

\section{\label{sec:ScatAmp}Scattering Amplitudes}

The scattering amplitude $\mathcal{A}$ for a neutron-sample collision process, where the neutron spatial-spin incoming state $\vert \vec{k_{i}} \nu_{i} \rangle$ is transfered into the spatial-spin outgoing channel $\vert \vec{k_{f}} \nu_{f} \rangle$ while the sample initial state $\vert \lambda_{i} \rangle$ is transformed into the final state $\vert \lambda_{f} \rangle$, is computed equal to the matrix element
\begin{equation}
\label{eq:scatampl}
\mathcal{A} \left[ \{(\vec{k_{i}} \nu_{i}) \lambda_{i}\} \rightsquigarrow \{(\vec{k_{f}} \nu_{f}) \lambda_{f})\} \right] = -\frac{m_{\mathrm n}}{2\pi \hbar^{2}} \left \langle (\vec{k_{f}} \nu_{f}) \lambda_{f}) \left \vert \mathrm{~{\bf V}~} \right \vert (\vec{k_{i}} \nu_{i}) \lambda_{i}) \right \rangle
\end{equation}
in the Born approximation \cite{Messiah}. $m_{\mathrm n}$ is the neutron mass \footnote{\label{neutronmass} $m_{\mathrm n}$ = 1.67492~10$^{-27}$ kg (939.57 MeV/$c^{2}$) = 1838.62 $m_{\mathrm e}$ = 1.001375 $m_{\mathrm p}$, where $m_{\mathrm e}$ = 0.910956~10$^{-30}$~kg (= 0.511 MeV/$c^{2}$) is the electron mass and $m_{\mathrm p}$ = 1.67261~10$^{-27}$~kg (= 938.28 MeV/$c^{2}$) the proton mass. c  = 2.997925~10$^{8}$~ms$^{-1}$ is the light speed and $\hbar = \frac{h}{2\pi}$ = 1.05459~10$^{-34}$~Js the Dirac constant (Planck constant/$2\pi$).}.  {\bf V} is the neutron-sample interaction potential. \medskip

A first contribution to {\bf V} is the neutron-nucleus nuclear interaction potential. Since its range ($\simeq 1.5 \times 10^{-15}$ m) and the nucleus radius  ($\simeq 10^{-14}$ m) are several orders of magnitude smaller than the wavelength of thermal neutrons, it essentially gives rise to isotropic s-wave neutron scattering and thus is well approximated by the Fermi-Dirac pseudo-potential
\begin{equation}
\label{eq:nucpot1}
\mathbf{V}_{\mathrm N} = \frac{2\pi \hbar^{2}}{m_{\mathrm n}} \sum_{\mathrm p} \mathbf{b}_{\mathrm p} ~ \delta (\vec{\mathbf{r}}_{\mathrm n}-\vec{\mathbf{R}_{\mathrm p}}) 
\end{equation}
where $\delta$ is the Dirac delta generalized function, $\vec{\mathbf{r}}_{\mathrm n}$ the neutron position operator and $\vec{\mathbf{R}_{\mathrm p}}$ the p-th nucleus position operator. {\bf b}$_{\mathrm p}$ = A$_{\mathrm p} \mathbf{~1}_{\iota_{\mathrm p} \otimes\sigma}$ + B$_{\mathrm p}~\vec{\boldsymbol{\iota}}_{\mathrm p} \cdot \vec{\boldsymbol{\sigma}}$, for the nuclear interaction between two nucleons depends on their spin. $\mathbf{1}_{\iota_{\mathrm p} \otimes\sigma}$ is the unit operator in the $p-th ~ nucleous ~ \otimes  ~ neutron$ spin space, $\vec{\boldsymbol{\iota}}_{\mathrm p}$ the p-th nucleus spin operator and $\frac{1}{2}\vec{\boldsymbol{\sigma}}$ the neutron spin operator. The scattering lengths A$_{\mathrm p}$ and B$_{\mathrm p}$ are complex numbers that vary from nucleus to nucleus and as a function of the incident neutron energy, with the imaginary part representing radiative capture~\cite{Marshall&Lovesey}. \medskip

A second contribution to {\bf V}, which naturally emerges from the non relativistic limit of the neutron Hamitonian in an electromagnetic field \cite{Messiah}, is the interaction potential
\begin{equation}
\label{eq:magpot1}
\mathbf{V}_{\mathrm M} =  -\vec{\boldsymbol{\mu}}_{\mathrm n} \cdot \vec{\mathbf{B}}(\vec{\mathbf{r}}_{\mathrm n}) =   -\gamma_{\mathrm n} ~ \mu_{\mathrm N}  ~ \vec{\boldsymbol{\sigma}} \cdot \vec{\mathbf{B}}(\vec{\mathbf{r}}_{\mathrm n})
\end{equation}
of the neutron magnetic moment $\vec{\boldsymbol{\mu}}_{\mathrm n} = \gamma_{\mathrm n} ~ \mu_{\mathrm N}  ~ \vec{\boldsymbol{\sigma}}$ with the magnetic induction $\vec{\mathbf{B}}(\vec{\mathbf{r}}_{\mathrm n})$ created by the sample at the neutron position $\vec{\mathbf{r}}_{\mathrm n}$. $\gamma_{\mathrm n} = -1.91348$ is the neutron gyromagnetic ratio and $\mu_{\mathrm N}$ the nuclear magneton~\footnote{\label{nuclearmagneton} $\mu_{\mathrm N} = \frac{\left \vert e \right \vert \hbar}{2 m_{\mathrm p}} = 5.05095~10^{-27} \mathrm{~JT}^{-1}$. e (electron charge) = $-1.60219~10^{19}$~C.}. If the current density $\vec {\mathbf{j}}(\vec{r})$ engendering the magnetic induction $\vec{\mathbf{B}}(\vec{\mathbf{r}}_{\mathrm n})$ is stationary then
\begin{equation}
\label{eq:magpot2}
\vec{\mathbf{B}}(\vec{\mathbf{r}}_{\mathrm n}) =  \frac{\mu_{0}}{4\pi} \int_{\Omega} ~ \vec {\mathbf{j}}(\vec{r}) \wedge \frac{\vec{\mathbf{r}}_{\mathrm n}-\vec{r}}{\vert ~ \vec{\mathbf{r}}_{\mathrm n}-\vec{r} ~ \vert ^{3}}  ~ d\vec{r}
\end{equation}
where $\mu_{0}$ is the vacuum permeability~\footnote{\label{magneticpermeability} $\mu_{0} = 4 \pi~10^{-7}~\frac{\mathrm{T}}{\mathrm{A~m}^{-1}}$} and $\Omega$ any volume of the space with a boundary $\partial\Omega$ strictly enclosing $\vec{\mathbf{j}}(\vec{r})$ (cf.~eq.~\eqref{eq:FT3} and eq.~\eqref{eq:FT4}). An external magnetic induction also influences the neutron spatial-spin state, which allows guiding or rotating the neutron polarization or else, in the case of strong gradients, channelling the neutron, but it affects the scattering by modifying the neutron statistical ensemble and not through the neutron-sample scattering amplitude. \medskip

A neutron also reacts to the electrostatic field $\vec {\mathbf{E}}(\vec{\mathbf{r}}_{\mathrm n})$ of the nuclear and electron charge densities through the first order relativistic corrections to the neutron Hamiltonian, namely the Spin - Orbit (SO) + Darwin Correction (DC) interaction potentials
\begin{equation}
\label{eq:magpot3}
\mathrm{ {\bf V}_{ SO} } + \mathrm{ {\bf V}_{DC} } = -\frac{\gamma_{\mathrm n} ~ \mu_{\mathrm N}}{m_{\mathrm n} ~ c^{2}} ~ \vec{\boldsymbol{\sigma}} \cdot (\vec {\mathbf{E}}(\vec{\mathbf{r}}_{\mathrm n}) \wedge \vec{\mathbf{p}}_{\mathrm n}) -\frac{\gamma_{\mathrm n} ~ \mu_{\mathrm N}}{2m_{\mathrm n} ~ c^{2}} ~ \hbar \vec{\nabla}_{\vec{\mathbf{r}}_{\mathrm n}} \cdot \vec {\mathbf{E}}(\vec{\mathbf{r}}_{\mathrm n})
\end{equation}
but these provide with negligibly small scattering amplitudes as compared to those nuclear ({\bf V}$_{\mathrm {N}}$) and magnetic ({\bf V}$_{\mathrm {M}}$) and can be safely ignored~\cite{Marshall&Lovesey}. A little attention should be paid only to the N-SO elastic interference, bringing about, for initial neutron polarization $\vec{\mathrm P}_{i}$, a scattered intensity $\propto\vec{\mathrm P}_{i} \cdot (\vec{k}_{f} \wedge \vec{k}_{i})$ and, for null $\vec{\mathrm P}_{i}$, a scattered polarization $\propto \vec{k}_{f} \wedge \vec{k}_{i}$, which might give rise to detectable effects when the samples shows small nuclear coherent scattering, contains strongly absorbing species or possesses a non centrosymmetric crystal structure \cite{Marshall&Lovesey}. \medskip
 
The integration over the neutron space variable in the scattering amplitudes is straightforward, since $\langle \vec{u}_{\mathrm n} \vert \vec{k_{i}} \rangle = \exp\{i (k_{i}\cdot\vec{u}_{\mathrm n})\}$ and $\langle \vec{k_{f}} \vert \vec{s}_{\mathrm n} \rangle = \exp\{-i (k_{f}\cdot\vec{s}_{\mathrm n})\}$ are plane waves. Appropriately inserting the closure relation $\int  \vert \vec{r}_{\mathrm n} \rangle \langle \vec{r}_{\mathrm n} \vert ~ d\vec{r}_{\mathrm n} = \int  \vert \vec{s}_{\mathrm n} \rangle \langle \vec{s}_{\mathrm n} \vert ~ d\vec{s}_{\mathrm n} = \int  \vert \vec{u}_{\mathrm n} \rangle \langle \vec{u}_{\mathrm n} \vert ~ d\vec{u}_{\mathrm n} = \cdots = \mathbf{1}$ and making use of the Kronecker symbol $\delta_{A, B}$ to express that $\delta_{A, B} = 1 \iff A = B$ and $\delta_{A, B} = 0 \iff A \neq B$, we get 
\begin{itemize}
\item for the nuclear scattering amplitude
\begin{multline}
\label{eq:nsf1}
-\frac{m_{\mathrm n}}{2\pi \hbar^{2}} \langle (\vec{k_{f}} \nu_{f}) \lambda_{f}) \vert \mathrm {~{\bf V}_{\mathrm N}~} \vert (\vec{k_{i}} \nu_{i}) \lambda_{i}) \rangle =
\\
-\sum_{\mathrm p} \left\{\mathrm {A}_{\mathrm p} ~\delta_{\lambda_{f}, \lambda_{i}} ~\delta_{\nu_{f},\nu_{i}} + \mathrm {B}_{\mathrm p} ~\langle \lambda_{f} \vert ~\vec{\boldsymbol{\iota}}_{\vec{R}}~ \vert \lambda_{i}\rangle \cdot \langle \nu_{f} \vert ~\vec{\boldsymbol{\sigma}}~  \vert \nu_{i} \rangle \right\} ~\exp\{i (\vec{\varkappa} \cdot \vec{\mathbf{R}_{\mathrm p}})\}  
\end{multline}
bearing in mind that $\langle \vec{s}_{\mathrm n} \vert ~\delta (\vec{\mathbf{r}}_{\mathrm n} - \vec{\mathbf{R}_{\mathrm p}})~ \vert \vec{u}_{\mathrm n} \rangle = \langle \vec{s}_{\mathrm n} \vert \vec{u}_{\mathrm n} \rangle  ~\delta (\vec{u}_{\mathrm n} - \vec{\mathbf{R}_{\mathrm p}}) = \delta (\vec{s}_{\mathrm n} - \vec{u}_{\mathrm n}) ~ \delta (\vec{u}_{\mathrm n} - \vec{\mathbf{R}_{\mathrm p}})$,
\item for the magnetic scattering amplitude
\begin{multline}
\label{eq:msf1}
-\frac{m_{\mathrm n}}{2\pi \hbar^{2}} \left \langle (\vec{k_{f}} \nu_{f}) \lambda_{f}) \left \vert \mathrm {~{\bf V}_{\mathrm M}~} \right \vert (\vec{k_{i}} \nu_{i}) \lambda_{i}) \right \rangle  =   \left \langle \nu_{f}  \left \vert ~\vec{\boldsymbol{\sigma}}~ \right \vert \nu_{i} \right \rangle \cdot  \vec{\mathcal E}_{\lambda_{f}, \lambda_{i}} (\vec{\varkappa})
\\
\vec{\mathcal E}_{\lambda_{f}, \lambda_{i}} (\vec{\varkappa}) = \gamma_{\mathrm n} ~ \mathrm r_{0} \frac{1}{2\mu_{\mathrm B} \varkappa^{2}} ~ i \vec{\varkappa} \wedge \left \langle \lambda_{f} \left \vert ~ \int_{\Omega} ~ \vec{\mathbf{j}}(\vec{r}) \exp\{i (\vec{\varkappa} \cdot \vec{r})\}  ~ d\vec{r} ~ \right \vert \lambda_{i} \right \rangle \quad
\end{multline}
by using the identity 
$\frac{\vec{r}}{r^{3}} = -\nabla_{\vec{r}} \left ( \frac{1}{r}\right ) = -\nabla_{\vec{r}} \left( \frac{1}{2\pi^{2}} \int \frac{1}{\varkappa^{2}} \exp\{i (\vec{\varkappa} \cdot \vec{r})\} ~ d\vec{\varkappa} \right)$ (cf.~eq.~\eqref{eq:FT2}).
\end{itemize}
$\vec{\varkappa} = \vec{k_{i}} - \vec{k_{f}}$ defines the scattering vector. $\mu_{\mathrm B}$ is the Bohr magneton and r$_{0}$ the classical radius of the electron~\footnote{\label{bohrmagneton} $\mu_{\mathrm B} = \frac{ e \hbar}{2 m_{\mathrm e}} = -9.27410~10^{-24} \mathrm{~JT}^{-1}$. r$_{0} = \frac{\mu_{0}}{4 \pi} \frac{e^{2}}{m_{\mathrm e}} = 2.81794~10^{-15}$~m.}. $\vec{\varkappa}$ and $\vec{r}$ are ordinary vectors so that 
\begin{equation}
i \vec{\varkappa} \wedge \langle \lambda_{f} \vert ~ \int_{\Omega} ~ \vec{\mathbf{j}}(\vec{r}) \exp\{i (\vec{\varkappa} \cdot \vec{r})\}  ~ d\vec{r} ~ \vert \lambda_{i} \rangle = i \vec{\varkappa} \wedge \int_{\Omega} ~ \langle \lambda_{f} \vert ~\vec{\mathbf{j}}(\vec{r})~ \vert \lambda_{i} \rangle \exp\{i (\vec{\varkappa} \cdot \vec{r})\}  ~d\vec{r}
\end{equation}

The vector field $\langle \lambda_{f} \vert ~\vec{\mathbf{j}} (\vec{r})~ \vert \lambda_{i} \rangle$ can always be split into the sum of an irrotational component $\vec{j}_{\lambda_{f}, \lambda_{i}}^{I} (\vec{r}) = -\vec{\nabla}_{\vec{r}} ~\langle \lambda_{f} \vert ~\mathbf{\Xi}(\vec{r})~ \vert \lambda_{i} \rangle$ and a solenoidal component $\vec{j}_{\lambda_{f}, \lambda_{i}}^{S} (\vec{r}) = \vec{\nabla}_{\vec{r}} \wedge \langle \lambda_{f} \vert ~\vec{\mathbf{M}}(\vec{r})~ \vert \lambda_{i} \rangle$ :
\begin{equation}
\langle \lambda_{f} \vert ~\vec{\mathbf{j}} (\vec{r})~ \vert \lambda_{i} \rangle = -\vec{\nabla}_{\vec{r}} ~\langle \lambda_{f} \vert ~\mathbf{\Xi}(\vec{r})~ \vert \lambda_{i} \rangle + \vec{\nabla}_{\vec{r}} \wedge \langle \lambda_{f} \vert ~\vec{\mathbf{M}}(\vec{r})~ \vert \lambda_{i} \rangle
\end{equation}
where
\begin{equation}
\begin{split}
\langle \lambda_{f} \vert ~\mathbf{\Xi}(\vec{r})~ \vert \lambda_{i} \rangle & = \frac{1}{4\pi}~ \int_{\Omega} \frac{\vec{\nabla}_{\vec{s}}~ \cdot \langle \lambda_{f} \vert ~\vec{\mathbf{j}}(\vec{s})~ \vert \lambda_{i} \rangle}{\vert~\vec{r}-\vec{s}~\vert} ~ d\vec{s}
\\
\langle \lambda_{f} \vert ~\vec{\mathbf{M}}(\vec{r})~ \vert \lambda_{i} \rangle & = \frac{1}{4\pi}~ \int_{\Omega} \frac{\vec{\nabla}_{\vec{s}}~ \wedge \langle \lambda_{f} \vert ~\vec{\mathbf{j}}(\vec{s})~ \vert \lambda_{i} \rangle}{\vert~\vec{r}-\vec{s}~\vert} ~ d\vec{s}
\end{split}
\end{equation}
(cf.~eq.~\eqref{eq:FT5}). The \emph{irrotational component} $\vec{j}_{\lambda_{f}, \lambda_{i}}^{I} (\vec{r})$ \emph{does not at all contribute} to the scattering amplitude, since $\vec{\nabla}_{\vec{r}} ~ \wedge ~ \vec{j}_{\lambda_{f}, \lambda_{i}}^{I} (\vec{r}) = 0$ so that $-i~\vec{\varkappa}~ \wedge \int_{\Omega} ~ \vec{j}_{\lambda_{f}, \lambda_{i}}^{I} (\vec{r}) \exp\{i (\vec{\varkappa} \cdot \vec{r})\}  ~d\vec{r} = 0$. On the contrary, the \emph{solenoidal component} $\vec{j}_{\lambda_{f}, \lambda_{i}}^{S} (\vec{r})$ is \emph{fully probed} by the neutron, since $\vec{\nabla}_{\vec{r}}~\cdot~\vec{j}_{\lambda_{f}, \lambda_{i}}^{S} (\vec{r}) = 0$ so that $\int_{\Omega} ~ \vec{j}_{\lambda_{f}, \lambda_{i}}^{S} (\vec{r}) \exp\{i (\vec{\varkappa} \cdot \vec{r})\}  ~d\vec{r}$ is orthogonal to $\vec{\varkappa}$.
\medskip

The substitution of $-\vec{\nabla}_{\vec{r}} ~\langle \lambda_{f} \vert ~\mathbf{\Xi}(\vec{r})~ \vert \lambda_{i} \rangle + \vec{\nabla}_{\vec{r}} \wedge \langle \lambda_{f} \vert ~\vec{\mathbf{M}}(\vec{r})~ \vert \lambda_{i} \rangle$ for $\langle \lambda_{f} \vert ~\vec{\mathbf{j}} (\vec{r})~ \vert \lambda_{i} \rangle$ in the magnetic scattering amplitude results in re-formulating $\vec{\mathcal E}_{\lambda_{f}, \lambda_{i}} (\vec{\varkappa})$ in the eq.~\eqref{eq:msf1} as  
\begin{equation}
\label{eq:msf2}
\vec{\mathcal E}_{\lambda_{f}, \lambda_{i}} (\vec{\varkappa}) = -\gamma_{\mathrm n} ~ \mathrm r_{0} \frac{1}{2\mu_{\mathrm B} \varkappa^{2}} ~ \vec{\varkappa} \wedge \left \langle \lambda_{f} \left \vert ~ \int_{\Omega} ~ \vec{\mathbf{M}}(\vec{r}) \exp\{i (\vec{\varkappa} \cdot \vec{r})\}  ~ d\vec{r} ~ \right \vert \lambda_{i} \right \rangle \wedge \vec{\varkappa} 
\end{equation}
$\vec{\mathbf{M}}(\vec{r})$ is formally interpreted as the \emph{magnetic density} operator since $\vec{j}_{\lambda, \lambda}^{S} (\vec{r}) = \vec{\nabla}_{\vec{r}} \wedge \langle \lambda \vert ~\vec{\mathbf{M}}(\vec{r})~ \vert \lambda \rangle$ is understood as the stationary current density in the sample in the state $\vert \lambda \rangle$. All the other contributions to the current density are necessarily irrotational, thus should be ascribed to external sources to comply with charge conservation. We can safely ignore them although they produce magnetic fields, for these might act solely on the neutron statistical ensemble. A magnetic field $\vec{\mathrm H}_{\vec{\mathrm M}} (\vec{r})$ of dipolar nature is also created by the magnetic density. Within the sample it is equal to the opposite of the irrotational component of the magnetic density. \medskip

$\langle \lambda \vert ~\vec{\mathbf{M}}(\vec{r})~ \vert \lambda \rangle$ behaves as the customary conceived magnetic dipole densities that emerge from the spatio-temporal coarse-graining of the Maxwell equations when the magnetic materials are approximated by continuous media. At the difference of these, by essence macroscopic but unequivocally defined, $\langle \lambda \vert ~\vec{\mathbf{M}}(\vec{r})~ \vert \lambda \rangle$ makes sense down to the subatomic scale but shows \emph{gauge invariance}, since 
\begin{equation}
\vec{\nabla}_{\vec{r}} \wedge \{ \langle \lambda \vert ~\vec{\mathbf{M}}(\vec{r})~ \vert \lambda \rangle + \vec{\nabla}_{\vec{r}} ~\langle \lambda \vert ~\Psi (\vec{r})~ \vert \lambda \rangle \} = \vec{\nabla}_{\vec{r}} \wedge \langle \lambda \vert ~\vec{\mathbf{M}}(\vec{r})~ \vert \lambda \rangle
\end{equation} 
whatever the scalar field operator $\Psi (\vec{r})$ with constant quantum average at the boundaries of the sample and everywhere outside. This gauge freedom is a source of ambiguities for the reconstruction of the magnetic densities from the neutron data \cite{Hirst}. It is not eliminated by constraining the magnetic density to vanish outside the sample, where the solenoidal current density is null, nor by imposing that its volume integration gives the magnetic moment of the sample. It also survives over all the other electromagnetic equations for continuous media. As an example, the magnetic field created by $\langle \lambda \vert ~\vec{\mathbf{M}}(\vec{r})~ \vert \lambda \rangle + \vec{\nabla}_{\vec{r}} ~\langle \lambda \vert ~\Psi (\vec{r})~ \vert \lambda \rangle$ is computed equal to $\vec{\mathrm H}_{\vec{\mathrm M}+\vec{\nabla} \Psi} (\vec{r}) = \vec{\mathrm H}_{\vec{\mathrm M}} (\vec{r}) - \vec{\nabla}_{\vec{r}}~ \langle \lambda \vert ~ \Psi (\vec{r})~ \vert \lambda \rangle$, which shows that the magnetic induction $\vec{\mathrm B}(\vec{r}) = \mu_{0} \{ \vec{\mathrm H}(\vec{r}) + \langle \lambda \vert ~\vec{\mathbf{M}}(\vec{r})~ \vert \lambda \rangle \}$ does not depend on the gauge choice for $\langle \lambda \vert ~\vec{\mathbf{M}}(\vec{r})~ \vert \lambda \rangle$. A gauge can be fixed with the equation $\vec{\nabla}_{\vec{r}} \cdot \langle \lambda \vert ~\vec{\mathbf{M}}(\vec{r})~ \vert \lambda \rangle = 0$ so as to retain only the solenoidal component of the magnetic density, which is fully probed by the neutron, but then the concept is often not intuitive. Other gauges might reveal themselves more insightful depending on the physics under concern. \medskip

Actually, the spin magnetism is naturally described by continuous distributions of point-like magnetic dipoles $\vec{\boldsymbol{\mu}}_{j}$ at positions $\vec{\mathbf{r}}_{j}$
\begin{equation}
\vec{\mathbf{M}}_{\mathrm S}(\vec{r}) = \sum_{j} \delta(\vec{\mathbf{r}}_{j} - \vec{r})  \vec{\boldsymbol{\mu}}_{j}
\end{equation}
for these do exist objectively. All the other contributions to the magnetic density should be considered through the convection current density operator
\begin{equation}
\vec{\mathbf{j}}_{\mathrm C}(\vec{r}) = \frac{1}{2} \sum_{j} q_{j} \left[ \delta(\vec{\mathbf{r}}_{j} - \vec{r}) \vec{\mathbf{v}}_{j} + \vec{\mathbf{v}}_{j} \delta(\vec{\mathbf{r}}_{j} - \vec{r}) \right]
\end{equation}
of moving particles $j$ of electric charge $q_{j}$. $\vec{\mathbf{r}}_{j}$ is the position operator and $\vec{\mathbf{v}}_{j}$ the velocity operator of the $j-$particle. $\vec{\mathbf{v}}_{j} = \frac{i}{\hbar} [\mathcal{H}, \vec{\mathbf{r}}_{j}] = \vec{\nabla}_{\vec{\mathbf{p}}_{j}}\mathcal{H}$, where $\mathcal{H}$ is the Hamiltonian of the sample.\medskip

Nuclear magnetism is proportional to $\mu_{\mathrm N}$ when electron magnetism is proportional to $\mu_{\mathrm B}$, so can be disregarded. As a matter of fact, nuclear magnetism is rather probed through the neutron-nucleus interaction potential $\mathbf{V}_{\mathrm N}$. We shall henceforth consider only pure electron parameters : $\vec{\boldsymbol{\mu}}_{j} = g_{\mathrm e} ~ \mu_{\mathrm B} ~ \vec{\mathbf{s}}_{j}$, where $g_{\mathrm e} = 2.00232 \approx 2$ is the electron gyromagnetic ratio, and $q_{j} = e ~ \forall j$. Assuming further that the $\vec{\mathbf{p}}-$dependent terms in the Hamiltonian are dominated by the non relativistic kinetic energy, $\vec{\mathbf{v}}_{j} = \tfrac{\vec{\mathbf{p}}_{j}}{m_{e}}$. Accordingly, $\vec{\mathcal E}_{\lambda_{f}, \lambda_{i}} (\vec{\varkappa})$ in the eq.~\eqref{eq:msf1}) writes
\begin{equation}
\label{eq:msf3}
\begin{split}
\vec{\mathcal E}_{\lambda_{f}, \lambda_{i}} (\vec{\varkappa}) & = -\gamma_{\mathrm n} ~ \mathrm r_{0}  ~\frac{1}{\varkappa^{2}} ~  \vec{\varkappa} \wedge \left \langle \lambda_{f} \left \vert ~ \sum_{j} \left[ \exp\{i (\vec{\varkappa} \cdot \vec{\mathbf{r}}_{j})\} ~ \vec{\mathbf{s}}_{j} \right] ~ \right \vert \lambda_{i} \right \rangle \wedge \vec{\varkappa} + {}
\\ 
& \quad + \gamma_{\mathrm n} ~ \mathrm r_{0} ~\frac{i}{2\hbar \varkappa^{2}} ~ \vec{\varkappa} \wedge \left \langle \lambda_{f} \left \vert ~ \sum_{j} \left[ \exp\{i (\vec{\varkappa} \cdot \vec{\mathbf{r}}_{j})\} ~ \vec{\mathbf{p}}_{j} + \vec{\mathbf{p}}_{j} ~ \exp\{i (\vec{\varkappa} \cdot \vec{\mathbf{r}}_{j})\} \right] \right \vert \lambda_{i} \right \rangle
\end{split}
\end{equation}

\section{\label{sec:FormFac}Form Factors}

Let us consider the concrete instance where the sample is a collection of kinematically independent ions $\mathbb{A}_{\mathrm p}$ at the positions defined by the eigenvalues of $\vec{\mathbf{R}}_{\mathrm p}$. Any electron $j$ then belongs to a single ion, which implies that any sample state $\vert \lambda_{f, i} \rangle$ is the tensor product $\vert \lambda_{f, i} \rangle = \bigotimes_{\mathrm p} \vert \lambda_{f, i}^{\mathrm p} \rangle$ of the ionic states $\vert \lambda_{f, i}^{\mathrm p} \rangle$ and that $\langle \lambda_{f} \vert \mathbf{O}_{j \in \mathbb{A}_{\mathrm p}} \vert \lambda_{i} \rangle = \langle \lambda_{f}^{\mathrm p} \vert \mathbf{O}_{j \in \mathbb{A}_{\mathrm p}} \vert \lambda_{i}^{\mathrm p} \rangle$ whatever the one-electron operator  $\mathbf{O}_{j \in \mathbb{A}_{\mathrm p}}$ associated to the $j-$electron of the ion $\mathbb{A}_{\mathrm p}$. We deduce
\begin{equation}
\label{eq:msf4}
\vec{\mathcal E}_{\lambda_{f}, \lambda_{i}} (\vec{\varkappa}) = -\gamma_{\mathrm n} ~ \mathrm r_{0} ~ \sum_{\mathrm p} ~ \left \{ \frac{1}{\varkappa^{2}} ~ \vec{\varkappa} \wedge \vec{\mathcal{F}}_{\lambda_{f}^{\mathrm p}, \lambda_{i}^{\mathrm p}}(\vec{\varkappa}) \wedge \vec{\varkappa} \right \} ~ \exp\{i (\vec{\varkappa} \cdot \vec{\mathbf{R}}_{\mathrm p})\}
\end{equation}
where
\begin{equation}
\vec{\mathcal{F}}_{\lambda_{f}^{\mathrm p}, \lambda_{i}^{\mathrm p}}(\vec{\varkappa}) = \frac{1}{2\mu_{\mathrm B}} ~ \left \langle \lambda_{f}^{\mathrm p} \left \vert ~ \int_{\Omega} ~ \vec{\mathbf{M}}(\vec{r}) \exp\{i (\vec{\varkappa} \cdot \vec{r})\}  ~ d\vec{r} ~ \right \vert \lambda_{i}^{\mathrm p} \right \rangle
\end{equation}
defines the \emph{magnetic vector form factor} of the ion $\mathbb{A}_{\mathrm p}$. Using the spherical tensor operator formalism, it may be computed exactly, but this necessitates defining a number of mathematical concepts and performing weighty algebraic manipulations, so will be detailed in the last section and in the appendix. In this section rather qualitative and simplified aspects will be discussed.\medskip

Let $\mathbf{T}$ be the time inversion operator. $\mathbf{T} = \mathbf{Y} ~ \mathbf{K}$, where the unitary operator $\mathbf{Y} = \exp\{ {-i~\pi~ \hat{e}_{y} \cdot \sum_{j} \vec{\mathbf{s}}_{j}} \}$ acts solely on the spin states and the antiunitary operator $\mathbf{K}$ of complex conjugation acts solely on the spatial states \cite{Messiah}. Let {\bf X} be a pure imaginary operator, that is fully reversed by the time inversion: $\mathbf{T} ~\mathbf{X}~ \mathbf{T}^{-1} = - \mathbf{X}$. It follows that $\langle \psi_{f} \vert ~ \mathbf{X} ~\vert \psi_{i} \rangle = \langle \psi_{f} \vert ~ \mathbf{T}^{-1} \mathbf{T} ~\mathbf{X}~ \mathbf{T}^{-1} \mathbf{T}~\vert \psi_{i} \rangle = -\langle \psi_{f} \vert ~ \mathbf{T}^{-1} ~\mathbf{X}~ \mathbf{T}~\vert \psi_{i} \rangle = -\langle \mathbf{T} \psi_{f}  \vert ~  \mathbf{X} ~\vert \mathbf{T}\psi_{i} \rangle^{\ast}$, since, {\bf T} being antiunitary, $\langle \psi_{f}  \vert  \mathbf{T}^{-1} \chi \rangle = \langle \psi_{f}  \vert  \mathbf{T}^{+} \chi \rangle = \langle  \mathbf{T} \psi_{f}  \vert  \chi \rangle^{\ast}$. Assuming that $\mathbf{X}$ acts solely on the spatial states and that $\vert \psi_{f, i} \vert \rangle$ are products of spatial states $\vert \xi_{f, i} \rangle$ by spin states $\vert \zeta_{f, i} \rangle$ or, if not, expanding over such basis states, the spin degrees of freedom can be ignored. We then may write $\langle \xi_{f} \vert ~ \mathbf{X} ~\vert \xi_{i} \rangle =  -\langle \mathbf{K} \xi_{f}  \vert ~  \mathbf{X} ~\vert \mathbf{K} \xi_{i} \rangle^{\ast} = -\langle \mathbf{K} \xi_{i}  \vert ~  \mathbf{X} ~\vert \mathbf{K} \xi_{f} \rangle = -\int d\vec{r} \int d\vec{s} ~ \langle \xi_{i} \vert \vec{r} \rangle^{\ast} \langle \vec{r} \vert ~\mathbf{X}~ \vert \vec{s} \rangle \langle \vec{s} \vert \xi_{f} \rangle^{\ast}$. If the Hamiltonian is Hermitian and if the states $\xi_{f, i}$ are non degenerate then the wavefunctions $\langle \vec{r} \vert \xi_{f, i} \rangle$ should be proportional to their complex conjugate, but $\mathbf{K}^{2} = \mathbf{1}$ so that $\langle \vec{r} \vert \xi_{f, i} \rangle^{\ast} = \exp\{i\varphi_{f, i}\} \langle \vec{r} \vert \xi_{f, i} \rangle$, in which case $\langle \xi_{f} \vert ~ \mathbf{X} ~\vert \xi_{i} \rangle = -\exp\{i(\varphi_{f}-\varphi_{i})\} \langle \xi_{f} \vert ~ \mathbf{X} ~\vert \xi_{i} \rangle^{\ast}$. If furthermore $\xi_{f} = \xi_{i}$ then the matrix element is an imaginary number. Using this, we may qualitatively infer that if the ionic states $\vert \lambda_{f, i}^{\mathrm p} \rangle$ belong to a same electronic configuration $(nl)$ then there should be no contribution to the matrix element of the convective current density from the radial degree of freedom of the electrons, since $\mathbf{T}~ \vec{\mathbf{j}}_{\mathrm C}(\vec{r}) ~\mathbf{T}^{-1} = - \vec{\mathbf{j}}_{\mathrm C}(\vec{r})$ but the diagonal elements of $\vec{\mathbf{j}}_{\mathrm C}(\vec{r})$ should be real measurable quantities. Explicit calculations confirm this by more precisely showing that if $\vec{\mathbf{v}}_{j} = \frac{\vec{\mathbf{p}}_{j}}{m_{e}}$ then $\langle nl \vert ~\vec{\mathbf{j}}_{\mathrm C}(\vec{r})~ \vert nl \rangle = \frac{R_{nl}^{2}(r)}{r} \langle l \vert ~\vec{\mathbf{j}}_{\mathrm C}(\vec{r}) ~\vert l \rangle$, where $R_{nl} = \langle r \vert n l \rangle$ is the radial wavefunction associated with the electronic configuration. If moreover the ionic states $\vert \lambda_{f, i}^{\mathrm p} \rangle$ are in the same orbitally quenched state then for the same reasons there also should be no contribution to the matrix element of the convection current density from the angular degree of freedom of the electrons, in which case we may set $\vec{\mathbf{M}}(\vec{r}) = \vec{\mathbf{M}}_{\mathrm S}(\vec{r}) =  g_{\mathrm e} ~ \mu_{\mathrm B} \sum_{j} \delta(\vec{\mathbf{r}}_{j} - \vec{r}) ~ \vec{\mathbf{s}}_{j}$ so that
\begin{equation}
\vec{\mathcal{F}}_{\lambda_{f}^{\mathrm p}, \lambda_{i}^{\mathrm p}}(\vec{\varkappa}) = \left \langle \lambda_{f}^{\mathrm p} \left \vert ~ \sum_{j \in \mathbb{A}_{\mathrm p}} \exp\{i (\vec{\varkappa} \cdot \vec{\mathbf{r}}_{j_{\mathrm p}})\} ~ \vec{\mathbf{s}}_{j}~ \right \vert \lambda_{i}^{\mathrm p} \right \rangle 
\end{equation}
Using the closure relation $\sum_{\eta} \left \vert \eta^{\mathrm p} \left \rangle \right \langle \eta^{\mathrm p} \right \vert  = \mathbf{1}$ over the ionic states, this rewrites
\begin{multline}
\vec{\mathcal{F}}_{\lambda_{f}^{\mathrm p}, \lambda_{i}^{\mathrm p}}(\vec{\varkappa}) = \left \langle \lambda_{f}^{\mathrm p} \left \vert ~ \sum_{j \in \mathbb{A}_{\mathrm p}} \exp\{i (\vec{\varkappa} \cdot \vec{\mathbf{r}}_{j_{\mathrm p}})\} ~ \sum_{\eta} \left \vert \eta^{\mathrm p} \left \rangle \right \langle \eta^{\mathrm p} \right \vert ~ \vec{\mathbf{s}}_{j}~ \right \vert \lambda_{i}^{\mathrm p} \right \rangle = {}
\\
= \left \langle \lambda_{f}^{\mathrm p} \left \vert ~  \sum_{\eta} \exp\{i (\vec{\varkappa} \cdot \vec{\mathbf{r}}_{j_{\mathrm p}})\} ~ \left \vert \eta^{\mathrm p} \left \rangle \right \langle \eta^{\mathrm p} \right \vert ~ \sum_{j \in \mathbb{A}_{\mathrm p}} \vec{\mathbf{s}}_{j}~ \right \vert \lambda_{i}^{\mathrm p} \right \rangle
\end{multline}
where the phase factor can be factorized because the matrix element of an one-electron operator over electrons states of a same configuration, which are linear combinations of antisymmetrized products of single electron orthogonal states, does not depend on which electron is chosen to compute it.  A pure spatial state is insensitive to pure spin operators, therefore, because of the matrix element $\langle \eta^{\mathrm p} \vert ~ \sum_{j \in \mathbb{A}_{\mathrm p}} \vec{\mathbf{s}}_{j}~ \vert \lambda_{i}^{\mathrm p} \rangle$, the closure relation $\sum_{\eta} \left \vert \eta^{\mathrm p} \left \rangle \right \langle \eta^{\mathrm p} \right \vert  = \mathbf{1}$ gets limited to the ionic states $\eta^{\mathrm p}$ with the same spatial component as the $\vert \lambda_{f, i}^{\mathrm p} \rangle$. It then is inferred that $\langle \lambda_{f}^{\mathrm p} \vert ~ \exp\{i (\vec{\varkappa} \cdot \vec{\mathbf{r}}_{j_{\mathrm p}})\} ~ \vert \eta^{\mathrm p} \rangle \neq 0 \iff \vert \lambda_{f}^{\mathrm p} \rangle = \vert \eta^{\mathrm p} \rangle$. Accordingly, 
\begin{equation}
\vec{\mathcal{F}}_{\lambda_{f}^{\mathrm p}, \lambda_{i}^{\mathrm p}}(\vec{\varkappa}) = \left \langle \lambda_{f}^{\mathrm p} \left \vert ~ \exp\{i (\vec{\varkappa} \cdot \vec{\mathbf{r}}_{j_{\mathrm p}})\} ~ \right \vert \lambda_{f}^{\mathrm p} \right \rangle \left \langle \lambda_{f}^{\mathrm p} \left \vert ~ \sum_{j \in \mathbb{A}_{\mathrm p}} \vec{\mathbf{s}}_{j}~ \right \vert \lambda_{i}^{\mathrm p} \right \rangle = f_{\lambda_{f}}^{\mathrm p}(\vec{\varkappa}) ~ \left \langle \lambda_{f}^{\mathrm p} \left \vert ~\vec{\mathbf{S}}^{p}~ \right \vert \lambda_{i}^{\mathrm p} \right \rangle
\end{equation}
where $\vec{\mathbf{S}}^{p}$ is the total spin operator of the ion $\mathbb{A}_{\mathrm p}$. $f_{\lambda_{f}}^{\mathrm p}(\vec{\varkappa}) = \langle \lambda_{f}^{\mathrm p} \vert ~ \exp\{i (\vec{\varkappa} \cdot \vec{\mathbf{r}}_{j_{\mathrm p}})\} ~ \vert \lambda_{f}^{\mathrm p} \rangle$ defines the so called \emph{extracted scalar form factor}. It is tempting to generalize this to non degenerate spatial states as
\begin{equation}
\left \langle \lambda_{f}^{\mathrm p} \left \vert ~ \int_{\Omega} ~ \vec{\mathbf{M}}(\vec{r}) \exp\{i (\vec{\varkappa} \cdot \vec{r})\}  ~ d\vec{r} ~ \right \vert \lambda_{i}^{\mathrm p} \right \rangle = f_{\lambda_{f}, \lambda_{i}}^{\mathrm p}(\vec{\varkappa}) \left \langle \lambda_{f}^{\mathrm p} \left \vert ~ \int_{\Omega} ~ \vec{\mathbf{M}}(\vec{r})~ d\vec{r} ~ \right \vert \lambda_{i}^{\mathrm p} \right \rangle
\end{equation}
which presupposes that the intra-atomic magnetic density $\vec{\mathbf{M}}(\vec{r})$ is collinear. This often is a good approximation widely used in practice, in particular when the physical interest is rather focussed at the inter-atomic magnetic correlation, but should be considered with some caution. Non collinear magnetic densities are naturally expected in the presence of an antiferromagnetic order or any non collinear magnetic order, for there is no reason to believe that the magnetic orientation would abruptly change at well-defined inter-atomic boundaries. The non collinear character of the intra-atomic magnetic densities even come out in collinear ferromagnets or in paramagnets under magnetic field owing to the relativistic spin-orbit interaction which couples the magnetic densities to the local environments. Clear experimental evidences are revealed in precise measurements \cite{Brown}. Strictly speaking, the concept of scalar form factors is limited. \medskip

According to the exact computations detailed in the last section, the spherical components of the quantity $\{ \frac{1}{\varkappa^{2}} ~ \vec{\varkappa} \wedge \vec{\mathcal{F}}_{\lambda_{f}^{\mathrm p}, \lambda_{i}^{\mathrm p}}(\vec{\varkappa}) \wedge \vec{\varkappa} \}$ in the eq.~\eqref{eq:msf4}, for ionic states $\vert \lambda_{f, i}^{\mathrm p} \rangle$ belonging to the same electronic configuration $(nl)$, can be put in the global form
\begin{equation}
\mathcal{F}_{\lambda_{f}^{\mathrm p}, \lambda_{i}^{\mathrm p}}^{\bot~ q} (\vec{\varkappa}) = \sum_{K} \langle  j_{K}(\varkappa) \rangle \sum_{K^{\prime}, Q^{\prime}} \mathcal{C}_{K, K^{\prime}, Q^{\prime}}^{q} (\lambda_{f}^{\mathrm p}, \lambda_{i}^{\mathrm p}) ~ Y_{Q^{\prime}}^{K^{\prime}}(\Omega_{\vec{\varkappa}})
\end{equation}
where $Y_{Q^{\prime}}^{K^{\prime}}(\Omega_{\vec{\varkappa}})$ are spherical harmonics over the spherical angles $\Omega_{\vec{\varkappa}} = (\theta_{\vec{\varkappa}}, \varphi_{\vec{\varkappa}})$ of the scattering vector and $\mathcal{C}_{K, K^{\prime}, Q^{\prime}}^{q} (\lambda_{f}^{\mathrm p}, \lambda_{i}^{\mathrm p})$ gather matrix elements between angular momentum states and geometrical factors.  At a first sight these appear algebraically complicated, involving coefficients of fractional parentage and summations over Wigner symbols, but in fact are formally not difficult to evaluate whatever the ionic states $\vert \lambda_{f, i}^{\mathrm p} \rangle$. $\langle  j_{K}(\varkappa) \rangle$ are index K radial integrals : 
\begin{equation}
\langle  j_{K}(\varkappa) \rangle = \int_{0}^{\infty} r^{2} ~R_{n l}^{2}(r) ~j_{K}(\varkappa r)~ dr \qquad (R_{nl} = \langle r \vert n l \rangle)
\end{equation}
where $R_{nl} = \langle r \vert n l \rangle$ is the radial wavefunction associated with the electronic configuration. These can be calculated at once and tabulated for each ion and valence by numerically solving the Hartree-Fock problem for the electrons interacting with the nucleus and between them in the one-electron central potential approximation. With the heavy ions the experimental measurements are enough precise to evidence significant deviation due to relativistic effects, for instance in the lanthanide or actinide series the f electrons are radially more expanded. Accordingly, the radial integrals for these ions are numerically calculated from the relativistic Dirac-Fock Hamiltonian \cite{Desclaux,Desclauxf}. This, in addition to the Dirac one-electron Hamiltonian and the electron-electron Coulomb interaction, includes the Breit interaction, which is treated as a first order perturbation, in order to partially take into account the relativistic character of the interaction between the electrons : the Coulomb interaction is not Lorentz covariant but emerges as the leading term in an expansion of the interaction energy in powers of the fine structure constant obtained by the methods of quantum electrodynamics. As to be consistent the scattering amplitude also should be calculated on relativistic states from the relativistic expression of the current density operator $\vec{\mathbf{j}}(\vec{r}) = \sum_{j} c~e~\vec{\boldsymbol{\alpha}}_{j} ~\delta(\vec{\mathbf{r}}_{j} - \vec{r})$, where $\vec{\boldsymbol{\alpha}}_{j}$ is the Dirac vector operator for the $j-$electron (notice that in the non relativistic limit this split back into the sum of a spin and a convection current density operators) \cite{Messiah}. We shall not get into the details of the formulation, which is too weighty to be account for in these notes but can be found elsewhere \cite{Stassis&Deckman-r1,Stassis&Deckman-r2}. An effective operator approach is considered in these works which allows expressing the matrix elements of relativistic operators over relativistic states as matrix elements of effective operators over non relativistic states. It has been argued that weak corrections might occur in addition to the radial one, answerable primarily to the relativistic mass-correction term, which decreases the convection current density in the high-kinetic-energy region near the nucleus thus producing a more spatially extended current density, and secondarily to the relativistic spin-orbit term, which slightly changes the effective electronic $g_{e}$ factor. In practice, these effects are ignored and the relativistic scattering amplitude is obtained from the non relativistic one by merely replacing the radial integrals by their relativistic counterpart.\medskip

When the inverse modulus $\varkappa^{-1}$ of the scattering vector $\vec{\varkappa}$ is much larger than the mean radius of the radial wavefunction of the unpaired electrons giving rise to the magnetic scattering the magnetic vector form factor can be approximated by the very simplified expression
\begin{equation}
\label{eq:dipoapprox}
\vec{\mathcal{F}}_{\lambda_{f}^{\mathrm p}, \lambda_{i}^{\mathrm p}}(\vec{\varkappa}) = \left \langle \lambda_{f}^{\mathrm p} \left \vert ~ \langle  j_{0}(\varkappa) \rangle \vec{\mathbf{S}} + \frac{1}{2} \left[\langle  j_{0}(\varkappa) \rangle + \langle  j_{2}(\varkappa) \rangle \right] \vec{\mathbf{L}} ~ \right \vert \lambda_{i}^{\mathrm p} \right \rangle = \left \langle \lambda_{f}^{\mathrm p} \left \vert ~ \vec{\boldsymbol{\mathcal{F}}}(\varkappa) ~ \right \vert \lambda_{i}^{\mathrm p} \right \rangle
\end{equation}
where $\vec{\mathbf{S}}$ and $\vec{\mathbf{L}}$ are the total spin and total orbital angular momentum operators of the ion. As discussed in the last section, this is inferred from the behavior of the spherical Bessel functions for small arguments and defines the \emph{dipole approximation} to the scattering amplitude. Notice that the form factor then is isotropic in the reciprocal space, that is it depends solely on the modulus of the scattering vector and not on its spherical angles. If the unpaired electrons are in an orbital singlet state then $\vec{\boldsymbol{\mathcal{F}}}(\varkappa) = \langle  j_{0}(\varkappa) \rangle~\vec{\mathbf{S}}$, which gives a scalar form factor $f(\vec{\varkappa}) \approx \langle  j_{0}(\varkappa) \rangle$ at small scattering angle. This often applies to the d electrons of the 3d transition metal ions, which, owing to their spatial extension, often experience strong crystalline electric field in the materials. Generally, the quenching is partially raised by the spin-orbit interaction. This acts as a perturbation weakly coupling the ground orbital state with excited orbital states, thus inducing a small orbital moment. The total magnetic moment of the ion then differs slightly from that of the spin contribution and comes out through a value of the electron gyromagnetic ratio $g$ different from that of the naked electron $g_{e}$. In order to keep the spin operator as the basic variable it is convenient so set $(\vec{\mathbf{L}}+g_{e}\vec{\mathbf{S}}) = g\vec{\mathbf{S}}$, which determines $g$, in which case $\vec{\boldsymbol{\mathcal{F}}}(\varkappa) = \frac{1}{2} \{ \langle  j_{0}(\varkappa) \rangle~g_{e}\vec{\mathbf{S}} + \left[\langle  j_{0}(\varkappa) \rangle + \langle  j_{2}(\varkappa) \rangle \right] \vec{\mathbf{L}} \}$ is expressed in the form
\begin{equation}
\vec{\boldsymbol{\mathcal{F}}}(\varkappa) = \frac{1}{2}gf(\varkappa)~\vec{\mathbf{S}} 
\end{equation} 
where
\begin{equation}
f(\varkappa) = \langle  j_{0}(\varkappa) \rangle + \left(1-\frac{g_{e}}{g}\right) \langle  j_{2}(\varkappa) \rangle
\end{equation} 
Notice that we systematically set $g_{e} \approx 2$. Unlike the d electrons of the 3d transition metal ions the f electrons of the lanthanide ions are much less extended, thus are subject to weaker crystalline electric field in the materials, whereas their spin-orbit interaction is strongly increased, becoming dominant. Accordingly, the total spin $\vec{\mathbf{S}}$ and total orbital $\vec{\mathbf{L}}$ moments are first coupled to form the angular moment $\vec{\mathbf{J}} = \vec{\mathbf{S}} + \vec{\mathbf{L}}$ and the energy spectrum gets structured in terms of the multiplets $\vert \tau L S J M \rangle$ on which the crystalline electric field potential will act as a perturbation. Within each multiplet the Wigner-Eckart theorem allows writing $2\vec{\mathbf{S}} = g_{S}\vec{\mathbf{J}}$, $\vec{\mathbf{L}} = g_{L}\vec{\mathbf{J}}$ and $2\vec{\mathbf{S}} + \vec{\mathbf{L}} = g_{J}\vec{\mathbf{J}}$, where $g_{J} = g_{S} + g_{L}$, $g_{S}$ and $g_{L}$ are geometrically determined from $g_{S}\vec{\mathbf{J}}^{2} = 2\vec{\mathbf{S}}\cdot\vec{\mathbf{J}} = \vec{\mathbf{S}}^{2} + 2\vec{\mathbf{S}}\cdot\vec{\mathbf{L}} = \vec{\mathbf{J}}^{2}-\vec{\mathbf{L}}^{2}+\vec{\mathbf{S}}^{2}$  and $g_{L}\vec{\mathbf{J}}^{2} = \vec{\mathbf{L}}\cdot\vec{\mathbf{J}} = \vec{\mathbf{L}}^{2} + \vec{\mathbf{L}}\cdot\vec{\mathbf{S}} = \frac{ \vec{\mathbf{J}}^{2}+\vec{\mathbf{L}}^{2}-\vec{\mathbf{S}}^{2} }{2}$, using the identity $\vec{\mathbf{J}}^{2} = (\vec{\mathbf{S}}+\vec{\mathbf{L}})^{2} = \vec{\mathbf{S}}^{2} + \vec{\mathbf{L}}^{2} + 2 \vec{\mathbf{S}}\cdot \vec{\mathbf{L}}$ : 
\begin{equation}
g_{S} = 1+\frac{S(S+1)-L(L+1)}{J(J+1)} \qquad g_{L} = \frac{1}{2}+\frac{L(L+1)-S(S+1)}{2J(J+1)} 
\end{equation}
Thus, if the initial $\vert \lambda_{i}^{\mathrm p} \rangle = \vert \tau L S J M_{i} \rangle$ and final $\vert \lambda_{f}^{\mathrm p} \rangle = \vert \tau L S J M_{f} \rangle$ states belong to the same multiplet $(J_{i} = J_{f} = J)$ then
\begin{equation}
\vec{\boldsymbol{\mathcal{F}}}(\varkappa) = \frac{1}{2}g_{J}f(\varkappa)~\vec{\mathbf{J}}
\end{equation} 
where
\begin{equation}
f(\varkappa) = \langle  j_{0}(\varkappa) \rangle + \frac{g_{L}}{g_{J}} \langle  j_{2}(\varkappa) \rangle
\end{equation} 
If the initial $\vert \lambda_{i}^{\mathrm p} \rangle = \vert \tau L S (J M)_{i} \rangle$ and final $\vert \lambda_{f}^{\mathrm p} \rangle = \vert \tau L S (J M)_{f} \rangle$ states belong to different multiplets $(J_{i} \neq J_{f})$ then $\langle \tau L S (J M)_{f} \vert \vec{\mathbf{J}} \vert \tau L S (J M)_{i} \rangle = 0$ so that $\langle \tau L S (J M)_{f} \vert \vec{\mathbf{S}} \vert \tau L S (J M)_{i} \rangle = -\langle \tau L S (J M)_{f} \vert \vec{\mathbf{L}} \vert \tau L S (J M)_{i} \rangle$. It in this case follows that 
\begin{equation}
\vec{\boldsymbol{\mathcal{F}}}(\varkappa) = \frac{1}{2} f_{(JM)_{f}, (JM)_{i}}(\varkappa)~\vec{\mathbf{S}}
\end{equation} 
where
\begin{equation}
f_{(J M)_{f}, (J M)_{i}}(\varkappa) = \langle  j_{0}(\varkappa) \rangle - \langle  j_{2}(\varkappa) \rangle
\end{equation} 
\medskip

So far the unpaired electrons giving rise to the magnetic scattering were assumed kinematically independent from atomic center to atomic center, but this, in the real magnetic materials, is more the exception than the rule : the states of neighboring centers get mixed by the inter-atomic exchange interactions, there may be covalent transfer from ligand ions surrounding the magnetic ions, modifying the states of these and creating spin polarization on the ligand ions, the electrons under concern may tend to participate to metallic bonding, et cetera. The magnetic scattering then contains multi-center matrix elements and its computation become more difficult. We shall not get into the details of the different instances, for these are quite varied, and shall only emphasize that the robustness of the one-center states is such that the use of the ionic-like form factors is often a good approximation. Thus, in most instances the intra-atomic exchange interactions are much larger than the inter-atomic exchange interactions, freezing out atomic or ionic angular momentum states on which these act as a perturbation, and the overlap mechanisms giving rise to these generally modify the spatial wavefunctions at large radial distance. Accordingly, the electrons preserve an atomic or ionic character and can be analyzed as if they were kinematically independent, at least at not too small modulus $\varkappa$ of the scattering vector. Actually, even when more significant effects on the scattering amplitude might be expected, for instance in the presence of strong ion-ligand covalent transfer, the single ion analysis is worth performing for it might allow assessing the relevance of the multi-center contributions. The case of itinerant electrons must be treated separately, using Wannier functions. These show strong atomic-like character in the case of 3d transition metals and 3d - 4f (or 5f) intermetallics and, ignoring the s electrons, spread essentially over the $z$ nearest neighbors. Labeling these with vectors $\vec{\rho}$, the magnetic scalar form factor $f_{W}(\vec{\varkappa}) = f_{A}(\vec{\varkappa}) \{ (1- zW^{2} + W^{2}\sum_{\vec{\rho}} \exp(i~\vec{\varkappa}\cdot\vec{\rho}) \}$ can be assigned to each atomic center, where W is a Wannier expansion coefficient and $f_{A}(\vec{\varkappa})$ an ionic scalar form factor. The interesting point is that if $\vec{\varkappa}$ is a reciprocal vector $\vec{K}$ then  $f_{W}(\vec{K}) = f_{A}(\vec{K})$, that is the itinerant nature will not be seen by elastic scattering \cite{Marshall&Lovesey}. In the case of 4d, 5d and 6d itinerant electrons, strong deviations from any ionic scalar form factor are generally found out even in the elastic scattering, signaling that the atomic integrity of these electrons then gets destroyed. 

\section{\label{sec:Experiment}Experimental Polarization Ratio Method}

Neutron experiments give access to the differential scattering cross section. This is given by the square modulus of the scattering amplitude $\mathcal{A}[ \{(\vec{k_{i}} \nu_{i}) \lambda_{i}\} \rightsquigarrow \{(\vec{k_{f}} \nu_{f}) \lambda_{f})\}]$, statistically averaged over the initial states of the sample and spin states of the incoming neutron and summed over all the final states of the sample and spin states of the outgoing neutron,
\begin{equation}
\frac{\partial^{2}}{\partial \Omega \partial E_{f}} = \frac{k_{f}}{k_{i}} \sum_{\lambda_{i}, \nu_{i}} p_{\lambda_{i}} p_{\nu_{i}} \sum_{\lambda_{f}, \nu_{f}}  \left \vert \mathcal{A} \left[ \{(\vec{k_{i}} \nu_{i}) \lambda_{i}\} \rightsquigarrow \{(\vec{k_{f}} \nu_{f}) \lambda_{f})\} \right] \right \vert^{2} \delta \left[ \frac{\hbar^{2} (k_{i}^{2}-k_{f}^{2})}{2m_{\mathrm n}} + E_{i}-E_{f} \right]
\end{equation}
where the Dirac delta generalized function expresses the constraint of energy conservation. The statistical weight $p_{\lambda_{i}}$ of the initial states $\vert \lambda_{i} \rangle$ of the sample is generally given by the Boltzman factor 
\begin{equation}
p_{\lambda_{i}} = \frac{\exp\{-E_{i}/k_{B}T\}}{\sum_{E_{i}}\exp\{-E_{i}/k_{B}T\}}
\end{equation}
The statistical weights $p_{\nu_{i}}$ of the spin states $\vert \nu_{i} \rangle$ of the incoming neutron depends on the incoming neutron beam polarization. This is evidenced by building up the density operator 
\begin{equation}
\boldsymbol{\rho} = \sum_{\nu_{i}} \vert \nu_{i} \rangle p_{\nu_{i}} \langle \nu_{i} \vert
\end{equation}
Its matrix representative over basis vectors in the neutron spin state space is a $2\times2$ non singular matrix. Any such matrix is a linear combination of the unit $2\times2$ matrix and the Pauli matrices (cf.~eq.~\eqref{eq:pma}), which altogether form a basis of the vector space of non singular $2\times2$ matrices. Accordingly, the density operator may write $\boldsymbol{\rho} = u \mathbf{1} + \vec{v} \cdot \vec{\boldsymbol{\sigma}}$, where $\mathbf{1}$ is the unit operator in the neutron spin space and $\vec{\boldsymbol{\sigma}}$ the Pauli vector operator. The Cartesian components $\boldsymbol{\sigma}_{m} ~ (m = x, y, z)$ of $\vec{\boldsymbol{\sigma}}$ satisfy the commutation relations 
\begin{center}
$[~\boldsymbol{\sigma}_{m}, ~\boldsymbol{\sigma}_{n}~] = i ~ \epsilon^{mnl} \boldsymbol{\sigma}_{l}$, 
\end{center}
where $\epsilon^{mnl} = 1$ if $(m, n, l)$ is co-cyclic to $(x, y, z)$, $= -1$ if $(m, n, l)$ is anti-cyclic to $(x, y, z)$ and $= 0$ otherwise, (cf.~eq.~\eqref{eq:com}) and the anticommutation relations
\begin{center}
$\boldsymbol{\sigma}_{m} \boldsymbol{\sigma}_{n} + \boldsymbol{\sigma}_{n} \boldsymbol{\sigma}_{m} = 2 \delta_{m,n} \mathbf{1}$. 
\end{center}
Combining these we may also write
$\boldsymbol{\sigma}_{m} \boldsymbol{\sigma}_{n} = \delta_{m, n} \mathbf{1}  + i \sum_{l} \epsilon^{mnl} \boldsymbol{\sigma}_{l}$, which, since $Tr[\boldsymbol{\sigma}_{m}] = 0$,  straightforwardly gives 
\begin{center}
$Tr[\boldsymbol{\sigma}_{m} \boldsymbol{\sigma}_{n}] = 2 \delta_{m,n}$ and $Tr[\boldsymbol{\sigma}_{m} \boldsymbol{\sigma}_{n} \boldsymbol{\sigma}_{l}] = 2i~\epsilon^{mnl}$.
\end{center}
Accordingly, $u = \frac{Tr[\rho]}{2}$ and $\vec{v} = \frac{Tr[\rho]}{2}Tr[\boldsymbol{\rho} \vec{\boldsymbol{\sigma}}]$, but $Tr[\rho] = \sum_{\nu_{i}} p_{\nu_{i}} = 1$ then $u = \frac{1}{2}$ and $\vec{v} = \frac{1}{2}Tr[\boldsymbol{\rho} \vec{\boldsymbol{\sigma}}]$, which precisely defines the polarization $\vec{\mathrm P}$ of the incident neutron beam. Thus
\begin{equation}
\boldsymbol{\rho} = \frac{1}{2} \mathbf{1} + \vec{\mathrm P} \cdot \vec{\boldsymbol{\sigma}}
\end{equation}
The scattering amplitude, in view of the expressions of the different contributions to the neutron-sample interaction potential (cf.~eq.~\eqref{eq:nucpot1}, eq.~\eqref{eq:magpot1} and eq.~\eqref{eq:magpot3}),  can always be put in the form
\begin{equation}
\mathcal{A} \left[ \{(\vec{k_{i}} \nu_{i}) \lambda_{i}\} \rightsquigarrow \{(\vec{k_{f}} \nu_{f}) \lambda_{f})\} \right] = \langle \nu_{f} ~\vert \alpha(\vec{\varkappa})_{f, i} \mathbf{~1}+ \vec{\beta}(\vec{\varkappa})_{f, i} \cdot \vec{\boldsymbol{\sigma}}~ \vert \nu_{i} \rangle
\end{equation}
where the quantities $\alpha(\vec{\varkappa})_{f, i}$ and $\vec{\beta}(\vec{\varkappa})_{f, i}$ are matrix elements of a scalar operator $\boldsymbol{\alpha}(\vec{\varkappa})$ and a vector operator $\vec{\boldsymbol{\beta}}(\vec{\varkappa})$, which refer solely to the sample. Symbolizing $\boldsymbol{\alpha}(\vec{\varkappa}) \mathbf{1}  + \vec{\boldsymbol{\beta}}(\vec{\varkappa}) \cdot \vec{\boldsymbol{\sigma}}$ by $\boldsymbol{\mathcal{O}}$, it follows that
\begin{multline}
\sum_{\nu_{i}} p_{\nu_{i}} \sum_{\nu_{f}}  \left \vert \mathcal{A} \left[ \{(\vec{k_{i}} \nu_{i}) \lambda_{i}\} \rightsquigarrow \{(\vec{k_{f}} \nu_{f}) \lambda_{f})\} \right] \right \vert^{2} =  \sum_{\nu_{f}, \nu_{i}} p_{\nu_{i}} \langle \nu_{i} \vert \boldsymbol{\mathcal{O}}^{+} \vert \nu_{f} \rangle \langle \nu_{f} \vert \boldsymbol{\mathcal{O}} \vert \nu_{i} \rangle = {}
\\
= \sum_{\nu_{f}} \langle \nu_{f} \vert \boldsymbol{\mathcal{O}} \left(\sum_{\nu_{i}} \vert \nu_{i} \rangle p_{\nu_{i}} \langle \nu_{i} \vert \right) \boldsymbol{\mathcal{O}}^{+} \vert \nu_{f} \rangle = Tr[\boldsymbol{\rho}~\boldsymbol{\mathcal{O}}^{+}\boldsymbol{\mathcal{O}}]
\end{multline}
but, since $Tr[\boldsymbol{\sigma}_{m}] = 0$, $Tr[\boldsymbol{\sigma}_{m} \boldsymbol{\sigma}_{n}] = 2 \delta_{m,n}$ and $Tr[\boldsymbol{\sigma}_{m} \boldsymbol{\sigma}_{n} \boldsymbol{\sigma}_{l}] = 2i~\epsilon^{mnl}$,
\begin{multline}
Tr[\boldsymbol{\rho}~\boldsymbol{\mathcal{O}}^{+}\boldsymbol{\mathcal{O}}] = Tr[\boldsymbol{\rho}~ \big\{\boldsymbol{\alpha}(\vec{\varkappa}) \mathbf{1} + \vec{\boldsymbol{\beta}}(\vec{\varkappa}) \cdot \vec{\boldsymbol{\sigma}} \big\}^{+} \big\{\boldsymbol{\alpha}(\vec{\varkappa}) \mathbf{1} + \vec{\boldsymbol{\beta}}(\vec{\varkappa}) \cdot \vec{\boldsymbol{\sigma}}\big\}] =  \boldsymbol{\alpha}^{+}(\vec{\varkappa})\boldsymbol{\alpha}(\vec{\varkappa}) + {}
\\
+ \vec{\boldsymbol{\beta}}^{+}(\vec{\varkappa}) \cdot \vec{\boldsymbol{\beta}}(\vec{\varkappa}) + \boldsymbol{\alpha}^{+}(\vec{\varkappa})(\vec{\mathrm P}\cdot \vec{\boldsymbol{\beta}}(\vec{\varkappa})) + (\vec{\mathrm P}\cdot \vec{\boldsymbol{\beta}}^{+}(\vec{\varkappa})) \boldsymbol{\alpha}(\vec{\varkappa}) + i~\vec{\mathrm P} \cdot (\vec{\boldsymbol{\beta}}^{+}(\vec{\varkappa}) \wedge \vec{\boldsymbol{\beta}}(\vec{\varkappa}))
\end{multline}
by understanding that the trace is to be taken over the neutron spin states only. Using the identity 
\begin{equation}
\delta \left[ \hbar\omega + E_{i}-E_{f} \right] = \frac{1}{2\pi\hbar\omega} \int_{-\infty}^{\infty} dt~\exp\{ -i~\omega t \} \exp\{ -i~\frac{(E_{i}-E_{f})}{\hbar}\}
\end{equation}
(cf.~eq.~\eqref{eq:FTC}), where $\hbar\omega = \frac{\hbar^{2} (k_{i}^{2}-k_{f}^{2})}{2m_{\mathrm n}}$, it finally is inferred that
\begin{multline}
\frac{\partial^{2}}{\partial \Omega \partial E_{f}} = \frac{k_{f}}{k_{i}} \int_{-\infty}^{\infty} dt~\exp\{ -i~\omega t \} \bigg\{ \left \langle \boldsymbol{\alpha}^{+}(\vec{\varkappa})\boldsymbol{\alpha}(\vec{\varkappa}, t) \right \rangle + \left \langle \vec{\boldsymbol{\beta}}^{+}(\vec{\varkappa}) \cdot \vec{\boldsymbol{\beta}}(\vec{\varkappa}, t) \right \rangle + {}
\\
+ \left \langle \boldsymbol{\alpha}^{+}(\vec{\varkappa})\big(\vec{\mathrm P}\cdot \vec{\boldsymbol{\beta}}(\vec{\varkappa}, t)\big) \right \rangle +\left \langle \big(\vec{\mathrm P}\cdot \vec{\boldsymbol{\beta}}^{+}(\vec{\varkappa})\big) \boldsymbol{\alpha}(\vec{\varkappa}, t) \right \rangle + i~\vec{\mathrm P} \cdot \left \langle\big(\vec{\boldsymbol{\beta}}^{+}(\vec{\varkappa}) \wedge \vec{\boldsymbol{\beta}}(\vec{\varkappa}, t) \big)\right \rangle \bigg\}
\end{multline}
by defining $\boldsymbol{\alpha}(\vec{\varkappa}, t) = \exp\left\{\frac{i~t\mathcal{H}}{\hbar} \right\} \boldsymbol{\alpha}(\vec{\varkappa})\exp\left\{\frac{-i~t\mathcal{H}}{\hbar} \right\}$ and $ \vec{\boldsymbol{\beta}}(\vec{\varkappa}, t) = \exp\left\{\frac{i~t\mathcal{H}}{\hbar} \right\} \vec{\boldsymbol{\beta}}(\vec{\varkappa}) \exp\left\{\frac{-i~t\mathcal{H}}{\hbar} \right\}$, where
$\mathcal{H}$ is the hamiltonian of the sample, and using the notation $\langle \cdots \rangle$ for the statistical average over the initial states of the sample (summation over the final states is made by closure).\medskip

The cross section for elastic scattering is obtained by taking the infinite time limit and that for coherent scattering by averaging over the nuclear isotope distributions and nuclear spin orientations. $\langle \boldsymbol{\alpha}^{+}(\vec{\varkappa}) \boldsymbol{\alpha}(\vec{\varkappa}, \infty) \rangle = \langle \boldsymbol{\alpha}^{+}(\vec{\varkappa}) \rangle ~ \langle \boldsymbol{\alpha}(\vec{\varkappa}, \infty) \rangle, ~ \cdots$, since processes well separated in time get uncorrelated while $\langle \boldsymbol{\alpha}(\vec{\varkappa}, \infty) \rangle = \langle \boldsymbol{\alpha}(\vec{\varkappa}) \rangle$ and $\langle \vec{\boldsymbol{\beta}}(\vec{\varkappa}, \infty) \rangle = \langle \vec{\boldsymbol{\beta}}(\vec{\varkappa}) \rangle$ under stationary conditions. If there is no net nuclear spin polarization, which amounts to assume that the temperature is not low enough to allow for nuclear spin order or that the hyperfine field is sufficiently weak so that the electronic magnetism does not polarize the nuclear spins or else if a magnetic field is present that this will not polarize the spins too importantly, then the nuclear scattering will contribute only to $\langle \boldsymbol{\alpha}(\vec{\varkappa}) \rangle$ by the interaction term A$_{\mathrm p}$ independent on the nuclear spins, averaged over the nuclear isotopes (cf.~eq.~\eqref{eq:nucpot1}). In the case of a crystal with nuclei $d$ at positions $\vec{r}_{d} + \vec{R}$, where the vectors $\vec{r}_{d}$ locate the nuclei in the unit cell and the vectors $\vec{R} = \sum_{j} n_{j} \vec{a}_{j}$, $(n_{j} \in \mathbb{Z})$ define the lattice translations, we get
\begin{equation}
\langle \boldsymbol{\alpha}(\vec{\varkappa}) \rangle = \sum_{\vec{R}} \left( \sum_{\mathrm d} {\mathrm A}_{\mathrm d} \exp\{i~ \vec{\varkappa} \cdot \vec{r}_{d} \} \right) \exp\{i~ \vec{\varkappa}  \cdot \vec{R} \} = \sum_{\vec{K}} F_{N}(\vec{K}) ~ \delta(\vec{\varkappa} - \vec{K})
\end{equation}
where $\vec{K}$ is a reciprocal lattice vector and $F_{N}(\vec{K})$ defines the unit cell nuclear structure factor. The magnetic contribution to the elastic scattering, according to the eq.~\eqref{eq:msf1} and the eq.~\eqref{eq:msf2}, is given by 
\begin{equation}
\langle \vec{\boldsymbol{\beta}}(\vec{\varkappa}) \rangle = -\gamma_{\mathrm n} ~ \mathrm r_{0} \frac{1}{2\mu_{\mathrm B} \varkappa^{2}} ~ \vec{\varkappa} \wedge \sum_{\lambda_{i}} p_{\lambda_{i}} \left \langle \lambda_{i} \left \vert ~ \int_{\Omega} ~ \vec{\mathbf{M}}(\vec{r}) \exp\{i (\vec{\varkappa} \cdot \vec{r})\}  ~ d\vec{r} ~ \right \vert \lambda_{i} \right \rangle \wedge \vec{\varkappa} 
\end{equation}
In the case of a crystal with spatially periodic magnetic density,
\begin{equation}
\sum_{\lambda_{i}} p_{\lambda_{i}} \left \langle \lambda_{i} \left \vert ~ \vec{\mathbf{M}}(\vec{r}) ~ \right \vert \lambda_{i} \right \rangle = \sum_{\vec{R}} \sum_{\vec{\tau}} U(\vec{\tau}) \exp\{i~ \vec{\tau} \cdot \vec{R} \} \sum_{\lambda_{i}} p_{\lambda_{i}} \left \langle \lambda_{i} \left \vert ~ \vec{\mathbf{M}}(\vec{r} - \vec{R}) ~ \right \vert \lambda_{i} \right \rangle
\end{equation}
we get 
\begin{equation}
\langle \vec{\boldsymbol{\beta}}(\vec{\varkappa}) \rangle = \sum_{\vec{K}} \sum_{\vec{\tau}} U(\vec{\tau}) ~\delta(\vec{\varkappa} - \vec{K} + \vec{\tau}) ~ \left( \vec{\varkappa} \wedge \vec{F}_{M}(\vec{K} - \vec{\tau}) \wedge \vec{\varkappa} \right)
\end{equation}
where
\begin{equation}
\vec{F}_{M}(\vec{\varkappa}) = -\gamma_{\mathrm n} ~ \mathrm r_{0} \frac{1}{2\mu_{\mathrm B}}  \sum_{\lambda_{i}} p_{\lambda_{i}} \left \langle \lambda_{i} \left \vert ~ \int_{\mathrm {U.C.}} ~ \vec{\mathbf{M}}(\vec{r}) \exp\{i (\vec{\varkappa} \cdot \vec{r})\}  ~ d\vec{r} ~ \right \vert \lambda_{i} \right \rangle
\end{equation}
defines the unit cell (U.C.) magnetic structure factor. If the magnetic periodicity is identical to the nuclear, $U(\vec{\tau}) \equiv \delta(\vec{\tau})$, then the coherent elastic scattering cross section reads
\begin{multline}
\frac{\partial\sigma}{\partial\Omega} = \Big\{F_{N}^{\ast}(\vec{K}) F_{N}(\vec{K}) + \vec{F}_{M}^{\bot \ast}(\vec{K}) \cdot \vec{F}_{M}^{\bot}(\vec{K}) + F_{N}^{\ast}(\vec{K}) (\vec{\mathrm P} \cdot \vec{F}_{M}^{\bot}(\vec{K})) + {} 
\\
+ (\vec{\mathrm P} \cdot \vec{F}_{M}^{\bot}(\vec{K}))^{\ast} F_{N}(\vec{K})  + i~\vec{\mathrm P} \cdot ( \vec{F}_{M}^{\bot \ast}(\vec{K}) \wedge \vec{F}_{M}^{\bot}(\vec{K})) \Big\} ~ \delta(\vec{\varkappa} - \vec{K})
\end{multline}
where $\vec{F}_{M}^{\bot}(\vec{K}) =  \vec{K} \wedge \vec{F}_{M}(\vec{K}) \wedge \vec{K}$. By Fourier inverting $\vec{F}_{M}(\vec{K})$, we find
\begin{equation}
\sum_{\lambda_{i}} p_{\lambda_{i}} \left \langle \lambda_{i} \left \vert ~ \vec{\mathbf{M}}(\vec{r})  ~ \right \vert \lambda_{i} \right \rangle = \frac{2\mu_{\mathrm B}}{-\gamma_{\mathrm n} ~ \mathrm r_{0}} \frac{1}{\mathrm{V_{U.C.}}} \sum_{\vec{K}} \vec{F}_{M}(\vec{K}) \exp\{-i (\vec{K} \cdot \vec{r})\}
\end{equation}
where V$_{\mathrm {U.C.}}$ is the unit cell volume, that is, up to gauge invariance, the magnetic density is fully determined from the knowledge of the scattering amplitude at the reciprocal lattice points only. \medskip

\noindent The polarization dependent magnetic chiral contribution $i~\vec{\mathrm P} \cdot ( \vec{F}_{M}^{\bot \ast}(\vec{K}) \wedge \vec{F}_{M}^{\bot}(\vec{K}))$ to the coherent elastic scattering cross section should be expected only for non collinear magnetic density without center of symmetry. If this case is excluded and if the $F_{N}(\vec{K})$ can be accurately determined from complementary measurements, for instance in the paramagnetic phase without magnetic field or by X-rays scattering, then the polarization dependent nuclear-magnetic interference contribution $\vec{\mathrm P} \cdot [ F_{N}^{\ast}(\vec{K}) \vec{F}_{M}^{\bot}(\vec{K}) + F_{N}(\vec{K}) \vec{F}_{M}^{\bot}(\vec{K})^{\ast} ]$ can be used to allow for precision measurements of the $ \vec{F}_{M}^{\bot}(\vec{K})$ and thus of form factors. It indeed is an evidence that weak $\vec{F}_{M}^{\bot}(\vec{K})$, amplified by $F_{N}(\vec{K})$, will much more accurately be detected than the non chiral magnetic contribution $\vec{F}_{M}^{\bot \ast}(\vec{K}) \cdot \vec{F}_{M}^{\bot}(\vec{K})$, especially at large scattering angle where the form factors tend to vanish out and where extreme sensitivity is often required to get insights about fine details of the magnetic density. A necessary condition of course is that the interference contribution itself does not cancel out, which is the case if $F_{N}(\vec{K})$ and $\vec{F}_{M}^{\bot}(\vec{K})$ are not in phase quadrature. The method applies to polarized paramagnets, to ferromagnets or ferrimagnets and to in-cell antiferromagnets with in-phase unit cell magnetic and nuclear structure factors.
\medskip

\noindent Experimentally, the so-called \emph{polarization ratios} $R(\vec{K})$ between the scattered intensities for two opposite initial polarizations $+\vec{\mathrm P}$ and $-\vec{\mathrm P}$ of the incident neutron beam are measured on a series of reciprocal lattice points $\vec{K}$ (Bragg scattering). The extraction from these data of the $\vec{F}_{M}^{\bot}(\vec{K})$ in general is not straightforward, for these are complex vector quantities, the direction, modulus and phase of which have to be determined, when only one real scalar quantity, $R(\vec{K})$, is measured. Models then must be worked out providing with calculated $\vec{F}_{M}(\vec{K})$ and, from these, calculated $R(\vec{K})$, to directly confront with the experimental values. A direct access to the $\vec{F}_{M}(\vec{K})$ nevertheless is possible in collinear magnets with a centric crystal structure. In this case, both $F_{N}(\vec{K})$ and $\vec{F}_{M}(\vec{K})$ are real. If furthermore the polarization $+\vec{\mathrm P}$ is set parallel to the magnetization vector then 
\begin{equation}
R(\vec{K}) = \frac{1 + 2 \sin^{2}\alpha ~ \gamma(\vec{K}) + \sin^{2}\alpha ~ (\gamma(\vec{K}))^{2}}{1 - 2 \sin^{2}\alpha ~ \gamma(\vec{K}) + \sin^{2}\alpha ~ (\gamma(\vec{K}))^{2}}, \qquad \gamma(\vec{K}) = \frac{F_{N}(\vec{K})}{F_{M}(\vec{K})}
\end{equation}
where $\alpha$ is the angle between the magnetization vector and the scattering vector $\vec{K}$. This is easily solved for $F_{M}(\vec{K})$, which, it should be emphasized, is not the modulus of $\vec{F}_{M}(\vec{K})$ but its algebraic value along the magnetization vector. In fact, we get two mathematical solutions, but, in practice, it most often is evident to recognize which of the two is physical and which is unphysical. The actual measurements are performed under strong magnetic fields, which in the case of paramagnets allows inducing the largest magnetization and overcome eventual significant magnetic anisotropy and in the case of a sample with a net global magnetic moment allows avoiding depolarization by selecting a single magnetic domain and orient this along the magnetic field direction. The experimental corrections to be considered are essentially extinction corrections, associated with amplitude coupling within perfect blocks (primary extinction) and between perfect blocks (secondary extinction), which, increasing proportionally to $\vert\vec{k}_{i}\vert^{-3}$, is accurately estimated from measurements at different incident neutron wavelength $2\pi \vert\vec{k}_{i}\vert^{-1}$. Corrections which are identical for the two incident neutron polarization (absorption, $\cdots$) are irrelevant and using filters the fractional wavelength contamination can be removed. Almost perfect incident neutron beam polarization can be produced whereas depolarization by the sample can be minimized by using samples elongated along the magnetization axis and cleanly polishing the surfaces. No description of specific experiments will be given in these notes, but overviews of investigations can be found in the literature \cite{Ressouche}. Also of interest to consult is the methodology inspired from the spherical neutron polarimetry to measure the form factor in in-cell antiferromagnets with magnetic and nuclear structure factors in phase quadrature \cite{BrownFT}.

\section{\label{sec:Algebra}Algebra of Ionic Form Factors}

An in-depth quantitative analysis is provided in this section of the magnetic vector form factor of atomic electrons, more precisely of the expression between the braces in the eq.~\eqref{eq:msf4}. Calling to mind that $\vec{\varkappa} \wedge [\vec{\mathbf{p}}, f(\vec{\mathbf{r}})] = \vec{\varkappa} \wedge \{ -i \hbar~ \vec{\nabla}_{\vec{\mathbf{r}}} f(\vec{\mathbf{r}}) \}$ is null for $f(\vec{\mathbf{r}}) =  \exp\{i (\vec{\varkappa} \cdot \vec{\mathbf{r}})\}$, this also writes
\begin{multline}
\frac{1}{\varkappa^{2}} ~ \vec{\varkappa} \wedge \vec{\mathcal{F}}_{\lambda_{f}^{\mathrm p}, \lambda_{i}^{\mathrm p}}(\vec{\varkappa}) \wedge \vec{\varkappa} = {} \\ = \left \langle \lambda_{f}^{\mathrm p} \left \vert ~ \sum_{j \in \mathbb{A}_{\mathrm p}} \left[ \exp\{i (\vec{\varkappa} \cdot \vec{\mathbf{r}}_{j_{\mathrm p}})\} ~  \left ( \frac{1}{\varkappa^{2}} ~ \vec{\varkappa} \wedge \vec{\mathbf{s}}_{j}  \wedge \vec{\varkappa} - \frac{i}{\hbar \varkappa^{2}} \vec{\varkappa} \wedge \vec{\mathbf{p}}_{j} \right )\right] ~ \right \vert \lambda_{i}^{\mathrm p} \right \rangle
\end{multline}
which distinguishes the contribution of the spin magnetic moments from the contribution of the convection current density. $\vec{\mathbf{r}}_{j_{\mathrm p}} = \vec{\mathbf{r}}_{j} - \vec{\mathbf{R}}_{\mathrm p}$ is the position operator of an electron $j$ of $\mathbb{A}_{\mathrm p}$ with respect to the position of the nucleus of $\mathbb{A}_{\mathrm p}$. Using the spherical tensor operator formalism \cite{Judd, Edmonds}, this quantity can be computed exactly \cite{Marshall&Lovesey, Stassis&Deckman, Lovesey&Rimmer, Lovesey, Balcar&Lovesey}. The method basically is not complex but the associated algebra is unwieldy. A fortunate fact is that we have to deal only with matrix elements of sums of one-electron operators, which ultimately  can be deduced from the matrix elements of these operators between uncoupled single-electron states $\vert \eta_{f, i}^{\mathrm p} \rangle = \vert (n lm_{l}sm_{s})_{f, i}^{\mathrm p} \rangle$.

\subsection{\label{sec:SingleElectronSpinFactor}Single Electron Spin Vector form Factor}

Let us focus our attention on the \emph{spin vector form factor}, which is associated with the operator $\sum_{j \in \mathbb{A}_{\mathrm p}} \exp\{i (\vec{\varkappa} \cdot \vec{\mathbf{r}}_{j_{\mathrm p}})\} ~ \vec{\mathbf{s}}_{j}$ and examine the matrix element between uncoupled single-electron states
\begin{equation}
\label{eq:msfuses1}
\vec{\mathfrak{S}}_{\eta_{f}^{\mathrm p}, \eta_{i}^{\mathrm p}}(\vec{\varkappa}) = \left \langle (n lm_{l}sm_{s})_{f}^{\mathrm p} \left \vert ~ \exp\{i (\vec{\varkappa} \cdot \vec{\mathbf{r}}_{j_{\mathrm p}})\} ~ \vec{\mathbf{s}}_{j} ~ \right \vert (n lm_{l}sm_{s})_{i}^{\mathrm p} \right \rangle
\end{equation}
We shall temporarily drop the indexes $p$ and $j$ for notation convenience (notice that the index $j$ is unnecessary for equivalent electrons since these cannot be distinguished from each other). The spatial wavefunction associated with $\vert \eta \rangle = \vert n lm_{l}sm_{s} \rangle$ will be denoted $\langle \vec{r}~ \vert~ n lm_{l}sm_{s} \rangle = \langle r~\Omega_{\vec{r}}~ \vert ~n lm_{l}sm_{s} \rangle = R_{n l}(r) ~Y_{m}^{l}(\Omega_{\vec{r}}) ~\vert sm_{s} \rangle = R_{n l}(r) ~\vert lm_{l}sm_{s} \rangle$, where $R_{n l}(r)$ is the radial component and $Y_{m}^{l}(\Omega_{\vec{r}}) = \langle \Omega_{\vec{r}} ~\vert~ lm_{l} \rangle$ the angular component. Appropriately inserting the closure relation $\int  \vert \vec{r} \rangle \langle \vec{r} \vert ~ d\vec{r} = \int_{0}^{\infty} \vert r \rangle \langle r \vert ~ dr  \int_{S_{2}} \vert \Omega_{\vec{r}} \rangle \langle \Omega_{\vec{r}} \vert ~ d\Omega_{\vec{r}} = \mathbf{1}$ and using the multipole expansion
\begin{equation}
\exp(i~\vec{\varkappa} \cdot \vec{r}) = 4\pi \sum_{KQ} i^{K} j_{K}(\varkappa r) \left[Y_{Q}^{K}(\Omega_{\vec{\varkappa}})\right]^{\ast} Y_{Q}^{K}(\Omega_{\vec{r}})
\end{equation}
to separate the variables $\vec{\varkappa} = (\varkappa, \Omega_{\vec{\varkappa}}) = (\varkappa, \theta_{\vec{\varkappa}}, \varphi_{\vec{\varkappa}})$ and $\vec{r} = (r, \Omega_{\vec{r}}) = (r, \theta_{\vec{r}}, \varphi_{\vec{r}})$ (cf.~eq.~\eqref{eq:ME2}) the spherical components of $\vec{\mathfrak{S}}_{\eta_{f}, \eta_{i}}(\vec{\varkappa})$ can be expressed in the form
\begin{equation}
\mathfrak{S}_{\eta_{f}, \eta_{i}}^{q}(\vec{\varkappa}) = 4 \pi \sum_{KQ} i^{K} \langle  j_{K}(\varkappa) \rangle_{f, i} \left[Y_{Q}^{K}(\Omega_{\vec{\varkappa}})\right]^{\ast} \left \langle (lm_{l}sm_{s})_{f} \left \vert ~ \mathbf{Y}_{Q}^{K} ~\mathbf{s}_{q} ~ \right \vert (lm_{l}sm_{s})_{i} \right \rangle
\end{equation}
where
\begin{equation}
\langle  j_{K}(\varkappa) \rangle_{f, i} = \int_{0}^{\infty} r^{2} ~R_{n l}^{f}(r) ~R_{n l}^{i}(r) ~j_{K}(\varkappa r)~ dr
\end{equation}
are dubbed index K \emph{radial integrals}.
The $2K+1$ operators $\mathbf{Y}_{Q}^{K}$ are the components of the spherical harmonic tensor operator $\mathbf{Y}_{K}$ of order $K$. $\langle \Theta_{\vec{r}} \vert ~ \mathbf{Y}_{Q}^{K} ~ \vert \Omega_{\vec{r}} \rangle \equiv \langle \Theta_{\vec{r}} \vert ~ Y_{Q}^{K}(\mathbf{\Omega}_{\vec{r}}) ~ \vert \Omega_{\vec{r}} \rangle= \delta_{\Theta_{\vec{r}}, \Omega_{\vec{r}}}  Y_{Q}^{K}(\Omega_{\vec{r}})$ by definition. ${\bf s}_{-1} = \frac{1}{\sqrt{2}}({\bf s}_{x} - i~{\bf s}_{y})$,  ${\bf s}_{0} = {\bf s}_{z}$ and ${\bf s}_{+1} = -\frac{1}{\sqrt{2}}({\bf s}_{x} + i~{\bf s}_{y})$ are the spherical components of the electron spin vector operator $\vec{\mathbf{s}}$. Applying the Wigner-Eckart theorem separately to $\mathbf{Y}_{K}$ and to $\vec{\mathbf{s}}$, 
\begin{equation}
\begin{split}
\left \langle (lm_{l}sm_{s})_{f} \left \vert  ~ \mathbf{Y}_{Q}^{K} ~\mathbf{s}_{q}~ \right \vert (lm_{l}sm_{s})_{i} \right \rangle = (-)^{l_{f}-m_{l_{f}}} \left ( \begin{array}{ccc}l_{f} & K & l_{i} \\-m_{l_{f}} & Q & m_{l_{i}}\end{array} \right )  \left ( l_{f} ~\Vert~ \mathbf{Y}_{K} ~\Vert~l_{i} \right ) \times {}
\\
\times (-)^{s_{f}-m_{s_{f}}} \left ( \begin{array}{ccc}s_{f} & 1 & s_{i} \\-m_{s_{f}} & q & m_{s_{i}}\end{array} \right )\left ( s_{f} ~\Vert~ \mathrm{{\bf s}} ~\Vert~s_{i} \right )
\end{split}
\end{equation}
(cf.~Appendix~\ref{WET}) with the help of the eq.~\eqref{eq:rms} and the eq.~\eqref{eq:rmy} for the reduced matrix elements, we get
\begin{equation}
\label{eq:msfuses2}
\begin{split}
\mathfrak{S}_{\eta_{f}, \eta_{i}}^{q}(\vec{\varkappa}) & = (6 \pi)^{\frac{1}{2}} \sum_{KQ} i^{K} \langle  j_{K}(\varkappa) \rangle_{f, i} \left[Y_{Q}^{K}(\Omega_{\vec{\varkappa}})\right]^{\ast} (-)^{s_{f}-m_{s_{f}}+m_{l_{f}}} \delta_{s_{f}, s_{i}} \left[ l_{f}, K, l_{i} \right] \times
\\
& \qquad \qquad \qquad \times \left ( \begin{array}{ccc}l_{f} & K & l_{i} \\0 & 0 & 0\end{array} \right ) \left ( \begin{array}{ccc}l_{f} & K & l_{i} \\-m_{l_{f}} & Q & m_{l_{i}}\end{array} \right ) \left ( \begin{array}{ccc}s_{f} & 1 & s_{i} \\-m_{s_{f}} & q & m_{s_{i}}\end{array} \right )
\end{split}
\end{equation}
where $\left[ l_{1}^{\epsilon_{1}},  l_{2}^{\epsilon_{2}}, \cdots \right]$ is a standard abbreviation for $\left \{ (2 l_{1}+1)^{\epsilon_{1}}(2 l_{2}+1)^{\epsilon_{2}}\cdots \right \}^{\frac{1}{2}}$. $\mathfrak{S}_{\eta_{f}, \eta_{i}}^{q}(\vec{\varkappa})$ is null unless $K$ satisfies the symmetry condition $l_{f} + K + l_{i} =$ even integer on the 3jm symbol with all m null (cf.~eq.~\eqref{eq:3s3}) and the triangular condition $K = l_{f} + l_{i}, l_{f} + l_{i} - 1, \cdots, \vert~ l_{f} - l_{i} ~\vert$.\medskip

\noindent Owing to the behavior of the spherical Bessel functions for small arguments (cf.~eq.~\eqref{eq:ME3}) it is inferred that for $\vec{\varkappa} \rightarrow 0$ the expansion of $\mathfrak{S}_{\eta_{f}, \eta_{i}}^{q}(\vec{\varkappa})$ over the spherical harmonics $[Y_{Q}^{K}(\Omega_{\vec{\varkappa}})]^{\ast}$ should be dominated by the $K=0,~Q=0$ contribution, that is,
\begin{equation*}
\text{since} \quad [Y_{0}^{0}(\Omega_{\vec{\varkappa}})]^{\ast} = (4\pi)^{-\frac{1}{2}} \quad \text{and} \quad \left ( \begin{array}{ccc}l_{f} & 0 & l_{i} \\-m_{l_{f}} & 0 & m_{l_{i}}\end{array} \right ) = (-)^{l_{i}-m_{l_{f}}} ~\delta_{l_{f}, l_{i}} ~\delta_{m_{l_{f}}, m_{l_{i}}} \left[l_{f}^{-1}\right],
\end{equation*}
\begin{equation}
\mathfrak{S}_{\eta_{f}, \eta_{i}}^{q}(\vec{\varkappa})_{\vec{\varkappa} \rightarrow 0} \approx \left \langle  j_{0}(\varkappa) \right \rangle_{f, i}~\delta_{l_{f}, l_{i}} ~\delta_{m_{l_{f}}, m_{l_{i}}} \left \langle (sm_{s})_{f} \left \vert ~\mathbf{s}_{q}~ \right \vert (sm_{s})_{i} \right \rangle
\end{equation} 
which defines the \emph{dipole approximation} to the spin vector form factor. Notice that this is null if $l_{f} \neq l_{i}$. Generalization to multi-electron ions is straightforward. \medskip

\noindent The spherical components $\mathfrak{S}_{\eta_{f}, \eta_{i}}^{q~\bot}(\vec{\varkappa})$ of the matrix element $\frac{1}{\varkappa^{2}}\vec{\varkappa} \wedge \vec{\mathfrak{S}}_{\eta_{f}^{\mathrm p}, \eta_{i}^{\mathrm p}}(\vec{\varkappa}) \wedge \vec{\varkappa}$ are given by 
\begin{equation}
\mathfrak{S}_{\eta_{f}, \eta_{i}}^{q~\bot}(\vec{\varkappa}) = \mathfrak{S}_{\eta_{f}, \eta_{i}}^{q}(\vec{\varkappa}) - \frac{1}{\varkappa^{2}} \varkappa_{q} \sum_{\bar{q}} (-)^{\bar{q}} \varkappa_{\bar{q}} \mathfrak{S}_{\eta_{f}, \eta_{i}}^{-\bar{q}}(\vec{\varkappa})
\end{equation}
where $\varkappa_{q}$ is the $q-$spherical component of the scattering vector. Calling to mind that  $Y_{0}^{1}(\Omega_{\vec{\varkappa}}) = (\frac{3}{4\pi})^{\frac{1}{2}} \cos \theta_{\vec{\varkappa}}$ and $Y_{\pm 1}^{1}(\Omega_{\vec{\varkappa}}) = \mp (\frac{3}{4\pi})^{\frac{1}{2}} (\frac{1}{2})^{\frac{1}{2}} \sin \theta_{\vec{\varkappa}} \exp\{\pm i \varphi_{\vec{\varkappa}}\}$, we may write $\varkappa_{q} = \varkappa ~ (\frac{4\pi}{3})^{\frac{1}{2}} Y_{q}^{1}(\Omega_{\vec{\varkappa}})$ whence
\begin{multline}
\frac{1}{\varkappa^{2}} \varkappa_{q} \sum_{\bar{q}} (-)^{\bar{q}} \varkappa_{\bar{q}} \mathfrak{S}_{\eta_{f}, \eta_{i}}^{-\bar{q}}(\vec{\varkappa}) = \left(\frac{4\pi}{3}\right) Y_{q}^{1}(\Omega_{\vec{\varkappa}}) \sum_{\bar{q}} (-)^{\bar{q}} Y_{\bar{q}}^{1}(\Omega_{\vec{\varkappa}}) \mathfrak{S}_{\eta_{f}, \eta_{i}}^{-\bar{q}}(\vec{\varkappa}) =  {} 
\\
= (4\pi)^{\frac{1}{2}} \sum_{\substack {\bar{q} \\ K, Q}} (5)^{\frac{1}{2}} (-)^{\bar{q}+Q}\left(\begin{array}{ccc}1 & 1 & K \\0 & 0 & 0\end{array}\right) \left(\begin{array}{ccc}1 & 1 & K \\q & \bar{q} & Q\end{array}\right) Y_{-Q}^{K}(\Omega_{\vec{\varkappa}}) \mathfrak{S}_{\eta_{f}, \eta_{i}}^{-\bar{q}}(\vec{\varkappa})
\end{multline}
(cf.~eq.~\eqref{eq:AM1}). The 3jm symbol with all m null is computed to $-\frac{1}{\sqrt{3}}$ for $K = 0$ and to $\frac{2}{\sqrt{15}}$ for $K = 2$ (cf.~eq.~\eqref{eq:AM0}). It is null by symmetry for $K = 1$ (cf.~eq.~\eqref{eq:3s3}) and by triangular conditions for any other $K$. It then follows, thanks to the eq.~\eqref{eq:3j0}, since $Y_{0}^{0}(\Omega_{\vec{\varkappa}}) = \left(\frac{1}{4\pi}\right)^{-\frac{1}{2}}$ and taking into account the zero-sum condition $q + \bar{q} + Q = 0$ on the second 3jm symbol, that
\begin{equation}
\label{eq:msfuses3}
\mathfrak{S}_{\eta_{f}, \eta_{i}}^{q~\bot}(\vec{\varkappa}) = \frac{2}{3}\mathfrak{S}_{\eta_{f}, \eta_{i}}^{q}(\vec{\varkappa}) + (-)^{1-q} \left(\frac{8\pi}{3} \right)^{\frac{1}{2}} \sum_{\bar{q}}  \left(\begin{array}{ccc}1 & 1 & 2 \\q & \bar{q} & -q-\bar{q}\end{array}\right) Y_{q+\bar{q}}^{2}(\Omega_{\vec{\varkappa}}) ~ \mathfrak{S}_{\eta_{f}, \eta_{i}}^{-\bar{q}} (\vec{\varkappa})
\end{equation}
Using for the 3jm symbol the algebraic form
\begin{equation}
\left(\begin{array}{ccc}1 & 1 & 2 \\q & \bar{q} & -q-\bar{q}\end{array}\right) = (-)^{-q-\bar{q}} \left[ \frac{(2+q+\bar{q})! (2-q-\bar{q})!}{30 (1-q)!(1+q)!(1-\bar{q})!(1+\bar{q})!} \right]^{\frac{1}{2}}\qquad 
\end{equation}
(cf.~eq.~\eqref{eq:3j2}) and for the spherical harmonics under concern the functional forms 
\begin{gather}
\label{eq:spharord2}
Y_{0}^{2}(\Omega_{\vec{\varkappa}}) = \left(\frac{5}{4\pi}\right)^{\frac{1}{2}} \left(1-\frac{3}{2} \sin^{2}\theta_{\vec{\varkappa}} \right)
\\ \nonumber
Y_{\pm 1}^{2}(\Omega_{\vec{\varkappa}}) = \mp \left(\frac{15}{8\pi}\right)^{\frac{1}{2}} \sin\theta_{\vec{\varkappa}}\cos\theta_{\vec{\varkappa}} ~ \exp\{\pm i \varphi_{\vec{\varkappa}}\} , \quad Y_{\pm 2}^{2}(\Omega_{\vec{\varkappa}}) = \left(\frac{15}{32\pi}\right)^{\frac{1}{2}} \sin^{2}\theta_{\vec{\varkappa}} ~ \exp\{\pm 2i \varphi_{\vec{\varkappa}}\}
\end{gather}
(cf.~eq.~\eqref{eq:MEB}), it finally is found that
\begin{gather}
\label{eq:msfuses4}
\mathfrak{S}_{\eta_{f}, \eta_{i}}^{0~\bot}(\vec{\varkappa}) = \sin^{2}\theta_{\vec{\varkappa}} ~ \mathfrak{S}_{\eta_{f}, \eta_{i}}^{0}(\vec{\varkappa}) + \frac{1}{\sqrt{2}} \sin\theta_{\vec{\varkappa}} \cos\theta_{\vec{\varkappa}} \left( e^{ i \varphi_{\vec{\varkappa}} } \mathfrak{S}_{\eta_{f}, \eta_{i}}^{1}(\vec{\varkappa}) - e^{ -i \varphi_{\vec{\varkappa}} } \mathfrak{S}_{\eta_{f}, \eta_{i}}^{-1}(\vec{\varkappa})\right)
\\ \nonumber
\mathfrak{S}_{\eta_{f}, \eta_{i}}^{\pm 1~\bot}(\vec{\varkappa}) = \left(1-\frac{1}{2}\sin^{2}\theta_{\vec{\varkappa}}\right) \mathfrak{S}_{\eta_{f}, \eta_{i}}^{\pm 1}(\vec{\varkappa}) 
\\ \nonumber
- \frac{1}{\sqrt{2}} \sin\theta_{\vec{\varkappa}} \cos\theta_{\vec{\varkappa}} ~ e^{ \pm i \varphi_{\vec{\varkappa}} } \mathfrak{S}_{\eta_{f}, \eta_{i}}^{0}(\vec{\varkappa}) + \frac{1}{2}  \sin^{2}\theta_{\vec{\varkappa}} ~ e^{\pm 2 i \varphi_{\vec{\varkappa}} } \mathfrak{S}_{\eta_{f}, \eta_{i}}^{\mp 1}(\vec{\varkappa})
\end{gather}

\subsection{\label{sec:SingleElectronOrbitalFactor}Single Electron Orbital Vector Form Factor}

No more difficulties arise when dealing with the \emph{orbital vector form factor} associated with the operator $\sum_{j \in \mathbb{A}_{\mathrm p}} \exp\{i (\vec{\varkappa} \cdot \vec{\mathbf{r}}_{j_{\mathrm p}})\} ~  ( -\frac{i}{\hbar \varkappa^{2}} \vec{\varkappa} \wedge \vec{\mathbf{p}}_{j} ) = \sum_{j \in \mathbb{A}_{\mathrm p}} [ -\frac{1}{\varkappa^{2}} \exp\{i (\vec{\varkappa} \cdot \vec{\mathbf{r}}_{j_{\mathrm p}})\} ~  ( \vec{\varkappa} \wedge \vec{\boldsymbol \nabla}_{\vec{\mathbf{r}}_{j_{\mathrm p}}} ) ]$ and, in particular, with the matrix element between uncoupled single-electron states
\begin{align}
\label{eq:msfuses5}
\vec{\mathfrak{L}}_{\eta_{f}^{\mathrm p}, \eta_{i}^{\mathrm p}}^{~\bot}(\vec{\varkappa}) & = \left \langle (n lm_{l}sm_{s})_{f}^{\mathrm p} \left \vert ~ \left[ -\frac{1}{\varkappa^{2}}~ \exp\{i (\vec{\varkappa} \cdot \vec{\mathbf{r}}_{j_{\mathrm p}})\} ~ \vec{\varkappa} \wedge \vec{\boldsymbol \nabla}_{\vec{\mathbf{r}}_{j_{\mathrm p}}} \right] ~ \right \vert (n lm_{l}sm_{s})_{i}^{\mathrm p} \right \rangle = {}
\nonumber \\
& =  -\delta_{(sm_{s})_{f}^{\mathrm p}, (sm_{s})_{i}^{\mathrm p}} \left \langle (n lm_{l})_{f}^{\mathrm p} \left \vert ~ \frac{1}{\varkappa^{2}}~ \exp\{i (\vec{\varkappa} \cdot \vec{\mathbf{r}}_{j_{\mathrm p}})\} ~ \vec{\varkappa} \wedge \vec{\boldsymbol \nabla}_{\vec{\mathbf{r}}_{j_{\mathrm p}}}~ \right \vert (n lm_{l})_{i}^{\mathrm p} \right \rangle
\end{align}
but the computations are a little lengthier. A detailed description of these when the states $\eta_{f}$ and $\eta_{i}$ belong to a same electronic configuration, that is for $(nl)_{f} \equiv (nl)_{i}$, is provided at several places in the literature \cite{Marshall&Lovesey, Lovesey&Rimmer, Lovesey, Balcar&Lovesey} with only faint differences in the formulations. Let us replicate one of the approaches and re-write the matrix element
\begin{gather}
\vec{\mathfrak{L}}_{m_{l_{f}}, m_{l_{i}}}^{~\bot}(\vec{\varkappa}) = -\left \langle nlm_{l_{f}} \left \vert ~ \frac{1}{\varkappa^{2}}~ \exp\{i (\vec{\varkappa} \cdot \vec{\mathbf{r}})\} ~ (\vec{\varkappa} \wedge \vec{\boldsymbol \nabla}_{\vec{\mathbf{r}}})~ \right \vert nlm_{l_{i}} \right \rangle
\\ \nonumber
= \frac{-1}{2\varkappa^{2}} \left \{ \left \langle nlm_{l_{f}} \left \vert \exp\{i (\vec{\varkappa} \cdot \vec{\mathbf{r}})\} ~ (\vec{\varkappa} \wedge \vec{\boldsymbol \nabla}_{\vec{\mathbf{r}}}) \right \vert nlm_{l_{i}} \right \rangle - \left \langle nlm_{l_{i}} \left \vert (\vec{\varkappa} \wedge \vec{\boldsymbol \nabla}_{\vec{\mathbf{r}}})~\exp\{-i (\vec{\varkappa} \cdot \vec{\mathbf{r}})\} \right \vert nlm_{l_{f}} \right \rangle^{\ast} \right \}
\\ \nonumber
= -\frac{1}{2\varkappa^{2}} \int d\vec{r}~ \exp\{i (\vec{\varkappa} \cdot \vec{\mathrm r})\} ~  R_{n l}^{2}(r)  ~ \times {}
\\ \nonumber
\times \left\{ [Y_{m_{l_{f}}}^{l}(\Omega_{\vec{r}})]^{\ast} \left ( (\vec{\varkappa} \wedge \vec{\nabla}_{\vec{\mathrm {r}}}) ~Y_{m_{l_{i}}}^{l}(\Omega_{\vec{r}}) \right ) - Y_{m_{l_{i}}}^{l}(\Omega_{\vec{r}}) \left ( (\vec{\varkappa} \wedge \vec{\nabla}_{\vec{\mathrm {r}}})~[Y_{m_{l_{f}}}^{l}(\Omega_{\vec{r}})]^{\ast} \right ) \right\}
\end{gather}
The indexes $p$ and $j$ are temporarily dropped and the spin quantum numbers temporarily ignored for notation convenience. The second equality gets obvious when calling to mind that $\forall \eta_{f}~ \forall \eta_{i}~ \forall\mathbf{O}~ \langle \eta_{f} \vert \mathbf{O} \vert \eta_{i} \rangle = \langle \eta_{i} \vert \mathbf{O}^{+} \vert \eta_{f} \rangle^{\ast}$ and $\forall \mathbf{O_{1}}~ \forall \mathbf{O_{2}}~ (\mathbf{O_{1}O_{2}})^{+} = \mathbf{O_{2}}^{+} \mathbf{O_{1}}^{+}$, where $\mathbf{O}^{+}$ symbolizes the adjoint of $\mathbf{O}$ and that $[ \exp\{i (\vec{\varkappa} \cdot \vec{\mathbf{r}})\} ~ \vec{\varkappa} \wedge \vec{\boldsymbol \nabla}_{\vec{\mathbf{r}}} ]^{+} = -\vec{\varkappa} \wedge \vec{\boldsymbol \nabla}_{\vec{\mathbf{r}}} ~ \exp\{-i (\vec{\varkappa} \cdot \vec{\mathbf{r}}) \}$, since $(\exp\{i (\vec{\varkappa} \cdot \vec{\mathbf{r}})\})^{+} = \exp\{-i (\vec{\varkappa} \cdot \vec{\mathbf{r}} \}$ and $(\vec{\boldsymbol \nabla}_{\vec{\mathbf{r}}})^{+} = -\vec{\boldsymbol \nabla}_{\vec{\mathbf{r}}}$. The third equality is derived by appropriate insertion of the closure relation $\int  \vert \vec{r} \rangle \langle \vec{r} \vert ~ d\vec{r} = \int_{0}^{\infty} \vert r \rangle \langle r \vert ~ dr  \int_{S_{2}} \vert \Omega_{\vec{r}} \rangle \langle \Omega_{\vec{r}} \vert ~ d\Omega_{\vec{r}} = \mathbf{1}$. Using the spherical components of the vector operator $\vec{\varkappa} \wedge \vec{\boldsymbol \nabla}_{\vec{\mathbf{r}}}$,
\begin{equation}
(\vec{\varkappa} \wedge \vec{\boldsymbol \nabla}_{\vec{\mathbf{r}}})_{q} = i (-)^{1+q}\sqrt{6} \sum_{q_{1}, q_{2}} \varkappa_{q_{1}} {\boldsymbol \nabla}_{q_{2}} \left(\begin{array}{ccc}1 & 1 & 1 \\q_{1} & q_{2} & -q\end{array}\right)
\end{equation}
(cf.~eq.~\eqref{eq:ST0}) and the multipole expansion 
\begin{equation}
\exp(i~\vec{\varkappa} \cdot \vec{r}) = 4\pi \sum_{KQ} i^{K} j_{K}(\varkappa r) \left[Y_{Q}^{K}(\Omega_{\vec{\varkappa}})\right]^{\ast} Y_{Q}^{K}(\Omega_{\vec{r}})
\end{equation} 
to separate the variables $\vec{\varkappa} = (\varkappa, \Omega_{\vec{\varkappa}}) = (\varkappa, \theta_{\vec{\varkappa}}, \varphi_{\vec{\varkappa}})$ and $\vec{r} = (r, \Omega_{\vec{r}}) = (r, \theta_{\vec{r}}, \varphi_{\vec{r}})$ (cf.~eq.~\eqref{eq:ME2}), the spherical components of $\vec{\mathfrak{L}}_{m_{l_{f}}, m_{l_{i}}}^{~\bot}(\vec{\varkappa})$ can be expressed in the form 
\begin{gather}
\mathfrak{L}_{m_{l_{f}}, m_{l_{i}}}^{q~\bot}(\vec{\varkappa}) = \frac{i (-)^{q}\sqrt{6} }{2\varkappa^{2}}  \sum_{q_{1}, q_{2}} \varkappa_{q_{1}} \left(\begin{array}{ccc}1 & 1 & 1 \\q_{1} & q_{2} & -q\end{array}\right) ~ 4\pi \sum_{KQ} i^{K} \left[Y_{Q}^{K}(\Omega_{\vec{\varkappa}})\right]^{\ast} \times {}
\\ \nonumber
\times \int d\vec{r}~ j_{K}(\varkappa r) ~ R_{n l}^{2}(r) ~ Y_{Q}^{K}(\Omega_{\vec{r}}) \left\{ [Y_{m_{l_{f}}}^{l}(\Omega_{\vec{r}})]^{\ast} \left ( \nabla_{q_{2}} ~Y_{m_{l_{i}}}^{l}(\Omega_{\vec{r}}) \right ) - Y_{m_{l_{i}}}^{l}(\Omega_{\vec{r}}) \left ( \nabla_{q_{2}}~[Y_{m_{l_{f}}}^{l}(\Omega_{\vec{r}})]^{\ast} \right ) \right\}
\end{gather}
As from the identity
\begin{gather}
\nabla_{q_{2}} = \frac{1}{2}~ [ ~\vec{\nabla}_{\vec{r}}^{2}, ~ r_{q_{2}}~]
\\ \nonumber
\vec{\nabla}_{\vec{\mathrm r}}^{2} = \frac{1}{r} ~\partial_{r} (\partial_{r} r) + \frac{1}{r^{2}} ~\vec{\nabla}_{\theta, \varphi}^{2} = \frac{1}{r} ~\partial_{r} (\partial_{r} r) +  \frac{1}{r^{2}} \left( \frac{1}{\sin \theta} \partial_{\theta} \left (\sin \theta ~\partial_{\theta} \right ) + \frac{1}{\sin\theta^{2}}\partial^{2}_{\varphi} \right)
\end{gather}
and since spherical harmonics are eigenstates of the angular laplacian $\vec{\nabla}_{\theta, \varphi}^{2}$ (cf.~Appendix~\ref{ME}), 
\begin{equation}
\vec{\nabla}_{\theta, \varphi}^{2} ~ Y_{m}^{l}(\Omega_{\vec{r}}) =  l(l+1) Y_{m}^{l}(\Omega_{\vec{r}})
\end{equation}
we may write
\begin{multline}
\left\{ [Y_{m_{l_{f}}}^{l}(\Omega_{\vec{r}})]^{\ast} \left ( \nabla_{q_{2}} ~Y_{m_{l_{i}}}^{l}(\Omega_{\vec{r}}) \right ) - Y_{m_{l_{i}}}^{l}(\Omega_{\vec{r}}) \left ( \nabla_{q_{2}}~[Y_{m_{l_{f}}}^{l}(\Omega_{\vec{r}})]^{\ast} \right ) \right\} = {}
\\
\frac{1}{2} \left\{ [Y_{m_{l_{f}}}^{l}(\Omega_{\vec{r}})]^{\ast} \left ( \vec{\nabla}_{\vec{r}}^{2} \left[ r_{q_{2}} ~Y_{m_{l_{i}}}^{l}(\Omega_{\vec{r}}) \right] \right )  - Y_{m_{l_{i}}}^{l}(\Omega_{\vec{r}}) \left ( \vec{\nabla}_{\vec{r}}^{2} \left[ r_{q_{2}}~[Y_{m_{l_{f}}}^{l}(\Omega_{\vec{r}})]^{\ast} \right]  \right ) \right\}
\end{multline}
but, substituting for $r_{q_{2}}$ its expression $r~ \left ( \frac{4\pi}{3} \right )^{\frac{1}{2}}  Y_{q_{2}}^{1}(\Omega_{\vec{r}})$ in spherical coordinates $(r, \theta_{\vec{r}}, \varphi_{\vec{r}})$ and taking into account the identities
\begin{multline}
\left ( \frac{4\pi}{3} \right )^{\frac{1}{2}} Y_{q_{2}}^{1}(\Omega_{\vec{r}}) ~Y_{m}^{l}(\Omega_{\vec{r}}) = \sum_{\overline{q}} (-)^{l+\overline{q}}~l^{\frac{1}{2}} Y_{-\overline{q}}^{l-1}(\Omega_{\vec{r}}) \left(\begin{array}{ccc}1 & l & l-1 \\q_{2} & m & \overline{q}\end{array}\right) + {} \qquad
\\
+ \sum_{\overline{q}} (-)^{l+1+\overline{q}} (l+1)^{\frac{1}{2}} Y_{-\overline{q}}^{l+1}(\Omega_{\vec{r}}) \left(\begin{array}{ccc}1 & l & l+1 \\q_{2} & m & \overline{q}\end{array}\right)
\end{multline}
(cf.~eq.~\eqref{eq:AM1}~and~eq.~\eqref{eq:AM0}) and
\begin{equation}
\vec{\nabla}_{\vec{r}}^{2}~ (r ~Y_{-\overline{q}}^{k}(\Omega_{\vec{r}})) = \left (\frac{1}{r} ~\partial_{r} (\partial_{r} r) - \frac{k(k+1)}{r^{2}} \right )\left( r ~Y_{-\overline{q}}^{k}(\Omega_{\vec{r}}) \right)= -\frac{1}{r} (k-1)(k+2) ~Y_{-\overline{q}}^{k}(\Omega_{\vec{r}})
\end{equation}
it is inferred that 
\begin{multline}
\left\{ [Y_{m_{l_{f}}}^{l}(\Omega_{\vec{r}})]^{\ast} \left ( \vec{\nabla}_{\vec{r}}^{2} \left[ r_{q_{2}} ~Y_{m_{l_{i}}}^{l}(\Omega_{\vec{r}}) \right] \right )  - Y_{m_{l_{i}}}^{l}(\Omega_{\vec{r}}) \left ( \vec{\nabla}_{\vec{r}}^{2} \left[ r_{q_{2}}~[Y_{m_{l_{f}}}^{l}(\Omega_{\vec{r}})]^{\ast} \right]  \right ) \right\}  = \frac{1}{r} (-)^{l+1+m_{l_{f}}} \times {}
\\
\shoveright {\times \sum_{\overline{q}} (-)^{\overline{q}}~ \Bigg\{~Y_{-m_{l_{f}}}^{l}(\Omega_{\vec{r}}) \bigg[~l^{\frac{1}{2}}(l-2)(l+1) ~Y_{-\overline{q}}^{l-1}(\Omega_{\vec{r}}) \left(\begin{array}{ccc}1 & l & l-1 \\q_{2} & m_{l_{i}} & \overline{q}\end{array}\right) - {} \qquad}
\\
\shoveright {- (l+1)^{\frac{1}{2}}l(l+3) ~Y_{-\overline{q}}^{l+1}(\Omega_{\vec{r}}) \left(\begin{array}{ccc}1 & l & l+1 \\q_{2} & m_{l_{i}} & \overline{q}\end{array}\right) \bigg] - {} }
\\
\shoveright {- Y_{m_{l_{i}}}^{l}(\Omega_{\vec{r}}) \bigg[~l^{\frac{1}{2}}(l-2)(l+1) ~Y_{-\overline{q}}^{l-1}(\Omega_{\vec{r}}) \left(\begin{array}{ccc}1 & l & l-1 \\q_{2} & -m_{l_{f}} & \overline{q}\end{array}\right) - {} \qquad}
\\
- (l+1)^{\frac{1}{2}}l(l+3) ~Y_{-\overline{q}}^{l+1}(\Omega_{\vec{r}}) \left(\begin{array}{ccc}1 & l & l+1 \\q_{2} & -m_{l_{f}} & \overline{q}\end{array}\right) \bigg] \Bigg\}
\end{multline}
The angular part of the integral over $\vec{r}$ in $\mathfrak{L}_{m_{l_{f}}, m_{l_{i}}}^{q~\bot}(\vec{\varkappa})$ thus is reduced to integrals of products of three spherical harmonics (cf.~eq.~\eqref{eq:AM2}) and therefore effortlessly found out to be 
\begin{multline}
\int d\Omega_{\vec{r}} ~Y_{Q}^{K}(\Omega_{\vec{r}}) \left\{ [Y_{m_{l_{f}}}^{l}(\Omega_{\vec{r}})]^{\ast} \left ( \nabla_{q_{2}} ~Y_{m_{l_{i}}}^{l}(\Omega_{\vec{r}}) \right ) - Y_{m_{l_{i}}}^{l}(\Omega_{\vec{r}}) \left ( \nabla_{q_{2}}~[Y_{m_{l_{f}}}^{l}(\Omega_{\vec{r}})]^{\ast} \right ) \right\} = {}
\\
= \frac{1}{2r} (-)^{l+1+m_{l_{f}}} \left( \frac{(2K+1)(2l+1)}{4\pi} \right)^{\frac{1}{2}} \sum_{\overline{q}} (-)^{\overline{q}} \times {}
\\
\times \Bigg\{ \left[l(2l-1)\right]^\frac{1}{2}(l-2)(l+1) \left(\begin{array}{ccc}K & l & l-1 \\0 & 0 & 0\end{array}\right) \times
\\
\shoveright {\bigg[ \left(\begin{array}{ccc}K & l & l-1 \\Q & -m_{l_{f}} & -\overline{q}\end{array}\right) \left(\begin{array}{ccc}1 & l & l-1 \\q_{2} & m_{l_{i}} & \overline{q}\end{array}\right) - \left(\begin{array}{ccc}K & l & l-1 \\Q & m_{l_{i}} & -\overline{q}\end{array}\right) \left(\begin{array}{ccc}1 & l & l-1 \\q_{2} & -m_{l_{f}} & \overline{q}\end{array}\right) \bigg] - {}}
\\
\qquad - \left[(l+1)(2l+3)\right]^\frac{1}{2}l(l+3) \left(\begin{array}{ccc}K & l & l+1 \\0 & 0 & 0\end{array}\right) \times
\\
\bigg[ \left(\begin{array}{ccc}K & l & l+1 \\Q & -m_{l_{f}} & -\overline{q}\end{array}\right) \left(\begin{array}{ccc}1 & l & l+1 \\q_{2} & m_{l_{i}} & \overline{q}\end{array}\right) - \left(\begin{array}{ccc}K & l & l+1 \\Q & m_{l_{i}} & -\overline{q}\end{array}\right) \left(\begin{array}{ccc}1 & l & l+1 \\q_{2} & -m_{l_{f}} & \overline{q}\end{array}\right) \bigg] \Bigg \} \\[-20pt]
\end{multline}
K must be an odd integer, owing to the the symmetry condition $K + l + l - 1 =$ even integer on the 3jm symbol with all m null (cf.~eq.~\eqref{eq:3s3}). Using the identity
\begin{multline}
\sum_{\overline{q}} (-)^{\overline{q}} \left(\begin{array}{ccc}l & 1 & l\pm1 \\x & q_{2} & \overline{q}\end{array}\right) \left(\begin{array}{ccc}K & l & l\pm1 \\Q & y & -\overline{q}\end{array}\right) = {}
\\
= \sum_{K^{\prime} Q^{\prime}} (-)^{l\pm1+K^{\prime}+Q^{\prime}} (2K^{\prime}+1) \left\{\begin{array}{ccc}l & 1 & l\pm1 \\K & l & K^{\prime}\end{array}\right\} \left(\begin{array}{ccc}K & 1 & K^{\prime} \\Q & q_{2} & -Q^{\prime}\end{array}\right) \left(\begin{array}{ccc}l & l & K^{\prime} \\x & y & Q^{\prime}\end{array}\right)
\end{multline}
(cf.~eq.~\eqref{eq:AM4}), where the triangular condition for the non vanishing of the $3jm$ symbols imposes that $K^{\prime} = K, K\pm1$, then finding out that $K^{\prime}$ must be an odd integer because of the factorization 
\begin{equation}
\left(\begin{array}{ccc}l & l & K^{\prime} \\m_{l_{i}} & -m_{l_{f}} & Q^{\prime}\end{array}\right) - \left(\begin{array}{ccc}l & l & K^{\prime} \\-m_{l_{f}} & m_{l_{i}} & Q^{\prime}\end{array}\right) = (1-(-)^{2l+K^{\prime}})\left(\begin{array}{ccc}l & l & K^{\prime} \\m_{l_{i}} & -m_{l_{f}} & Q^{\prime}\end{array}\right)
\end{equation}
so that $K^{\prime} = K$ only, and finally taking advantage of the algebraic expression for the $3jm$ symbols with all m null  (cf.~eq.~\eqref{eq:AM0}) and the following formula for the $6j$ symbol (cf.~eq.~\eqref{eq:6j2})
\begin{multline}
\left\{\begin{array}{ccc}l & K & z \\1 & z-1 & K\end{array}\right\} = (-)^{l+K+z} \left[ \frac{2(l+K+z+1)(K+z-l)(l+z-K)(l+K-z+1)}{2K(2K+1)(2K+2)(2z-1)2z(2z+1)} \right]^{\frac{1}{2}} \\[-16pt]
\end{multline}
to notice that
\begin{align}
\mathfrak{A}(K, l) &= \left[l(2l-1)\right]^\frac{1}{2} \left(\begin{array}{ccc}K & l & l-1 \\0 & 0 & 0\end{array}\right) \left\{\begin{array}{ccc}l & K & l \\1 & l-1 & K\end{array}\right\} = {}
\nonumber \\
& = \left[(l+1)(2l+3)\right]^\frac{1}{2} \left(\begin{array}{ccc}K & l & l+1 \\0 & 0 & 0\end{array}\right) \left\{\begin{array}{ccc}l & K & l+1 \\1 & l & K\end{array}\right\}
\end{align}
we get the more compact expression 
\begin{multline}
\int d\Omega_{\vec{r}} ~Y_{Q}^{K}(\Omega_{\vec{r}}) \left\{ [Y_{m_{l_{f}}}^{l}(\Omega_{\vec{r}})]^{\ast} \left ( \nabla_{q_{2}} ~Y_{m_{l_{i}}}^{l}(\Omega_{\vec{r}}) \right ) - Y_{m_{l_{i}}}^{l}(\Omega_{\vec{r}}) \left ( \nabla_{q_{2}}~[Y_{m_{l_{f}}}^{l}(\Omega_{\vec{r}})]^{\ast} \right ) \right\} = {}
\\
\frac{2}{r}~ \left( \frac{1}{4\pi} \right)^{\frac{1}{2}} \left[ K^{3}, l^{2} \right] ~(-)^{l+1}~\mathfrak{A}(K, l) \sum_{Q^{\prime}} (-)^{Q^{\prime}} \left \langle KQ^{\prime}lm_{l_{i}} ~\vert~ lm_{l_{f}} \right \rangle \left(\begin{array}{ccc}K & 1 & K \\Q & q_{2} & -Q^{\prime}\end{array}\right)
\end{multline}
The radial part of the integration over $\vec{r}$ in the expression of $\mathfrak{L}_{m_{l_{f}}, m_{l_{i}}}^{q~\bot}(\vec{\varkappa})$ is more straightforwardly found out to be
\begin{equation}
\int dr~ \frac{1}{r}j_{K}(\varkappa r) ~ R_{n l}^{2}(r) = \frac{\varkappa}{2 K + 1} \left \{ \left \langle j_{K-1}(\varkappa) \right \rangle +  \left \langle j_{K+1}(\varkappa) \right \rangle \right \}
\end{equation}
where
\begin{equation}
\langle  j_{K}(\varkappa) \rangle = \int_{0}^{\infty} r^{2} ~R_{n l}^{2}(r) ~j_{K}(\varkappa r)~ dr
\end{equation}
thanks to a recursion relation for the spherical Bessel functions (cf.~eq.~\eqref{eq:ME4}). 
Gathering the equations and using the multipole expansion 
\begin{multline}
\varkappa_{q_{1}} \left[Y_{Q}^{K}(\Omega_{\vec{\varkappa}})\right]^{\ast} = \varkappa (-)^{Q}  \sum_{\overline{K}, \overline{Q}} (-)^{\overline{Q}} \left[(2K+1)(2\overline{K}+1)\right]^{\frac{1}{2}} \times {}
\\
\times \left(\begin{array}{ccc}1 & K & \overline{K} \\0 & 0 & 0\end{array}\right) \left(\begin{array}{ccc}1 & K & \overline{K} \\q_{1} & -Q & -\overline{Q}\end{array}\right) Y_{\overline{Q}}^{\overline{K}}(\Omega_{\vec{\varkappa}})
\end{multline}
(cf.~eq.~\eqref{eq:AM1}) and the identity
\begin{multline}
\sum_{q_{1}, q_{1}, Q} (-)^{q+Q} \left(\begin{array}{ccc}K & 1 & K \\Q & q_{2} & -Q^{\prime}\end{array}\right) \left(\begin{array}{ccc}1 & K & \overline{K} \\q_{1} & -Q & -\overline{Q}\end{array}\right) \left(\begin{array}{ccc}1 & 1 & 1 \\q_{1} & q_{2} & -q\end{array}\right) = {}
\\
= (-)^{K} \left(\begin{array}{ccc}K & \overline{K} & 1 \\Q^{\prime} & \overline{Q} & -q\end{array}\right) \left\{\begin{array}{ccc}K & \overline{K} & 1 \\1 & 1 & K\end{array}\right\}
\end{multline}
(cf.~eq.~\eqref{eq:AM3}) it finally is inferred, exploiting the symmetries and triangle conditions of the involved $3jm$ symbols, that
\begin{multline}
\label{eq:msfuses6}
\mathfrak{L}_{m_{l_{f}}, m_{l_{i}}}^{q~\bot}(\vec{\varkappa}) = (8\pi)^{\frac{1}{2}} (-)^{l+1} \left[l^{2}\right] \sum_{\overline{K}, \overline{Q}} \Bigg\{ \sum_{K, Q^{\prime}} i^{K+1} \left \{ \left \langle j_{K-1}(\varkappa) \right \rangle +  \left \langle j_{K+1}(\varkappa) \right \rangle \right \} \left[K^{2}\right] \mathfrak{A}(K, l) \times {}
\\
\times \left(\begin{array}{ccc}1 & \overline{K} & K \\0 & 0 & 0\end{array}\right) \left\{\begin{array}{ccc}K & \overline{K} & 1 \\1 & 1 & K\end{array}\right\} \left \langle KQ^{\prime}lm_{l{i}} ~\vert~ lm_{l{f}} \right \rangle \left \langle \overline{K}\overline{Q}KQ^{\prime} ~\vert~ 1q \right \rangle \Bigg\}  ~  \left[\overline{K}\right] Y_{\overline{Q}}^{\overline{K}}(\Omega_{\vec{\varkappa}})
\end{multline}

$\overline{K} = K\pm1$ must be an even integer. \medskip

\noindent When $\vec{\varkappa} \rightarrow 0$ the expansion may be limited to the lowest $K = 1$ order, in which case $\overline{K} = 0, 2$ and
\begin{multline}
\label{eq:msfuses7}
\mathfrak{L}_{m_{l_{f}}, m_{l_{i}}}^{q~\bot}(\vec{\varkappa})_{\vec{\varkappa} \rightarrow 0} = (8\pi)^{\frac{1}{2}} (-)^{l+1} \left[l^{2}\right] i^{2} \left \{ \left \langle j_{0}(\varkappa) \right \rangle +  \left \langle j_{2}(\varkappa) \right \rangle \right \} ~ 3 ~ \mathfrak{A}(1, l) \times {}
\\
\times \Bigg[ \left(\begin{array}{ccc}1 & 0 & 1 \\0 & 0 & 0\end{array}\right) \left\{\begin{array}{ccc}1 & 0 & 1 \\1 & 1 & 1\end{array}\right\} \sum_{Q^{\prime}}\left \langle 1Q^{\prime}lm_{l{i}} ~\vert~ lm_{l{f}} \right \rangle \left \langle 001Q^{\prime} ~\vert~ 1q \right \rangle 
~ Y_{0}^{0}(\Omega_{\vec{\varkappa}}) + {}
\\
+ \left(\begin{array}{ccc}1 & 2 & 1 \\0 & 0 & 0\end{array}\right) \left\{\begin{array}{ccc}1 & 2 & 1 \\1 & 1 & 1\end{array}\right\} \sum_{\overline{Q}, Q^{\prime}} \left \langle 1Q^{\prime}lm_{l{i}} ~\vert~ lm_{l{f}} \right \rangle \left \langle 2\overline{Q}1Q^{\prime} ~\vert~ 1q \right \rangle  ~  \sqrt{5} Y_{\overline{Q}}^{2}(\Omega_{\vec{\varkappa}})  \Bigg]
\end{multline}
It is a matter of elementary algebra to compute
\begin{gather*}
\mathfrak{A}(1, l) = \sqrt{l(2l-1)} \left(\begin{array}{ccc}1 & l & l-1 \\0 & 0 & 0\end{array}\right) \left\{\begin{array}{ccc}l & 1 & l \\1 & l-1 & 1\end{array}\right\} = (-)^{l+1} \left[ l^{-2} \right] \sqrt{\frac{l(l+1)}{6}}
\\
\left(\begin{array}{ccc}1 & 0 & 1 \\0 & 0 & 0\end{array}\right) = -\frac{1}{\sqrt{3}} \quad \left\{\begin{array}{ccc}1 & 0 & 1 \\1 & 1 & 1\end{array}\right\} = -\frac{1}{3} \quad \left(\begin{array}{ccc}1 & 2 & 1 \\0 & 0 & 0\end{array}\right) = \sqrt{\frac{2}{15}} \quad \left\{\begin{array}{ccc}1 & 2 & 1 \\1 & 1 & 1\end{array}\right\} = \frac{1}{6}
\\
\langle 001Q^{\prime} ~\vert~ 1q \rangle = \delta_{Q^{\prime}, q} \quad \langle 2\overline{Q}1Q^{\prime} ~\vert~ 1q \rangle = (-)^{1-Q^{\prime}} \left[ \frac{(2+Q^{\prime}-q)!(2-Q^{\prime}+q)!}{10 (1-Q^{\prime})!(1+Q^{\prime})!(1-q)!(1+q)!} \right] \delta_{\overline{Q}, q-Q^{\prime}}
\end{gather*}
taking advantage of the symmetry properties given in the eqs.~\eqref{eq:3s1}-\eqref{eq:3s2} and \eqref{eq:6s1}-\eqref{eq:6s2} and using the eqs.~\eqref{eq:3j2}, \eqref{eq:AM0} and \eqref{eq:6j2} to get the algebraic expression for the required 3jm symbols and 6j symbols as well as for the quantity $\mathfrak{A}(1, l)$. Considering then the functional form of the spherical harmonics displayed in the eq.~\eqref{eq:spharord2}, and calling to mind the Wigner-Eckart theorem to show, with the help of the eq.~\eqref{eq:rml} for the reduced matrix element, that 
\begin{equation}
\langle 1Q^{\prime} lm_{l{i}} ~\vert~ lm_{l{f}} \rangle = - \frac{\langle lm_{l{f}} \vert ~\mathbf{l}_{Q^{\prime}}~ \vert  lm_{l{i}} \rangle}{\sqrt{l(l+1)} }
\end{equation}
where $\mathbf{l}$ is the spatial angular momentum operator of the electron, we get, by comparison with the eq.~\eqref{eq:msfuses4}, 
\begin{equation}
\label{eq:msfuses8}
\vec{\mathfrak{L}}_{\eta_{f}^{\mathrm p}, \eta_{i}^{\mathrm p}}^{~\bot}(\vec{\varkappa})_{\vec{\varkappa} \rightarrow 0} = \frac{1}{2} \left \{ \left \langle j_{0}(\varkappa) \right \rangle +  \left \langle j_{2}(\varkappa) \right \rangle \right \}  \langle~ lm_{l{f}} \vert ~\frac{1}{\varkappa^{2}} \vec{\varkappa} \wedge \mathbf{l} \wedge \vec{\varkappa}~ \vert  lm_{l{i}} ~\rangle
\end{equation}
which defines the \emph{dipole approximation} to the orbital vector form factor. The generalization of this approximation to more than one electron is straightforward and gathering this with the dipole approximation for the spin vector form factor immediately leads to the eq.~\eqref{eq:dipoapprox}. This often is used for an estimation of the orbital contribution to the experimentally measured magnetic density in magnetic materials and is the origin of the so-called $\langle j_{0}\rangle - \langle j_{2} \rangle$ analysis. An exact computation, for instance for the lanthanide ions in the ground multiplet level of maximum azimuthal quantum number, shows that the obtained values may roughly be good in some instances, as for $Nd^{3+}$ or $Pr^{3+}$, but not always, as for $Dy^{3+}$ or $Ho^{3+}$, owing to the additional contributions to the coefficient of $\langle j_{2} \rangle$, from the quadrupolar term ($K = 2$) in the expansion of $\vec{\mathfrak{S}}_{\eta_{f}^{\mathrm p}, \eta_{i}^{\mathrm p}}^{~\bot}(\vec{\varkappa})$ and the octupolar term ($K = 3, \overline{K} = 2$) in the expansion of $\vec{\mathfrak{L}}_{\eta_{f}^{\mathrm p}, \eta_{i}^{\mathrm p}}^{~\bot}(\vec{\varkappa})$.

\subsection{\label{sec:MultiElectronSpinFactor}Multi Electron Magnetic Vector form Factor}

Let now us generalize to correlated electrons within a same electronic configuration, that is to equivalent electrons. Adopting the $\vert \varsigma_{f, i}^{\mathrm p} \rangle = \vert (\upsilon LM_{L}SM_{S})_{f, i}^{\mathrm p} \rangle$ quantization scheme, associated with the total orbital $\vec{\mathbf{L}} = \sum_{j} \vec{\mathbf{l}}_{j}$ and total spin $\vec{\mathbf{S}} = \sum_{j} \vec{\mathbf{s}}_{j}$ kinetic moments of the equivalent electrons, we are led to compute the matrix elements
\begin{multline*}
\frac{1}{\varkappa^{2}} ~ \vec{\varkappa} \wedge \vec{\mathcal{F}}_{\varsigma_{f}^{\mathrm p}, \varsigma_{i}^{\mathrm p}}(\vec{\varkappa}) \wedge \vec{\varkappa} =  \left \langle (\upsilon LM_{L}SM_{S})_{f}^{\mathrm p} \left \vert ~ \vec{\mathbf{F}}^{\bot} ~ \right \vert (\upsilon LM_{L}SM_{S})_{i}^{\mathrm p} \right \rangle = {} \\ = \left \langle (\upsilon LM_{L}SM_{S})_{f}^{\mathrm p} \left \vert ~ \sum_{j \in \mathbb{A}_{\mathrm p}} \left[ \exp\{i (\vec{\varkappa} \cdot \vec{\mathbf{r}}_{j_{\mathrm p}})\} ~  \left ( \frac{1}{\varkappa^{2}} ~ \vec{\varkappa} \wedge \vec{\mathbf{s}}_{j}  \wedge \vec{\varkappa} - \frac{i}{\hbar \varkappa^{2}} \vec{\varkappa} \wedge \vec{\mathbf{p}}_{j} \right )\right] ~ \right \vert (\upsilon LM_{L}SM_{S})_{i}^{\mathrm p} \right \rangle
\end{multline*}
$\upsilon$ groups additional quantum numbers, to be precise irreducible representations of certain groups, discriminating between the states with identical $LM_{L}SM_{S}$ quantum numbers. The operator $\vec{\mathbf{F}}^{\bot}$ is the sum $\vec{\mathbf{F}}^{\bot}= \sum_{j=1}^{\mathcal{N}} \vec{\mathbf{F}}^{\bot}_{j}$ of $\mathcal{N}$ one-electron operators. Using the concept of fractional parentage coefficients that emerge when building up a correlated n-electron state from correlated (n-1)-electron states and single electron states, its matrix elements may be calculated from the single electron matrix elements. The methodology details, too lenghty are provided in the Appendix~\ref{STO} and the Appendix~\ref{MEE}. Use must be made of the eq.~\eqref{eq:emo} for the convection current density contribution since we have to consider a tensor operator acting only on the orbital state space and of the eq.~\eqref{eq:ems} for the spin current density contribution because then we have to deal with double tensors, which behaves as a rank 1 tensor in the spin space and as a rank $K$ tensor in the orbital space owing to the spherical harmonic tensor that emerge from the mulipole expansion of the exponential operator to separate the spatial electron variables from the scattering vector parameter. The spin contribution reads
\begin{multline}
\langle (\upsilon LM_{L}SM_{S})_{f}^{\mathrm p} \vert~ \sum_{j \in \mathbb{A}_{\mathrm p}} \exp\{i (\vec{\varkappa} \cdot \vec{\mathbf{r}}_{j_{\mathrm p}})\} ~ (\mathbf{s}_{j})_{q} \vert (\upsilon LM_{L}SM_{S})_{i}^{\mathrm p} \rangle = (6\pi)^{\frac{1}{2}} (-)^{\frac{3}{2} + M_{Sf} - M_{Lf}} \left[ l^{2} \right] \times {}
\\
\times \sum_{KQ} i^{K} \langle j_{K}(\varkappa) \rangle Y_{Q}^{K\ast}(\Omega_{\vec{\varkappa}}) \left[K\right] \left(\begin{array}{ccc}l & K & l \\0 & 0 & 0\end{array}\right) \left(\begin{array}{ccc}S_{f} & 1 & S_{i} \\-M_{Sf} & q & M_{Si}\end{array}\right)  \left(\begin{array}{ccc}L_{f} & K & L_{i} \\-M_{Lf} & Q & M_{Li}\end{array}\right) \times {}
\\
\times  \mathcal{N} \sum_{\overline{\theta}} (\theta_{f} \{ \vert \overline{\theta}) (\overline{\theta} \vert \} \theta_{i} ) (-)^{\overline{S}+\overline{L}}  \left[ S_{f}S_{i} L_{f}L_{i} \right]  \left\{\begin{array}{ccc}S_{f} & 1 & S_{i} \\\frac{1}{2} & \overline{S} & \frac{1}{2}\end{array}\right\}  \left\{\begin{array}{ccc}L_{i} & K & L_{f} \\l & L_{f}  & l\end{array}\right\} 
\end{multline}
while the orbital contribution
\begin{multline}
\left \langle (\upsilon LM_{L}SM_{S})_{f}^{\mathrm p} \left \vert ~ \left[ \sum_{j \in \mathbb{A}_{\mathrm p}}  - \frac{i}{\hbar \varkappa^{2}} \vec{\varkappa} \wedge \vec{\mathbf{p}}_{j} )\right]_{q}~ \right \vert (\upsilon LM_{L}SM_{S})_{i}^{\mathrm p} \right \rangle = {}
\\
= (8\pi)^{\frac{1}{2}} (-)^{l+1} \left[l^{3} L_{f} \L_{i} \right] \delta_{S_{f}, S_{i}}\delta_{M_{Sf}, M_{Si}} (-)^{l-L_{f}+L_{i}-M_{Li}} \times {} 
\\
\times \sum_{\widetilde{K}, \widetilde{Q}} \Bigg\{ \sum_{K, Q^{\prime}} i^{K+1} \left \{ \left \langle j_{K-1}(\varkappa) \right \rangle +  \left \langle j_{K+1}(\varkappa) \right \rangle \right \} \left[K^{2}\right] \mathfrak{A}(K, l) (-)^{-K+Q^{\prime}}  \left \langle \widetilde{K} \widetilde{Q}KQ^{\prime} ~\vert~ 1q \right \rangle \times {}
\\
\times \left(\begin{array}{ccc}1 & \widetilde{K} & K \\0 & 0 & 0\end{array}\right) \left\{\begin{array}{ccc}K & \widetilde{K} & 1 \\1 & 1 & K\end{array}\right\} \left(\begin{array}{ccc}L_{i} & K & L_{f} \\M_{Li} & Q^{\prime} & -M_{Lf}\end{array}\right)  \times {}
\\
\times  \mathcal{N} \sum_{\overline{\theta}} (-)^{\overline{L}}(\theta_{f} \{ \vert \overline{\theta}) (\overline{\theta} \vert \} \theta_{i} ) \left\{\begin{array}{ccc}L{i} & K & L_{f} \\l & \overline{L} & l\end{array}\right\}
\Bigg\}  ~  \left[\widetilde{K}\right] Y_{\widetilde{Q}}^{\widetilde{K}}(\Omega_{\vec{\varkappa}})
\end{multline}
These matrix elements can be computed in the $\vert (\upsilon LSJM_{J})_{f, i}^{\mathrm p} \rangle$ quantization scheme, either directly or through the base transformation 
\begin{equation}
\vert (\upsilon LSJM_{J})_{f, i}^{\mathrm p} \rangle = \sum_{M_{L}M_{S}} \langle LM_{L}SM_{S} \vert JM_{J} \rangle \vert (\upsilon LM_{L}SM_{S})_{f, i}^{\mathrm p} \rangle
\end{equation} 
What matters at this stage is actually not the exact and full expression of the different matrix elements in the different quantization schemes, as these can be found at several places in the literature \cite{Marshall&Lovesey, Lovesey&Rimmer, Lovesey, Balcar&Lovesey}, but rather to get aware that these essentially depend on the single electron matrix elements and the few coefficients of fractional parentage. Although the calculation of these coefficients may not be easy, it has, at any rate, to be done only once. Finally, it is in principle not difficult to generalize the computations to the case where the scattering event induces inter-configuration electron transfer.

\Appendix

\section{\label{FT}Fourier Transforms}

Let T$^{(k)}$ be a tensor field of rank $k$ over the 3-D real space $\mathbb{R}^{3}$ with integrable components T$^{(k)}_{l}~(l = -k, -k+1, \cdots, k)$. Its Fourier transform $\mathcal{F}\{\mathrm{T}^{(k)}\}$ is the tensor field of rank $k$ over $\mathbb{R}^{3}$  $-$ in fact over the dual of $\mathbb{R}^{3}$ but identified with $\mathbb{R}^{3}$ $-$ the components  $\mathcal{F}\{\mathrm{T}^{(k)}\}_{l}~(l = -k, -k+1, \cdots, k)$ of which are the Fourier transforms $\mathcal{F}\{\mathrm{T}^{(k)}_{l}\}$ of the T$^{(k)}_{l}$, that is defined by
\begin{gather}
\label{eq:FT1}
\mathcal{F}\{\mathrm{T}^{(k)}_{l}\} (\vec{q}) = \int_{\mathbb{R}^{3}} \mathrm{T}^{(k)}_{l}(\vec{r})~ \exp\{ -i (\vec{q} \cdot \vec{r})\} ~ d\vec{r}
\\ \nonumber
\mathrm{T}^{(k)}_{l}(\vec{r}) = \frac{1}{(2 \pi)^{3}} \int_{\mathbb{R}^{3}} \mathcal{F}\{\mathrm{T}^{(k)}_{l}\} (\vec{q})~ \exp\{ i (\vec{q} \cdot \vec{r})\} ~ d\vec{q}
\end{gather}
\medskip

\noindent $\mathcal{F}\{\mathrm{T}^{(k)}_{l}\}$ is angular independent in the dual space, $\mathcal{F}\{\mathrm{T}^{(k)}_{l}\}(\vec{q}) = \mathcal{F}\{\mathrm{T}^{(k)}_{l}\}(q = \vert \vec{q} \vert) ~ \forall \vec{q}$, if and only if T$^{(k)}_{l}$ is angular independent in the direct space, $\mathrm{T}^{(k)}_{l}(\vec{r}) = \mathrm{T}^{(k)}_{l}(r = \vert ~\vec{r}~ \vert) ~ \forall \vec{r}$, in which case
\begin{equation*}
\mathcal{F}\{\mathrm{T}^{(k)}_{l}\}(q) = \frac{4 \pi}{q} \int_{0}^{\infty} r \mathrm{T}^{(k)}_{l}(r) \sin(qr) ~ dr \quad \vert \quad \mathrm{T}^{(k)}_{l}(r) =  \frac{1}{2 \pi^{2} r} \int_{0}^{\infty} q \mathcal{F}\{\mathrm{T}^{(k)}_{l}\}(q) \sin(qr) ~ dq.
\end{equation*}
As an example,
\begin{gather*}
\mathcal{F}\{\frac{1}{r}\}(\vec{q}) = \lim_{~\epsilon \rightarrow 0}~ \mathcal{F}\{\frac{\exp(-\epsilon r)}{r}\}(\vec{q}) = \frac{4 \pi}{q} \lim_{~\epsilon \rightarrow 0}~ \int_{0}^{\infty} \exp(-\epsilon r) \sin(qr) ~ dr = { }
\\
= \frac{4 \pi}{q} \lim_{~\epsilon \rightarrow 0}~ \int_{0}^{\infty} \exp(-\epsilon r) \frac{\exp(iqr) - \exp(-iqr)}{2i} ~ dr = \frac{4 \pi}{q} \lim_{~\epsilon \rightarrow 0}~(\frac{1}{\epsilon-iq} - \frac{1}{\epsilon+iq})\frac{1}{2i} = \frac{4\pi}{q^{2}},
\end{gather*}
that is
\begin{equation}
\label{eq:FT2}
\mathcal{F}\{\frac{1}{r}\}(\vec{q}) = \frac{4\pi}{q^{2}} \qquad \vert \qquad
\frac{1}{r} = \frac{1}{2\pi^{2}}  \int_{\mathbb{R}^{3}} \frac{1}{q^{2}} \exp\{ i (\vec{q} \cdot \vec{r})\} ~ d\vec{q} \qquad
\end{equation}
\medskip

\noindent $\Phi, \Xi, \cdots$ symbolizing scalar fields $(k = 0)$, it is an easy matter to demonstrate that 
\begin{equation}
\mathcal{F}\{\Phi \ast \Xi\} = \mathcal{F}\{\Phi\} \mathcal{F}\{\Xi\} \qquad \mathcal{F}\{\Phi \Xi\} = \frac{1}{(2\pi)^{3}}\mathcal{F}\{\Phi\} \ast \mathcal{F}\{\Xi\}
\end{equation}
where
\begin{equation*}
(\Phi \ast \Xi)(\vec{r}) = \int_{\mathbb{R}^{3}} \Phi(\vec{r}-\vec{s}) \Xi(\vec{s}) ~ d\vec{s} \quad \vert \quad
\mathcal{F}\{\Phi\} \ast \mathcal{F}\{\Xi\}(\vec{q}) = \int_{\mathbb{R}^{3}} \mathcal{F}\{\Phi\}(\vec{q}-\vec{\kappa})\mathcal{F}\{\Xi\}(\vec{\kappa}) ~ d\vec{\kappa}
\end{equation*}
The neutral element for the field convolution $\ast$ is the Dirac generalized function in the $\mathbb{R}^{3}$ space since by definition $\forall \chi ~\int_{\mathbb{R}_{3}} \delta (\vec{r}-\vec{s}) \chi (\vec{s}) ~ d\vec{s} = \chi (\vec{r})$, that is $\forall \chi ~ \delta \ast \chi = \chi$. We deduce 
\begin{equation}
\label{eq:FTC}
\mathcal{F}\{\delta(\vec{r})\} = 1\quad \text{and} \quad \delta(\vec{q}) = \frac{1}{(2\pi)^{3}}\mathcal{F}\{1\} = \frac{1}{(2\pi)^{3}}\int_{\mathbb{R}^{3}} \exp\{ -i (\vec{q} \cdot \vec{r})\} ~ d\vec{r}.
\end{equation}
\medskip

\noindent $\vec{\mathrm{G}}, \vec{\mathrm{H}}, \cdots$ symbolizing vector fields $(k = 1)$ with components $\mathrm{G}_{l}, \mathrm{H}_{l}, \cdots (l = -1, 0,1)$,
\begin{align}
\label{eq:FTA}
\mathcal{F}\{\Phi \ast \vec{\mathrm{G}}\} & = \mathcal{F}\{\Phi\} \mathcal{F}\{\vec{\mathrm{G}}\} & \mathcal{F}\{\Phi \vec{\mathrm{G}}\} & = \frac{1}{(2\pi)^{3}}\mathcal{F}\{\Phi\} \ast \mathcal{F}\{\vec{\mathrm{G}}\}
\\ \nonumber
\mathcal{F}\{\sum_{l} \mathrm{G}_{l} \ast \mathrm{H}_{l}\} & =  \mathcal{F}\{\vec{\mathrm{G}}\} \cdot \mathcal{F}\{\vec{\mathrm{H}}\} & \mathcal{F}\{\vec{\mathrm{G}} \cdot \vec{\mathrm{H}}\} & =  \frac{1}{(2\pi)^{3}} \sum_{l} \mathcal{F}\{\mathrm{G}_{l}\} \ast \mathcal{F}\{\mathrm{H}_{l}\}
\\ \nonumber
\sum_{l, m, n} \epsilon^{lmn} \mathcal{F}\{\mathrm{G}_{l} \ast \mathrm{H}_{m}\} \vec{e}_{n} & =  \mathcal{F}\{\vec{\mathrm{G}}\} \wedge \mathcal{F}\{\vec{\mathrm{H}}\} & \mathcal{F}\{\vec{\mathrm{G}} \wedge \vec{\mathrm{H}} \} & =  \frac{1}{(2\pi)^{3}} \sum_{l, m, n} \epsilon^{lmn} \mathcal{F}\{\mathrm{G}_{l}\} \ast \mathcal{F}\{\mathrm{H}_{m}\} \vec{e}_{n}
\end{align}
where $\epsilon^{lmn} = 1$ if $(l, m, n)$ is co-cyclic to $(1, 2, 3)$, $= -1$ if $(l, m, n)$ is anti-cyclic to $(1, 2, 3)$ and $= 0$ otherwise.
\medskip

\noindent $\partial_{r_{\hat{e}}}\mathrm{T}^{(k)}_{l}(\vec{r})$ symbolizing the derivation of $\mathrm{T}^{(k)}_{l}(\vec{r})$ with respect to $r_{\hat{e}} = \hat{e}\cdot\vec{r} \quad (e.g. ~r_{\hat{e}} = x, y, z)$ and assuming that $\mathrm{T}^{(k)}_{l}(\vec{r})$ tends algebraically to zero at infinity, 
\begin{gather*}
\mathcal{F}\{\partial_{r_{\hat{e}}}\mathrm{T}^{(k)}_{l}\} (\vec{q}) = \int_{\mathbb{R}^{3}} \left[ \partial_{r_{\hat{e}}}\mathrm{T}^{(k)}_{l}(\vec{r}) \right]~ \exp\{ -i (\vec{q} \cdot \vec{r})\} ~ d\vec{r} = {}
\\
= \left[ \mathrm{T}^{(k)}_{l}(\vec{r})~ \exp\{ -i (\vec{q} \cdot \vec{r})\} \right]_{\mathbb{R}^{3}} - \int_{\mathbb{R}^{3}} \mathrm{T}^{(k)}_{l}(\vec{r})~ \left[\partial_{r_{\hat{e}}}\exp\{ -i (\vec{q} \cdot \vec{r})\}\right] ~ d\vec{r} = (i~\hat{e}\cdot\vec{q}) ~\mathcal{F}\{\mathrm{T}^{(k)}_{l}\} (\vec{q}).
\end{gather*}
We find in particular that
\begin{gather}
\label{eq:FTB}
\mathcal{F}\{\vec{\nabla}_{\vec{r}}~ \Phi\} = i~\vec{q}~\mathcal{F}\{\Phi\}, \quad \mathcal{F}\{\triangle_{\vec{r}}~ \Phi\} = -q^{2}~\mathcal{F}\{\Phi\},
\\ \nonumber
 \mathcal{F}\{\vec{\nabla}_{\vec{r}} \cdot \vec{\mathrm{G}}\} = i~\vec{q} \cdot \mathcal{F}\{\vec{\mathrm{G}}\}, \quad \mathcal{F}\{\vec{\nabla}_{\vec{r}} \wedge \vec{\mathrm{G}}\} = i~\vec{q} \wedge \mathcal{F}\{\vec{\mathrm{G}}\}
\end{gather}
\medskip

\noindent These formulas are helpful for solving a number of partial differential equations. An example is $\triangle_{\vec{r}} ~\Phi(\vec{r}) = -4\pi \delta(\vec{r})$, which is immediately solved as $\Phi(\vec{r}) = \frac{1}{r}$. Another example is the pair of maxwell equations
\begin{equation}
\label{eq:FT3}
\vec{\nabla}_{\vec{r}} \wedge \vec{\mathrm {B}}(\vec{r}) = \mu_{0} \vec{\mathrm {j}}(\vec{r}) \qquad \qquad \vec{\nabla}_{\vec{r}} \cdot \vec{\mathrm {B}}(\vec{r}) = 0
\end{equation}
which must be solved to find the magnetic induction $\vec{\mathrm {B}}(\vec{r})$ created by a time independent current density $\vec{\mathrm {j}}(\vec{r})$. We get in the dual space 
\begin{equation*}
i~\vec{q} \wedge \mathcal{F}\{\vec{\mathrm{B}}\} = \mu_{0} \mathcal{F}\{\vec{\mathrm{j}}\} \qquad \qquad i~\vec{q} \cdot \mathcal{F}\{\vec{\mathrm{B}}\} = 0.
\end{equation*} 
(cf.~eq.~\eqref{eq:FTB}) The second equation, which merely is gauge fixing for $\vec{\mathrm {B}}(\vec{r})$, tells that there always exists a vector field $\vec{\mathrm {A}}(\vec{r})$ such that $\mathcal{F}\{\vec{\mathrm{B}}\} = i~\vec{q} \wedge \mathcal{F}\{\vec{\mathrm{A}}\}$, or equivalently $\vec{\mathrm{B}}(\vec{r}) = \vec{\nabla}_{\vec{r}} \wedge \vec{\mathrm {A}}(\vec{r})$. The first equation then gives 
\begin{equation*}
i~\vec{q} \wedge (i~\vec{q} \wedge \mathcal{F}\{\vec{\mathrm{A}}\}) = (i~\vec{q} \cdot  \mathcal{F}\{\vec{\mathrm{A}}\})~ i~\vec{q} + q^{2}~\mathcal{F}\{\vec{\mathrm{A}}\} = \mu_{0} \mathcal{F}\{\vec{\mathrm{j}}\}, 
\end{equation*}
since $ \forall (\vec{u}, \vec{v},\vec{w}) \quad \vec{u} \wedge (\vec{v} \wedge \vec{w}) = (\vec{u} \cdot \vec{w})\vec{v} -(\vec{u} \cdot \vec{v})\vec{w}$. $\vec{\mathrm{A}}(\vec{r})$ by definition is subject to gauge invariance and we are free to fix this as $i~\vec{q} \cdot \mathcal{F}\{\vec{\mathrm{A}}\} = 0$, or equivalently $\vec{\nabla}_{\vec{r}} \cdot \vec{\mathrm {A}}(\vec{r}) = 0$. We deduce 
\begin{equation*}
\mathcal{F}\{\vec{\mathrm{A}}\} = \mu_{0} \frac{1}{q^{2}} \mathcal{F}\{\vec{\mathrm{j}}\} = \mu_{0} \frac{1}{4\pi} \mathcal{F}\{\frac{1}{r}\} \mathcal{F}\{\vec{\mathrm{j}}\}
\end{equation*}
\begin{equation*}
\mathcal{F}\{\vec{\mathrm{B}}\} = \frac{\mu_{0}}{4\pi}~ i~\vec{q} \wedge \mathcal{F}\{\frac{1}{r}\} \mathcal{F}\{\vec{\mathrm{j}}\} = \frac{\mu_{0}}{4\pi}~ i~\vec{q} \wedge \mathcal{F}\{\frac{1}{r} \ast \vec{\mathrm{j}}\} = \frac{\mu_{0}}{4\pi}~ \mathcal{F}\{\vec{\nabla}_{\vec{r}} \wedge [ ~\frac{1}{r} \ast \vec{\mathrm{j}}~ ] \} 
\end{equation*}
(cf.~eq.~\eqref{eq:FT2}, eq.~\eqref{eq:FTA} and eq.~\eqref{eq:FTB}), that is
\begin{equation}
\label{eq:FT4}
\vec{\mathrm{B}}(\vec{r}) = \frac{\mu_{0}}{4\pi} ~ \vec{\nabla}_{\vec{r}} \wedge \left[ \int_{\mathbb{R}^{3}} \frac{\vec{\mathrm {j}}(\vec{s})}{\vec{r}-\vec{s}}  ~ d\vec{s} \right] = \frac{\mu_{0}}{4\pi} \int_{\mathbb{R}^{3}} ~ \vec {\mathbf{j}}(\vec{s}) \wedge \frac{\vec{r}-\vec{s}}{\vert ~ \vec{r}-\vec{s} ~ \vert ^{3}}  ~ d\vec{s}
\end{equation}

\noindent Any vector $\vec{u}$ is the unique sum of its longitudinal component $\frac{1}{q^{2}} (\vec{q} \cdot \vec{u})~\vec{q}$ and its transverse component $\frac{1}{q^{2}} (\vec{q} \wedge \vec{u} \wedge \vec{q})$ over the vector $\vec{q}$. On transposing this to the Fourier transform of a vector field $\vec{\mathrm{G}}$, we find 
\begin{gather*}
\mathcal{F}\{\vec{\mathrm{G}}\} = \frac{1}{q^{2}} [(\vec{q} \cdot \mathcal{F}\{\vec{\mathrm{G}}\})~\vec{q} + (\vec{q} \wedge \mathcal{F}\{\vec{\mathrm{G}}\} \wedge \vec{q})] 
\\
= \frac{1}{4\pi}~\mathcal{F}\{\frac{1}{r}\} [ -i~\vec{q}~(i~\vec{q} \cdot \mathcal{F}\{\vec{\mathrm{G}}\}) + (i~\vec{q} \wedge (i~\vec{q} \wedge \mathcal{F}\{\vec{\mathrm{G}}\}))] 
\\
= -i~\vec{q}~(\frac{1}{4\pi}~\mathcal{F}\{\frac{1}{r} \ast \vec{\nabla}_{\vec{s}} \cdot \vec{\mathrm{G}}\}) + i~\vec{q}~ \wedge (\frac{1}{4\pi}~\mathcal{F}\{\frac{1}{r} \ast \vec{\nabla}_{\vec{s}} \wedge \vec{\mathrm{G}}\})
\\
= \mathcal{F}\{ \vec{\nabla}_{\vec{r}} ~( \frac{1}{4\pi}~ \frac{1}{r} \ast \vec{\nabla}_{\vec{s}} \cdot \vec{\mathrm{G}}) + \vec{\nabla}_{\vec{r}} \wedge (\frac{1}{4\pi}~ \frac{1}{r} \ast  \vec{\nabla}_{\vec{s}} \wedge \vec{\mathrm{G}}) \}, 
\end{gather*}
that is $\vec{\mathrm{G}}$ is the unique sum $\vec{\mathrm{G}} = \vec{\mathrm{G}}_{I} + \vec{\mathrm{G}}_{S}$ of an irrotational component($\vec{\nabla}_{\vec{r}} \wedge \vec{\mathrm{G}}_{I}(\vec{r}) = 0)$ and a solenoidal component ($\vec{\nabla}_{\vec{r}} \cdot \vec{\mathrm{G}}_{S}(\vec{r}) = 0)$
\begin{equation}
\label{eq:FT5}
\vec{\mathrm{G}}(\vec{r}) = \vec{\mathrm{G}}_{I}(\vec{r}) + \vec{\mathrm{G}}_{S}(\vec{r}) = -\vec{\nabla}_{\vec{r}} ~\Upsilon (\vec{r}) + \vec{\nabla}_{\vec{r}} \wedge \vec{\mathrm{S}} (\vec{r}) 
\end{equation}
where
\begin{equation*}
\Upsilon (\vec{r}) = \frac{1}{4\pi}~ \int_{\mathbb{R}^{3}} \frac{\vec{\nabla}_{\vec{s}} \cdot \vec{\mathrm{G}}(\vec{s})}{\vert~\vec{r}-\vec{s}~\vert} ~ d\vec{s} \qquad \vert \qquad \vec{\mathrm{S}} (\vec{r}) = \frac{1}{4\pi}~ \int_{\mathbb{R}^{3}} \frac{\vec{\nabla}_{\vec{s}} \wedge \vec{\mathrm{G}}(\vec{s})}{\vert~\vec{r}-\vec{s}~\vert} ~ d\vec{s}
\end{equation*}
$\Upsilon$ is determined up to a real constant and $\vec{\mathrm{S}}$ up to the gradient of an arbitrary scalar field.

\section{\label{ME}Multipole Expansions}

A multipole expansion by definition is performed over the collection $\{Y_{m}^{l}\}_{-l \leq m \leq l, ~l = 0, 1, 2, \cdots}$ of spherical harmonics and by essence is substantiated from the Sturm-Liouville spectral problem
\begin{equation}
\{\vec{\nabla}_{\theta, \varphi}^{2}  + \Lambda \} f(\theta, \varphi) = 0
\end{equation} 
on the unit 2-sphere $S_{2}$ with the boundary condition that $f$ must be finite everywhere on $S_{2}$. This belongs to a wide class of mathematically solved problems, but  can also be approached more intuitively by interpreting the operator
\begin{equation*}
-\vec{\nabla}_{\theta, \varphi}^{2} = -\left( \frac{1}{\sin \theta} \partial_{\theta} \left (\sin \theta ~\partial_{\theta} \right ) + \frac{1}{\sin\theta^{2}}\partial^{2}_{\varphi} \right)
\end{equation*}
as the square $\vec{\mathbf{L}}^{2}$ of an orbital moment $\vec{\mathbf{L}}$ and the operators
\begin{equation}
\mathbf{L}_{z} = -i \partial_{\varphi} \quad \text{and} \quad \mathbf{L}_{\pm} = \pm \exp(\pm i \varphi) \left \{\partial_{\theta} \pm i \frac{\cos\theta}{\sin\theta} \partial_{\varphi} \right \}
\end{equation}
as the components of $\vec{\mathbf{L}}$ on the polar axis (z) of $S_{2}$ and on the two associated orthogonal helicity axes $(\pm)$. $[\vec{\nabla}_{\theta, \varphi}^{2}, \mathbf{L}_{z}] = 0$, which implies that $\exists~ Y_{m}^{\Lambda} :$ $\vec{\nabla}_{\theta, \varphi}^{2} Y_{m}^{\Lambda} = -\Lambda Y_{m}^{\Lambda}$ and $\mathbf{L}_{z} Y_{m}^{\Lambda} = m Y_{m}^{\Lambda}$. $[\mathbf{L}_{z}, \mathbf{L}_{\pm}] = \pm \mathbf{L}_{\pm}$ so that $\mathbf{L}_{z}(\mathbf{L}_{\pm}Y_{m}^{\Lambda}) = (m \pm 1)(\mathbf{L}_{\pm}Y_{m}^{\Lambda})$, which suggests that $\mathbf{L}_{\pm}$ are indeed ladder operators. $[\vec{\nabla}_{\theta, \varphi}^{2}, \mathbf{L}_{\pm}] = 0$, which tells that $Y_{m}^{\Lambda}$ and $\mathbf{L}_{\pm}Y_{m}^{\Lambda}$ belong to the same eigenspace labeled by $\Lambda$. $[\mathbf{L}_{+}, \mathbf{L}_{-}] = 2 \mathbf{L}_{z}$ and $-\vec{\nabla}_{\theta, \varphi}^{2} = \frac{1}{2} (\mathbf{L}_{+}\mathbf{L}_{-} + \mathbf{L}_{-}\mathbf{L}_{+}) + \mathbf{L}_{z}^{2}$ then $\mathbf{L}_{\mp}\mathbf{L}_{\pm} = -\vec{\nabla}_{\theta, \varphi}^{2} - \mathbf{L}_{z}^{2} \mp \mathbf{L}_{z}$ and
\begin{multline*}
0 \leq \int_{0}^{\pi} \sin \theta~d\theta \int_{0}^{2\pi} d\varphi ~ \left[ \mathbf{L}_{\pm} Y_{m}^{\Lambda} (\theta, \varphi) \right]^{\ast} \mathbf{L}_{\pm} Y_{m}^{\Lambda} (\theta, \varphi) = \langle \mathbf{L}_{\pm} Y_{m}^{\Lambda} \vert \mathbf{L}_{\pm} Y_{m}^{\Lambda} \rangle = \langle Y_{m}^{\Lambda} \vert  \mathbf{L}_{\mp} \mathbf{L}_{\pm} \vert Y_{m}^{\Lambda} \rangle = {}
\\
= \langle Y_{m}^{\Lambda} \vert  -\vec{\nabla}_{\theta, \varphi}^{2} - \mathbf{L}_{z}^{2} \mp \mathbf{L}_{z} \vert Y_{m}^{\Lambda} \rangle = (\Lambda-m(m\pm1)) \int_{0}^{\pi} \sin \theta~d\theta \int_{0}^{2\pi} d\varphi ~ \left[ Y_{m}^{\Lambda} (\theta, \varphi) \right]^{\ast} Y_{m}^{\Lambda} (\theta, \varphi),
\end{multline*}
which indicates that if $Y_{m}^{\Lambda}$ is normalized then so are $Y_{m \pm 1}^{\Lambda}$ and that once $\Lambda$ is fixed $m$ is bounded. If $l$ is the upper positive bound then $\langle \mathbf{L}_{+} Y_{m = l}^{\Lambda} \vert \mathbf{L}_{+} Y_{m = l}^{\Lambda} \rangle = 0$ so $\Lambda = l(l+1)$ and the lower negative bound is necessarily $-l$, hence the more appropriate notation $Y_{m}^{l} ~ (-l\leq m\leq l)$. An outcome is that $\forall ~ Y_{m}^{l} ~ \exists ~ p\in\mathbb{N} ~ \exists ~ q\in\mathbb{N} : \mathbf{L}_{+}^{p}Y_{m}^{l} = 0 \text{ and } \mathbf{L}_{-}^{q}Y_{m}^{l} = 0$, that is $m + p = l$ and $m - q = -l$, which implies that $2m = p - q\in \mathbb{N}$ and $2l = p + q \in \mathbb{N}$. Since $\mathbf{L}_{z} Y_{m}^{l} = -i\partial_{\varphi} Y_{m}^{l} = m Y_{m}^{l}$ the functional form of $Y_{m}^{l}$ should be $Y_{m}^{l}(\theta,  \varphi) = A_{m}^{l}(\theta) \exp(i m \varphi)$. Accordingly, if we further require that $Y_{m}^{l}$ must be a $2\pi-$periodic function in the angle $\varphi$ then $m$ itself must be an integer and so l as well : the allowed values of $\Lambda$ are $l(l+1), ~l = 0, 1, 2, \cdots$ and for each $l$ the allowed values of $m$ are $-l \leq m = 0, \pm 1, \pm 2, \cdots \leq l$. Negative l are irrelevant because $l(l+1)$ is invariant under the $l \rightarrow -(l+1)$ transformation. Each $(l,m)$ distinguishes a single $Y_{m}^{l}$, which means that the operators $\vec{\nabla}_{\theta, \varphi}^{2}$ and $\mathbf{L}_{z}$ form a complete set of commuting operators. Now from $\mathbf{L}_{\pm} Y_{\pm l}^{l} = 0$ it is inferred that $\partial_{\theta} A_{\pm l}^{l}(\theta) = l\frac{\cos\theta}{\sin\theta} A_{\pm l}^{l}(\theta)$ so $Y_{m = \pm l}^{l} (\theta, \varphi) = C_{\pm l} \sin^{l}\theta \exp(\pm i l \varphi)$, where $\vert C_{\pm l} \vert^{2}$ is fixed from the normalization condition $\langle Y_{\pm l}^{l} \vert Y_{\pm l}^{l} \rangle = 2\pi \vert C_{\pm l} \vert^{2} \int_{0}^{\pi} \sin^{2l+1} \theta ~ d\theta = 1$ and $C_{\pm l}$ from the phase convention $C_{l} = (-)^{l}\vert C_{\pm l} \vert$. It is then advantageous to make use of the identity
\begin{equation}
\label{eq:MEA}
2\int_{0}^{\frac{\pi}{2}} \cos^{2z_{1}+1} \theta ~ \sin^{2z_{2}+1} \theta ~ d\theta = \frac{z_{1}! ~ z_{2}!} {(z_{1} + z_{2} + 1)!} \quad \forall~z_{j}\in\mathbb{C} : ~ \Re(z_{j}) \geq -1 \quad (j = 1, 2)
\end{equation}
which is easily deduced from the double integral $\int_{0}^{\infty} \int_{0}^{\infty} x^{\zeta}y^{\xi} \exp(-(x^{2}+y^{2}))~dxdy$ by performing at first the $(u = x^{2}, ~v = y^{2})$ variable change then the $(x = r\cos\theta, ~ y = r\sin\theta)$ variable change : we are led to the equality $\frac{1}{2}\int_{0}^{\infty} u^{\frac{\zeta-1}{2}}\exp(-u)~du ~\frac{1}{2}\int_{0}^{\infty} v^{\frac{\xi-1}{2}}\exp(-v)~dv = \int_{0}^{\infty} r^{\zeta+\xi+1}\exp(-r^{2})~dr ~\int_{0}^{\frac{\pi}{2}} \cos^{\zeta}\theta \sin^{\xi}\theta~d\theta$ from which the identity immediately follows, with $\zeta = 2z_{1}+1$ and $\xi = 2z_{2}+1$, by calling to mind that $z \mapsto \int_{0}^{\infty} t^{z} \exp(-t)~dt = z!$ defines the factorial function : $t^{z} = t^{\Re(z)\exp\Im(z)\log(t)}$ and $\vert \int_{0}^{\infty} t^{z} \exp(-t)~dt \vert \leqslant \int_{0}^{\infty} t^{\Re(z)} \exp(-t)~dt$ so $ \int_{0}^{\infty} t^{z} \exp(-t)~dt$ makes sense for all complex numbers $z$ with real part $\Re(z) \geq -1$ while integrating by parts the functional relation $z! = z(z-1)!$ is recovered and in fact might be used to extend by analytic continuation the concept of factorial to the whole field $\mathbb{C}$ of complex numbers. When $z_{1} = -z_{2} = z$ with $-1 < \Re(z) < 1$, the identity \eqref{eq:MEA} reads
\begin{equation*}
z!~(-z)! = 2\int_{0}^{\frac{\pi}{2}} (\cot^{2} \theta)^{z} ~ d(\sin^{2} \theta) = \int_{0}^{\infty} \frac{w^{z}}{(1+w)^{2}} ~dw = \mathcal{I}
\end{equation*}
where $w = \cot^{2}\theta$. $\mathcal{I}$ can be evaluated by integration over the complex plane : the analytic function $s \in \mathbb{C} \mapsto g(s) = \frac{s^{z}}{(1+s)^{2}}$ admits $-1$ as a pole of second order and $0$ as a branching point. $g(s)$ is made uniform by cutting the complex plane along the positive real axis so that $0 < arg(s) < 2\pi$ and $s^{z} =\vert s \vert^{z} \exp\{ iz~arg(s) \}$ with $arg(s) = 0^{+}$ at the upper adherence to the cut. Let $\mathfrak{D}$ be the domain formed by a disk of radius $R$ centered about $s=0$ from which the cut along the positive real axis and a small disk of radius $r$ centered about $s = -1$ are removed and let $\mathfrak{C}$ be the external and $\mathfrak{c}$ the internal closed contours limiting $\mathfrak{D}$, oriented anticlockwise, then  
\begin{gather*}
\forall R ~ \forall ~ r < R ~ \int_{\mathfrak{C}} g(s) ~ds = \int_{\mathfrak{c}} g(s) ~ds
\\
\text{but} \quad \int_{\mathfrak{C}} g(s) ~ds \xrightarrow[ ]{R \rightarrow \infty} \mathcal{I}  ~(1 - \exp\{i ~ 2\pi ~ z\}) \quad \text{and} \quad \frac{1}{i ~2\pi}\int_{\mathfrak{c}} g(s) ~ds ~ \xrightarrow[ ]{r \rightarrow 0} (\exp\{i ~ \pi ~ z\}) (-z)
\end{gather*}
since $\vert g(s) \vert \approx \vert s \vert^{\Re(z)-2} \ll \frac{1}{\vert s \vert} ~(\vert\Re(z)\vert < 1)$ when $R \rightarrow \infty$ and $g(s = -1 + \epsilon) = \exp\{i ~ \pi ~ z\} \frac{(1-\epsilon)^{z}}{\epsilon^{2}} = \exp\{i ~ \pi ~ z\} (\frac{1}{\epsilon^{2}}- \frac{z}{\epsilon} + \cdots)$ when $r \rightarrow 0$. As a result
\begin{equation}
z!~(-z)! = \frac{\pi z}{\sin \pi z}
\end{equation}
true for all z in $\mathbb{C}$ by analytic continuation. We in particular have $\left(\frac{1}{2}\right)! \left(-\frac{1}{2}\right)! = \frac{\pi}{2}$, which implies that $\left(-\frac{1}{2}\right)! = \sqrt{\pi}$ and $\left(\frac{1}{2}\right)! = \frac{\sqrt{\pi}}{2}$ since $\left(\frac{1}{2}\right)! = \frac{1}{2} \left(-\frac{1}{2}\right)!$ and  $\left(-\frac{1}{2}\right)! > 0$. When $z_{1} = z_{2} = z$ with $\Re(z) \geq -1$, the identity \eqref{eq:MEA} reads
\begin{equation*}
2\int_{0}^{\frac{\pi}{2}} (\sin \theta \cos \theta)^{2z+1}~d\theta = \frac{z!^{2}}{(2z+1)!} =2\frac{1}{2^{2z+1}} \int_{0}^{\frac{\pi}{2}} (\sin \varphi)^{2z+1}~d \varphi = \frac{1}{2^{2z+1}} \frac{\sqrt{\pi}~z!}{(z+\frac{1}{2})!}
\end{equation*}
from which it is inferred that
\begin{equation}
\label{eq:ME0}
\sqrt{\pi} \left(2z+1\right)! = 2^{2z+1}~z!~\left(z + \frac{1}{2} \right)! \quad \text{or} \quad \sqrt{\pi} \left(2z \right)!~2^{2z}~z!~\left(z - \frac{1}{2} \right)! 
\end{equation}
true for all z in $\mathbb{C}$ by analytic continuation.
\medskip

\noindent Let us focus our attention back to the functional form of the spherical harmonics. Up to phase convention
\begin{equation*}
\mathbf{L}_{\pm} Y_{m}^{l} = [(l(l+1)-m(m\pm1)]^{\frac{1}{2}} Y_{m\pm1}^{l} = [(l \mp m)(l \pm m +1)]^{\frac{1}{2}} Y_{m\pm1}^{l}
\end{equation*}
since $\mathbf{L}_{z}(\mathbf{L}_{\pm}Y_{m}^{l}) = (m \pm 1)(\mathbf{L}_{\pm}Y_{m}^{l})$ and $\langle \mathbf{L}_{\pm} Y_{m}^{l} \vert \mathbf{L}_{\pm} Y_{m}^{l} \rangle = (l(l+1)-m(m\pm1) \langle Y_{m}^{l} \vert Y_{m}^{l} \rangle$, while 
\begin{gather*}
\mathbf{L}_{\pm} Y_{m}^{l} (\theta, \varphi) = \pm \exp\{ \pm i \varphi \} (\partial_{\theta} \pm i \frac{\cos \theta}{\sin \theta} \partial_{\varphi})~ A_{m}^{l}(\theta) \exp(i m \varphi) = {}
\\
= \pm \exp\{ \pm i \varphi \} (\partial_{\theta} \mp m \frac{\cos \theta}{\sin \theta})~ A_{m}^{l}(\theta) \exp(i m \varphi) = \pm \exp\{ \pm i \varphi \} (\sin^{\pm m}\theta ~\partial_{\theta} \sin^{\mp m}\theta)~ Y_{m}^{l} (\theta, \varphi)
\end{gather*}
Accordingly, 
\begin{gather}
Y_{m \pm p}^{l} (\theta, \varphi) = \sqrt{\frac{(l \pm m)! (l \mp m - p)!}{(l \mp m)! (l \pm m + p)!}} ~ \overset{\text{p times}}{\overbrace{\mathbf{L}_{\pm} \mathbf{L}_{\pm} \cdots \mathbf{L}_{\pm}}} ~ Y_{m}^{l} (\theta, \varphi) = {}
\nonumber \\
= \sqrt{\frac{(l \pm m)! (l \mp m - p)!}{(l \mp m)! (l \pm m + p)!}} ~ \left\{ (\mp)^{p} \sin^{\pm (m \pm p)} \theta ~\partial_{\cos \theta}^{p} \sin^{\mp m} \theta \exp\{ \pm i p \varphi \} \right\}  Y_{m}^{l} (\theta, \varphi)
\end{gather}
where $\partial_{\cos \theta} = \frac{1}{\partial_{\theta} \cos\theta} \partial_{\theta} = -\frac{1}{\sin\theta} \partial_{\theta}$. We deduce $Y_{0}^{l} (\theta, \varphi) = \frac{1}{\sqrt{(2l)!}} \frac{(-)^{l}}{2^{l}l!}\sqrt{\frac{(2l+1)!}{4\pi}}~\partial_{\cos \theta}^{l} \sin^{2l} \theta$ and $Y_{p}^{l} (\theta, \varphi) = \sqrt{\frac{(l-p)!}{(l+p)!}}~(-)^{p}\sin^{p} \theta ~\partial_{\cos \theta}^{p}  \exp\{ i p \varphi \} ~Y_{0}^{l} (\theta, \varphi)$. A more standard expression is
\begin{equation}
\label{eq:MEB}
Y_{m}^{l}(\theta, \varphi) = (-)^{m} \sqrt{\frac{2l+1}{4\pi} \frac{(l-m)!}{(l+m)!}} P_{m}^{l} (\cos \theta) \exp(i m \varphi)
\end{equation}
where
\begin{equation}
P_{m}^{l} (x) = \frac{(1-x^{2})^{\frac{m}{2}}}{2^{l}l!} \partial_{x}^{l+m} (x^{2}-1)^{l}
\end{equation}
are dubbed order $m$ associated functions of the Legendre polynomial $P_{0}^{l}$ of degree $l$.
\begin{gather*}
\partial_{x}^{l+m} (x^{2}-1)^{l} = \sum_{k = 0}^{l} \frac{(l+m)!}{k!(l+m-k)!} \left[ ~\partial_{x}^{k} (x-1)^{l}~ \right] \left[ ~\partial_{x}^{l+m-k} (x+1)^{l}~ \right] = {}
\\
= \sum_{k = 0}^{l} \frac{(l+m)! (l!)^{2}(x-1)^{l-k}(x+1)^{l-k+m}}{k!(l+m-k)!(l-k)!(k-m)!} \overset{(k \rightarrow h+m)}{=} (x^{2}-1)^{-m}\frac{(l+m)!}{(l-m)!} ~\partial_{x}^{l-m} (x^{2}-1)^{l}
\end{gather*}
thus
\begin{equation}
P_{-m}^{l} (x) = (-)^{m}\frac{(l-m)!}{(l+m)!} P_{m}^{l} (x) \quad \vert \quad \left[ Y_{m}^{l} (\zeta, \psi) \right]^{\ast} = (-)^{m} Y_{-m}^{l} (\zeta, \psi)
\end{equation}
Argument symmetry properties : $Y_{m}^{l} (\zeta, -\psi) = \left[ Y_{m}^{l} (\zeta, \psi) \right]^{\ast}$, $Y_{m}^{l} (-\zeta, \psi) = (-)^{m} Y_{m}^{l} (\zeta, \psi)$, $Y_{m}^{l} (-\zeta, -\psi) = Y_{-m}^{l} (\zeta, \psi)$, $Y_{m}^{l} (\pi - \zeta, \psi) = (-)^{l+m} Y_{m}^{l} (\zeta, \psi)$, $Y_{m}^{l} (\zeta, \pi + \psi) = (-)^{m} Y_{m}^{l} (\zeta, \psi)$.
\medskip

\noindent $\vec{\nabla}_{\theta, \varphi}^{2}$ and $\mathbf{L}_{z}$ are self-ajoint operators, which implies that the $Y_{m}^{l}$ are orthonormal, namely

\begin{equation}
\label{eq:MEF}
\int_{0}^{\pi} \sin \theta~d\theta \int_{0}^{2\pi} d\varphi ~ \left[ Y_{m}^{l} (\theta, \varphi) \right]^{\ast} Y_{q}^{k} (\theta, \varphi) = \delta_{lk} \delta_{mq}
\end{equation}
\medskip

\noindent Let $\mathcal{L}_{S_{2}}^{2}$ be the vector space of square integrable functions on $S_{2}$, $\mathcal{L}_{S_{2}}^{2}(\mathrm{N})$ the subspace engendered by $\{Y_{m}^{l}\}_{-l \leq m \leq l, ~l = 0, 1, 2, \cdots, \mathrm{N}}$ and $\mathcal{L}_{S_{2}}^{2}(\mathrm{N})^{\bot}$ its supplement : $\mathcal{L}_{S_{2}}^{2} = \mathcal{L}_{S_{2}}^{2}(\mathrm{N}) \bigoplus \mathcal{L}_{S_{2}}^{2}(\mathrm{N})^{\bot}$. 
\begin{gather*}
\forall g \in \mathcal{L}_{S_{2}}^{2} \quad h_{\mathrm{N}} = g - \sum_{l = 0}^{\mathrm{N}} \sum_{m = - l}^{l} \langle Y_{m}^{l} \vert g \rangle ~ Y_{m}^{l} \in \mathcal{L}_{S_{2}}^{2}(\mathrm{N})^{\bot} \quad 
\\[-6pt] 
\text{so} \quad \frac{\langle h_{\mathrm{N}} \vert -\vec{\nabla}_{\theta, \varphi}^{2} h_{\mathrm{N}} \rangle}{\langle h_{\mathrm{N}} \vert h_{\mathrm{N}} \rangle}  \geq (\mathrm{N}+1)(\mathrm{N}+2) \geq 0 \quad \text{since } \min_{~f \in \mathcal{L}_{S_{2}}^{2}(\mathrm{N})^{\bot}} \frac{\langle f \vert -\vec{\nabla}_{\theta, \varphi}^{2} f \rangle}{\langle f \vert f \rangle} = (\mathrm{N}+1)(\mathrm{N}+2)
\\[-6pt] 
\text{but} \quad \langle h_{\mathrm{N}} \vert -\vec{\nabla}_{\theta, \varphi}^{2} h_{\mathrm{N}} \rangle = \langle g \vert -\vec{\nabla}_{\theta, \varphi}^{2} g \rangle -  \sum_{l = 0}^{\mathrm{N}} \sum_{m = - l}^{l} \vert \langle Y_{m}^{l} \vert g \rangle \vert ^{2} l(l+1) \leq \langle g \vert -\vec{\nabla}_{\theta, \varphi}^{2} g \rangle
\\[-6pt] 
\text{then,} \quad \langle g \vert -\vec{\nabla}_{\theta, \varphi}^{2} g \rangle \text{ being N-independent,} \quad 0 \leq \lim_{~\mathrm{N} \rightarrow \infty}~\langle h_{\mathrm{N}} \vert h_{\mathrm{N}} \rangle \leq  \lim_{~\mathrm{N} \rightarrow \infty}~ \frac{\langle g \vert \vec{\nabla}_{\theta, \varphi}^{2} g \rangle}{ (\mathrm{N}+1)(\mathrm{N}+2)} = 0,
\end{gather*}
which establishes the completeness of  the semi-infinite set $\{Y_{m}^{l}\}_{-l \leq m \leq l, ~l = 0, 1, 2, \cdots}$ in $\mathcal{L}_{S_{2}}^{2}$, to be precise
\begin{equation}
\forall g\in \mathcal{L}_{S_{2}}^{2} \quad g(\theta, \varphi) = \sum_{l = 0}^{\infty} \sum_{m = -l}^{l} \left \{ \int_{0}^{\pi} \sin \zeta~d\zeta \int_{0}^{2\pi} d\psi ~ \left[ Y_{m}^{l} (\zeta, \psi) \right]^{\ast} g(\zeta, \psi) \right \} ~ Y_{m}^{l}(\theta, \varphi)
\end{equation}
\medskip

\noindent An equivalent form of the completeness relation which avoids specifying any function g in $\mathcal{L}_{S_{2}}^{2}$ is 
\begin{equation}
\sum_{l = 0}^{\infty} \sum_{m = -l}^{l} \left[ Y_{m}^{l} (\zeta, \psi) \right]^{\ast} ~ Y_{m}^{l}(\theta, \varphi) = \frac{1}{\sin\theta}\tilde{\delta} (\theta-\zeta) ~\tilde{\delta} (\varphi-\psi)
\end{equation}
where $\tilde{\delta}$ is the Dirac generalized function defined on the $0 \leq \theta, ~\zeta\leq \pi$ and $0 \leq \varphi, ~\psi \leq 2 \pi$ compact domains. Notice that $\frac{1}{\sin\theta}\tilde{\delta} (\theta-\zeta) = \tilde{\delta} (\cos \theta-\cos \zeta)$ and more generally  if $f$ is a uniform function with  derivative $\partial_{x}f$ and if $\{a_{i}\}$ are the roots of the equation $f(x) = 0$ then $\tilde{\delta} [f(x)] = \sum_{i} \frac{1}{\vert \partial_{x}f (a_{i}) \vert} \tilde{\delta} (x-a_{i})$. $\tilde{\delta}$ in compact domains $\mathcal{D}_{1}$ of the real space $\mathbb{R}$ and $\delta$ in compact domains $\mathcal{D}_{3}$ of the 3-D real space $\mathbb{R}^{3}$ are defined such that $\forall f ~\int_{\mathcal{D}_{1}} \tilde{\delta} (x-u) f (x) ~ dx = f (u)$ and $\forall \Phi ~\int_{\mathcal{D}_{3}} \delta (\vec{r}-\vec{s}) \Phi (\vec{r}) ~ d\vec{r} = \Phi (\vec{s})$. If $(r, \theta, \varphi)$ are the spherical coordinates of $\vec{r}$ and  $(s, \zeta, \psi)$ those of $\vec{s}$ then $d\vec{r} = -r^{2}~dr~d\cos(\theta)~d\varphi$ and $\delta (\vec{r}-\vec{s}) = \frac{1}{r^{2}} ~\tilde{\delta} (r-s) ~\tilde{\delta} (\cos \theta-\cos \zeta) ~\tilde{\delta} (\varphi-\psi)$. We immediately deduce that 
\begin{equation}
\label{eq:ME1}
\delta (\vec{r}-\vec{s}) =  \frac{1}{r^{2}} ~\tilde{\delta} (r-s) ~ \sum_{l = 0}^{\infty} \sum_{m = -l}^{l} \left[ Y_{m}^{l} (\zeta, \psi) \right]^{\ast} ~ Y_{m}^{l}(\theta, \varphi) 
\end{equation}

\noindent On more general grounds it can be shown that the solutions of any Sturm-Liouville spectral problem always form an orthonormal and complete set in the background vector space of the problem. Thus, the semi-infinite set of the Legendre polynomials $\{P_{0}^{l}\}_{l = 0, 1, 2, \cdots}$ are the solutions for the eigenvalues $\Lambda = l(l+1)$ of the Sturm-Liouville spectral problem 
\begin{equation}
\{\partial_{x} [(1-x^{2}) \partial_{x}]  + \Lambda \} f(x) = 0
\end{equation}
on the real interval $[-1, 1]$ with the boundary condition that $f$ must be finite everywhere on $[-1, 1]$. As such this set then is orthonormal and complete in the vector space $\mathcal{L}_{[-1, 1]}^{2}$ of square integrable functions on $[ -1, 1]$ and allows for series expansion in this space. We more concretely have
\begin{equation}
\int_{-1}^{1} P_{0}^{l} (x) P_{0}^{k} (x)~\left(l+\frac{1}{2}\right)~dx = \delta_{lk} \quad \vert \quad \sum_{l = 0}^{\infty}\left(l+\frac{1}{2}\right) P_{0}^{l} (x)P_{0}^{l} (y) = \tilde{\delta}(x-y)
\end{equation}
\medskip

\noindent When dealing with the functions $f(\vec{r}_{1}, \vec{r}_{2})$, which depend on two vectors $\vec{r}_{1} (r_{1}, \theta_{1}, \varphi_{1})$ and $\vec{r}_{2} (r_{2}, \theta_{2}, \varphi_{2})$, the multipole expansions must be performed over the collection of bi-polar spherical harmonics. These are obtained by irreducible tensor product  of the spherical harmonics with different arguments
\begin{equation}
(\mathbb{Y}^{l_{1}}(\theta_{1}, \varphi_{1}) \otimes (\mathbb{Y}^{l_{2}}(\theta_{2}, \varphi_{2}) )_{LM} = \sum_{m_{1}, m_{2}} \langle  l_{1}m_{1} l_{2}m_{2} \vert LM \rangle ~ Y_{m_{1}}^{l_{1}}(\theta_{1}, \varphi_{1})Y_{m_{2}}^{l_{2}}(\theta_{2}, \varphi_{2}) 
\end{equation}
and form a complete orthonormal set in $\mathcal{L}_{S_{2} \times {S_{2}}}^{2}$. $ \langle  l_{1}m_{1} l_{2}m_{2} \vert LM \rangle$ is a Clebsch-Gordan coefficient (cf.~Appendix~\ref{AMC}). Generalization to the functions of more than two vectors and to multi-polar spherical harmonics is straightforward \cite{Varshalovich}. If $f(\vec{r}_{1}, \vec{r}_{2})$ is invariant under rotation of coordinate systems then it depends only on $r_{1}$, $r_{2}$ and $\vec{r}_{1} \cdot \vec{r}_{2} = r_{1} r_{2} \cos \omega_{12}$, where $\cos \omega_{12} = \cos \theta_{1} \cos \theta_{2} + \sin \theta_{1} \sin \theta_{2} \cos (\varphi_{1}-\varphi_{2})$, and its multipole expansion contains only the zero rank bi-polar spherical harmonic $(\mathbb{Y}_{l_{1}}(\theta_{1}, \varphi_{1}) \otimes \mathbb{Y}_{l_{2}}(\theta_{2}, \varphi_{2}))_{00} = \frac{(-)^{l_{1}}}{\sqrt{2l_{1}+1}} (\mathbb{Y}_{l_{1}}(\theta_{1}, \varphi_{1}) \cdot \mathbb{Y}_{l_{1}}(\theta_{2}, \varphi_{2})) ~\delta_{l_{1}, l_{2}}$, where the scalar product
\begin{equation}
\label{eq:MED}
\mathbb{Y}_{l}(\theta_{1}, \varphi_{1}) \cdot \mathbb{Y}_{l}(\theta_{2}, \varphi_{2}) = \sum_{m = -l}^{l}  \left[Y_{m}^{l}(\theta_{1}, \varphi_{1})\right]^{\ast} Y_{m}^{l}(\theta_{2}, \varphi_{2}) = \frac{2l+1}{4\pi} P_{0}^{l}(\cos \omega_{12})
\end{equation}
The last equality expresses the \emph{addition theorem} for the spherical harmonics, proven as follows: 
\begin{gather*}
P_{0}^{l}(\cos \omega_{12}) = \sum_{m = -l}^{l}  a_{lm} Y_{m}^{l}(\theta_{2}, \varphi_{2}) 
\\[-6pt] 
\text{where} \quad a_{lm} = \left ( \frac{4\pi}{2l+1} \right )^{\frac{1}{2}} \int_{0}^{\pi} \sin \theta_{2}~d\theta_{2} \int_{0}^{2\pi} d\varphi_{2} \left[Y_{m}^{l}(\theta_{2}, \varphi_{2})\right]^{\ast} Y_{0}^{l}(\omega_{12}, \gamma_{12})
\\[-6pt]
\text{but} \quad Y_{m}^{l}(\theta_{2}, \varphi_{2}) = \sum_{q = -l}^{l}  b_{lq} Y_{q}^{l}(\omega_{12}, \gamma_{12}) \xrightarrow[\omega_{12 }\rightarrow 0]{} Y_{m}^{l} (\theta_{1}, \varphi_{1}) \underset {Y_{k\neq 0}^{l}(0, \gamma_{12}) = 0}{=} b_{l0} \left ( \frac{2l+1}{4\pi} \right )^{\frac{1}{2}} \quad \text{and} 
\\[-6pt]
b_{l0} = \int_{0}^{\pi} \sin\omega_{12}~d\omega_{12} \int_{0}^{2\pi} d\gamma_{12} \left[Y_{m}^{l}(\omega_{12}, \gamma_{12})\right]^{\ast} Y_{0}^{l}(\theta_{2}, \varphi_{2}) \quad \text{so} \quad a_{lm} =  \left ( \frac{4\pi}{2l+1} \right ) \left[Y_{m}^{l}(\theta_{1}, \varphi_{1})\right]^{\ast}
\end{gather*}
the angular variables $(\omega_{12}, \gamma_{12})$ and $(\theta_{2}, \varphi_{2})$ in the evaluation of $b_{l0}$ being exchangeable.
\medskip

\noindent Let now us focus our attention on the series expansion of $\exp(ixy)$ over the Legendre polynomials $P_{0}^{l}(y)$ :
\begin{gather}
\exp(ixy) = \sum_{l = 0}^{\infty} c_{l}(x) P_{0}^{l}(y)
\\ \nonumber
c_{l}(x) = \int_{-1}^{1} \exp(ixy)P_{0}^{l} (y)\left(l+\frac{1}{2}\right)~dy = \frac{(2l+1)}{2^{l+1}l!} \int_{-1}^{1} \exp(ixy) [ \partial_{y}^{l} (y^{2}-1)^{l}]~dy
\end{gather}
Integration l times by parts we get
\begin{gather*}
c_{l}(x) = \frac{(2l+1)}{2^{l+1}l!} (ix)^{l}\int_{-1}^{1} \exp(ixy) (1-y^{2})^{l} ~dy = i^{l}(2l+1)\sqrt{\frac{\pi}{2x}} ~ J_{l+\frac{1}{2}}(x)
\end{gather*}
where
\begin{equation}
J_{\nu}(x) = \frac{1}{\pi^{\frac{1}{2}}(\nu-\frac{1}{2})!}\left(\frac{x}{2}\right)^{\nu} \int_{-1}^{1} \exp(ixy) (1-y^{2})^{\nu-\frac{1}{2}} ~dy
\end{equation}
\medskip

\noindent $J_{\nu}(x)$ defines the Bessel function of index $\nu$ and is a particular solution of the Bessel equation
\begin{equation}
\left[ ~\partial_{x}^{2} + \frac{1}{x} \partial_{x} + \left(1-\frac{\nu^{2}}{x^{2}} \right)~ \right] f(x) = 0
\end{equation}
The general solution $Z_{\nu}(x)$ can be apprehended on representing the function $g(x) = x^{-\nu} f(x)$ by the generalized Laplace integral 
\begin{equation*}
g(x) = \int_{\mathcal{P}} \exp(xz)~v(z)~dz
\end{equation*}
and searching for analytic functions $v(z)$ and paths $\mathcal{P}$ in the complex plane that verify the equation $\left[ x \partial_{x}^{2} + (2\nu + 1) \partial_{x} + x \right] g(x) = 0$, that is such that
\begin{gather*}
0 = (2\nu + 1) \int_{\mathcal{P}} \exp(xz)~zv(z)~dz +  \int_{\mathcal{P}} \exp(xz)~x(1+z^{2})v(z)~dz
\\
= (2\nu + 1) \int_{\mathcal{P}} \exp(xz)~zv(z)~dz - \int_{\mathcal{P}} \exp(xz)~\partial_{z}\left[(1+z^{2})v(z)\right]~dz + \vert~\exp(xz)~(1+z^{2})v(z)~\vert_{\mathcal{P}}
\end{gather*}
or equivalently
\begin{equation*}
\int_{\mathcal{P}} \exp(xz)~[ (2\nu + 1) \frac{zu(z)}{z^{2}+1} - \partial_{z} u(z)]~dz - \vert~\exp(xz)~u(z)~\vert_{\mathcal{P}} = 0
\end{equation*}
where $u(z) = (1+z^{2})v(z)$. It suffices that the integral and the integrated term in this last expression are separately null. This implies that $(2\nu + 1) \frac{zu(z)}{z^{2}+1} - \partial_{z} u(z) = 0$, that is $u(z) = C (1+z^{2})^{\nu+\frac{1}{2}}$. So 
\begin{equation}
Z_{\nu}(x) = Cx^{\nu} \int_{\mathcal{P}} \exp(xz)~(1+z^{2})^{\nu-\frac{1}{2}}~dz
\end{equation}
with a path ${\mathcal{P}}$ such that the variation of $\exp(xz)~(1+z^{2})^{\nu+\frac{1}{2}}$ over it is null. The solution $Z_{\nu}(x) = J_{\nu}(x)$ is deduced by observing that $(1+z^{2})^{\nu+\frac{1}{2}} = 0$ for $z = \pm i$ provided that $\Re (\nu) > -\frac{1}{2}$, which suggests to take for ${\mathcal{P}}$ the straight line from $-i$ to $i$ and to perform the $z = iy$ variable change. A recipe to retain is that the method of Laplace integral in the complex plane works for all the homogeneous linear differential equations where the coefficients are linear functions of the variable. 

\noindent $x^{-\nu} J_{\nu}(x)$ is an analytic and even function of $x$, which thus admits a series expansion everywhere convergent over even integer powers of $x$. This allows forming a series for $J_{\nu}(x)$, which necessarily is convergent, and by analytic continuation defining $J_{\nu}(x)$ for any $\nu$. Concretely,
\begin{gather*}
\int_{-1}^{1} \exp(ixy) (1-y^{2})^{\nu-\frac{1}{2}} ~dy = 2 \int_{0}^{1} \cos(xy) (1-y^{2})^{\nu-\frac{1}{2}} ~dy 
= {} 
\\
= 2\sum_{k=0}^{\infty} (-)^{k} \int_{0}^{1} \frac{x^{2k}y^{2k}}{(2k)!} (1-y^{2})^{\nu-\frac{1}{2}} ~dy \underset{y = \sin\theta}{=} 2 \sum_{k=0}^{\infty} (-)^{k} \frac{x^{2k}}{(2k)!} \int_{0}^{\frac{\pi}{2}} \cos^{2\nu}\theta \sin^{2k}\theta~d \theta = {}
\\[6pt]
= \sum_{k=0}^{\infty} (-)^{k} \frac{x^{2k}(\nu-\frac{1}{2})! \sqrt{\pi}}{2^{2k}k!(\nu+k)!} 
\end{gather*}
(cf.~eq.~\eqref{eq:MEA} and eq.~\eqref{eq:ME0}), therefore
\begin{equation}
\label{eq:MEC}
J_{\nu}(x) = \sum_{k=0}^{\infty} (-)^{k}\frac{1}{k!(\nu+k)!}\left(\frac{x}{2}\right)^{\nu + 2k}
\end{equation}
\medskip

\noindent Let $\vec{r}$ be a vector in the real space $\mathbb{R}^{3}$ with spherical coordinates $(r, \theta_{r}, \varphi_{r})$, $\vec{q}$ a vector in the dual space $\mathbb{R}^{3}$ with spherical coordinates $(q, \theta_{q}, \varphi_{q})$ and $\omega$ the angle between them. If $qr$ is substituted for $x$ and $\cos \omega$ for $y$ in the expansion of $\exp(ixy)$ over $\{P_{0}^{l}\}_{l = 0, 1, 2, \cdots}$ then, using the addition theorem for the spherical harmonics (cf.~eq.~\eqref{eq:MED}), we find 
\begin{equation}
\label{eq:ME2}
\exp(i~\vec{q} \cdot \vec{r}) = 4\pi \sum_{l = 0}^{\infty} i^{l} j_{l}(qr) \sum_{m = -l}^{l}  \left[Y_{m}^{l}(\theta_{q}, \varphi_{q})\right]^{\ast} Y_{m}^{l}(\theta_{r}, \varphi_{r})
\end{equation}
where $j_{l}(x) = \sqrt{\frac{\pi}{2x}} ~ J_{l+\frac{1}{2}}(x)$ defines the spherical Bessel function of index $l$. As from the series expansion of $x^{-\nu} J_{\nu}(x)$ over even integer powers of $x$ (cf.~eq.~\eqref{eq:MEC}) and by making use of the identity (cf.~eq.~\eqref{eq:ME0})
\begin{equation*}
\sqrt{\pi} (2z+1)! = 2^{2z+1} z! \left(z+\frac{1}{2}\right)!
\end{equation*}
it is inferred that 
\begin{equation}
\label{eq:ME3}
j_{0}(x) = \frac{\sin x}{x} \quad \text{and} \quad j_{l}(x)_{x \rightarrow 0} \approx \frac{2^{l}l!}{(2l+1)!} x^{l} \end{equation}
Also useful is the recursion relation
\begin{equation}
\label{eq:ME4}
j_{l-1}(x) + j_{l+1}(x) = \left ( \frac{2l+1}{x} \right) j_{l}(x), \qquad l\geqslant1
\end{equation}
which also is deduced from the series expansion of $x^{-\nu} J_{\nu}(x)$ over even integer powers of $x$ (cf.~eq.~\eqref{eq:MEC}) by merely observing that
\begin{multline*}
\partial_{x} (x^{\nu}J_{\nu}(x)) = \sum_{k=0}^{\infty} (-)^{k}\frac{2^{\nu}}{k!(\nu+k)!}~ \partial_{x} \left( \frac{x}{2} \right)^{2\nu + 2k} = {}
\\
\shoveright{= \sum_{k=0}^{\infty} (-)^{k}\frac{1}{k!(\nu-1+k)!} \left( \frac{x}{2} \right)^{\nu -1+ 2k} = x^{\nu} J_{\nu-1}(x)}
\\
\shoveleft{\partial_{x} (x^{\nu}J_{\nu}(x)) = \sum_{k=0}^{\infty} (-)^{k} \frac{1}{2^{\nu}} \frac{1}{k!(\nu+k)!}~ \partial_{x} \left( \frac{x}{2} \right)^{2k} = {}}
\\
= \sum_{h=0}^{\infty} (-)^{h+1} \frac{1}{2^{\nu}} \frac{1}{(h)!(\nu+1+h)!}\left( \frac{x}{2} \right)^{1+ 2h} = -x^{-\nu} J_{\nu+1}(x)
\end{multline*}
so that
\begin{equation*}
\partial_{x} (J_{\nu}(x)) + \frac{\nu}{x} J_{\nu}(x) = J_{\nu-1}(x) \quad \text{and} \quad \partial_{x} (J_{\nu}(x)) - \frac{\nu}{x} J_{\nu}(x) = -J_{\nu+1}(x)
\end{equation*}

\section{\label{AM}Angular momenta}

\subsection{\label{AMA}Definition and Properties}

Angular momenta by definition are infinitesimal generators of the state transformations associated with the spatial rotations. We shall symbolize these
\begin{equation}
\vert \psi \rangle \rightarrow \vert \psi^{\prime} \rangle = \mathbf{U}(\omega, \hat{u}) \vert \psi \rangle
\end{equation}
for a quantum system subject to the rotation $R(\omega, \hat{u})$ about the unit vector $\hat{u}$ through an angle $\omega$. $\mathbf{U}(\omega, \hat{u})$ materializes a mapping of the state space of the quantum system over itself which is bijective and preserves the modulus of the scalar product so, by virtue of a Wigner theorem, is either a unitary or an anti-unitary operator up to phase factors. We can exclude the eventuality of the anti-unitarity for this changes the sign of the commutators, which would be inconsistent with the infinitesimal transformations, and we can fix the phase factors on imposing that the set of the operators $\mathbf{U}(\omega, \hat{u})$ equipped with the composition law for the state transformations form a group $\mathcal{G}$ homomorphic to the group $G$ of the spatial rotations. We call to mind that the kernel $\mathcal{K}$ of the homomorphism is an invariant subgroup of $\mathcal{G}$ and that the quotient group $\mathcal{G} / \mathcal{K}$ is isomorphic to $G$ \cite{Messiah}. Anyway, no phase factor should be expected without contradiction for the infinitesimal rotations $R(\delta\omega, \hat{u})$, so that $\lim_{\omega \rightarrow 0} \mathbf{U}(\omega, \hat{u}) = \mathbf{1} ~ \forall \hat{u}$ which in turn implies that 
\begin{equation}
\exists ~ \mathbf{J}_{\hat{u}} : \mathbf{U}(\delta\omega, \hat{u}) = \mathbf{1} - i ~ \delta\omega ~ \mathbf{J}_{\hat{u}} + O(\delta\omega)^{2}
\end{equation}
$\mathbf{J}_{\hat{u}}$ is self-adjoint since $\mathbf{U}(\delta\omega, \hat{u})$ is unitary. 
$\quad \forall \vec{v} ~ R(\delta\omega, \hat{u}) ~ \vec{v} = \vec{v} + \delta\omega (\hat{u} \wedge \vec{v}) + O(\delta\omega^{2})$ so that 
\begin{center} 
$R(\delta\rho, \hat{n}) R(\delta\omega, \hat{u}) ~\vec{v} = \vec{v} + \delta\omega (\hat{u} \wedge \vec{v}) + O(\delta\omega^{2}) + \delta\rho (\hat{n} \wedge \vec{v}) + O(\delta\rho \delta\omega) + O(\delta\rho)^{2}$
\end{center}
\begin{center}
and, $\hat{e}_{i}$ being basis vectors, $R(\delta\omega, \hat{u}) = R(\delta\omega, \sum_{i} (\hat{e}_{i} \cdot \hat{u}) \hat{e}_{i} ) = \Pi_{i} [R((\hat{e}_{i} \cdot \hat{u}) \delta\omega, \hat{e}_{i})]  + O(\delta\omega)^{2}$.
\end{center} 
We deduce that 
\begin{equation}
\mathbf{J}_{\hat{u}} = \sum_{i}  (\hat{e}_{i} \cdot \hat{u}) \mathbf{J}_{\hat{e}_{i}} = \hat{u} \cdot \vec{\mathbf{J}}
\end{equation}
by interpreting the operators $\mathbf{J}_{\hat{e}_{i}}$ as the components $\hat{e}_{i} \cdot \vec{\mathbf{J}}$ of a vector $\vec{\mathbf{J}}$ over the basis ($\hat{e}_{i}$). $\vec{\mathbf{J}}$ defines the total angular momentum of the quantum system under concern. A finite rotation can be built up from successive infinitesimal rotations about the same axis, to be precise 
$R(\omega, \hat{u}) = \lim_{N\rightarrow\infty} R(\frac{\omega}{N}, \hat{u})^{N}$. Accordingly,
\begin{equation}
\mathbf{U}(\omega, \hat{u}) = \lim_{N\rightarrow\infty} \left( \mathbf{1} - i ~ \frac{\omega}{N} ~  (\hat{u} \cdot \vec{\mathbf{J}}) + O\left( \frac{\omega}{N} \right)^{2} \right)^{N} = \exp\{ - i ~ \omega ~ ( \hat{u} \cdot \vec{\mathbf{J}} ) \}
\end{equation}
Notice that $\mathbf{U}(\omega + \delta \omega, \hat{u}) = \mathbf{U}(\delta \omega, \hat{u}) \mathbf{U}(\omega, \hat{u}) = [ \mathbf{1} - i ~ \delta\omega ~ ( \hat{u} \cdot \vec{\mathbf{J}} ) ] \mathbf{U}(\omega, \hat{u})$ so $\partial_{\omega} \mathbf{U}(\omega, \hat{u}) = - i ~ ( \hat{u} \cdot \vec{\mathbf{J}} ) \mathbf{U}(\omega, \hat{u})$, which, together with $\mathbf{U}(0, \hat{u}) =  \mathbf{1}$, is solved as $\mathbf{U}(\omega, \hat{u}) =  \exp\{ - i ~ \omega ~ ( \hat{u} \cdot \vec{\mathbf{J}} ) \}$. 
\medskip

\noindent A finite rotation is often described as the succession $R(\omega, \hat{u}) = R(\gamma, \hat{Z})R(\beta, \hat{y}^{\prime})R(\alpha, \hat{z})$ of a rotation about the $\hat{z}-$axis through an angle $\alpha = (\hat{y}, \hat{y}^{\prime})$, a rotation about the new $\hat{y}^{\prime}-$axis  through an angle $\beta = (\hat{z}, \hat{Z})$ and a rotation about the new $\hat{Z}-$axis through an angle $\gamma = (\hat{y}^{\prime}, \hat{Y})$. $\alpha, \beta, \gamma$ are called the Euler angles of the rotation, which then is denoted $R(\alpha, \beta, \gamma)$. $\mathbf{U}(\omega, \hat{u})$ as a function of these angles writes $\mathbf{U}(\alpha, \beta, \gamma) = \exp\{ - i ~ \gamma ~ ( \hat{Z} \cdot \vec{\mathbf{J}} ) \} \exp\{ - i ~ \beta ~ ( \hat{y}^{\prime} \cdot \vec{\mathbf{J}} ) \} \exp\{ - i ~ \alpha ~ ( \hat{z} \cdot \vec{\mathbf{J}} ) \}$. The same rotation is obtained by first performing a rotation about the $\hat{z}-$axis through an angle $\gamma$, then a rotation about the \emph{initial} $\hat{y}-$axis through an angle $\beta$ and finally a rotation about the \emph{initial} $\hat{z}-$axis through an angle $\alpha$, that is we also have $R(\omega, \hat{u}) = R(\alpha, \hat{z})R(\beta, \hat{y})R(\gamma, \hat{z})$ and 
\begin{equation}
\mathbf{U}(\alpha, \beta, \gamma) = \exp\{ - i ~ \alpha ~ ( \hat{z} \cdot \vec{\mathbf{J}} ) \} \exp\{ - i ~ \beta ~ ( \hat{y} \cdot \vec{\mathbf{J}} ) \} \exp\{ - i ~ \gamma ~ ( \hat{z} \cdot \vec{\mathbf{J}} ) \}
\end{equation}
Let {\bf O} be an operator acting on the states $\vert \psi \rangle$ of the quantum system. If this is submitted to the rotation $R(\omega, \hat{u})$ then $\forall ~\vert \psi \rangle ~ \langle \psi \vert \mathbf{O} \vert \psi \rangle = \langle \psi^{\prime} \vert [\mathbf{U}(\omega, \hat{u})] ~ \mathbf{O} ~ \mathbf{U}(\omega, \hat{u})^{+} \vert \psi^{\prime} \rangle = \langle \psi^{\prime} \vert \mathbf{O}^{\prime} \vert \psi^{\prime} \rangle$. We deduce that the operator transformations associated with $R(\omega, \hat{u})$ are given by
\begin{equation}
\mathbf{O} \rightarrow \mathbf{O}^{\prime} = \mathbf{U}(\omega, \hat{u}) ~ \mathbf{O} ~ [\mathbf{U}(\omega, \hat{u})]^{+}
\end{equation}
$\mathbf{S}$ is a scalar operator if and only if $\forall \omega ~ \forall \hat{u} ~ \mathbf{U}(\omega, \hat{u}) ~ \mathbf{S} ~ [\mathbf{U}(\omega, \hat{u})]^{+} = \mathbf{S}$ and $\mathbf{V}_{\hat{e}_{i}}$ is the component $\hat{e}_{i} \cdot \vec{\mathbf{V}}$ of a vector operator $\vec{\mathbf{V}}$ if and only if $\forall \omega ~ \forall \hat{u} ~ \mathbf{U}(\omega, \hat{u}) ~ \mathbf{V}_{\hat{e}_{i}} ~ [\mathbf{U}(\omega, \hat{u})]^{+}  = (R(\omega, \hat{u})\hat{e}_{i}) \cdot \vec{\mathbf{V}} = \hat{e}^{\prime}_{i} \cdot \vec{\mathbf{V}}$. When the rotation is infinitesimal the operator transformations take the form
\begin{equation}
\mathbf{O} \rightarrow {\mathbf{O}}^{\prime} = \mathbf{O} - i ~ \delta\omega~ [ ~\hat{u} \cdot \vec{\mathbf{J}}, ~\mathbf{O}~ ] + O(\delta\omega)^{2}
\end{equation}
We deduce that $\vec{\mathbf{V}}$ is a vector operator if and only if $\forall \hat{u} ~\forall \hat{v} ~ [~\hat{u} \cdot \vec{\mathbf{J}}, ~ \hat{v}\cdot\vec{\mathbf{V}}~ ] = i ~(\hat{u} \wedge \hat{v}) \cdot \vec{\mathbf{V}}$, since $\hat{v}\cdot\vec{\mathbf{V}} \rightarrow \hat{v}\cdot\vec{\mathbf{V}} - i ~ \delta\omega~ [ ~\hat{u} \cdot \vec{\mathbf{J}}, ~ \hat{v}\cdot\vec{\mathbf{V}}~ ] + O(\delta\omega)^{2}$ but also $\hat{v}\cdot\vec{\mathbf{V}} \rightarrow (R(\omega, \hat{u})~ \hat{v}) \cdot \vec{\mathbf{V}} = (\hat{v} + \delta\omega (\hat{u} \wedge \hat{v}) + O(\delta\omega^{2})) \cdot \vec{\mathbf{V}}$. An alternative definition of angular momenta is extracted from this, namely as vector operators $\vec{\mathbf{J}}$ satisfying the commutation relations
\begin{equation}
[ ~\vec{u}\cdot\vec{\mathbf{J}}, ~\vec{v}\cdot\vec{\mathbf{J}}~] = i ~ \left( \vec{u}\wedge\vec{v} \right)\cdot\vec{\mathbf{J}}
\end{equation}
where $(\vec{u}, \vec{v})$ is any pair of ordinary vectors or of vector operators that commute with each other and with $\vec{\mathbf{J}}$. Notice that this is a more general concept, which coincide with that of the total angular momentum of the quantum system under concern solely if for all the associated scalar $\mathbf{S}$ and vector $\vec{\mathbf{V}}$ operators we further have $[~\vec{u}\cdot\vec{\mathbf{J}}, ~\mathbf{S}~] = 0$ and $\forall \vec{u} ~ \forall \vec{v}~ [~\vec{u} \cdot \vec{\mathbf{J}}, ~\vec{v}\cdot\vec{\mathbf{V}}~] = i ~(\vec{u} \wedge \vec{v}) \cdot \vec{\mathbf{V}}$.
\medskip

\noindent Using the covariant $\hat{e}_{m}$ or contravariant $\hat{e}^{m} = \hat{e}_{m}$ Cartesian basis vectors in place of the vectors $\vec{u}$ and $\vec{v}$ the commutation relations defining $\vec{\mathbf{J}}$ are merely expressed in the form
\begin{equation}
\label{eq:com}
[~\mathbf{J}_{m}, ~\mathbf{J}_{n}~] = i ~ \epsilon^{mnl} \mathbf{J}_{l}
\end{equation}
where $\mathbf{J}_{m} = \hat{e}_{m} \cdot\vec{\mathbf{J}}  \quad (m = x, y, z)$ are the Cartesian components of $\vec{\mathbf{J}}$ and where $\epsilon^{mnl} = 1$ if $(m, n, l)$ is co-cyclic to $(x, y, z)$, $= -1$ if $(m, n, l)$ is anti-cyclic to $(x, y, z)$ and $= 0$ otherwise. Using the covariant spherical basis vectors $\hat{e}_{-1} = \frac{1}{\sqrt{2}} (\hat{e}_{x} - i ~ \hat{e}_{y})$, $\hat{e}_{0} = \hat{e}_{z}$, $\hat{e}_{+1} = -\frac{1}{\sqrt{2}} (\hat{e}_{x} + i ~ \hat{e}_{y})$ we get
\begin{equation}
[~\mathbf{J}_{\mu}, ~\mathbf{J}_{\nu}~] = -\sqrt{2} ~ \langle1\mu1\nu\vert1\lambda\rangle ~ \mathbf{J}_{\lambda} \quad (\mu, \nu, \lambda = \pm1, 0)
\end{equation}
where
\begin{gather*}
\mathbf{J}_{-1} = \hat{e}_{-1} \cdot\vec{\mathbf{J}} = \frac{1}{\sqrt{2}} (\mathbf{J}_{x} - i ~ \mathbf{J}_{y}), \quad \mathbf{J}_{0} = \hat{e}_{0} \cdot\vec{\mathbf{J}} = \mathbf{J}_{z}, \quad \mathbf{J}_{+1} = \hat{e}_{+1} \cdot\vec{\mathbf{J}} = -\frac{1}{\sqrt{2}} (\mathbf{J}_{x} + i ~ \mathbf{J}_{y}) \\  \langle1\mu1\nu\vert1\lambda\rangle = (\mu-\nu)\sqrt{\frac{(1+\mu+\nu)!(1-\mu-\nu)!}{2(1+\mu)!(1-\mu)!(1+\nu)!(1-\nu)!}} \delta_{\mu + \nu, \lambda} 
\end{gather*} 
are the covariant spherical components of $\vec{\mathbf{J}}$ and a Clebsch-Gordan coefficient (cf.~Appendix~\ref{AMC}). Using the contravariant spherical basis vectors $\hat{e}^{\mu} = \hat{e}_{\mu}^{\ast} = (-)^{\mu} \hat{e}_{-\mu}$ we rather get 
\begin{equation}
[~\mathbf{J}^{\mu}, ~\mathbf{J}^{\nu}~] = \sqrt{2} ~ \langle1\mu1\nu\vert1\lambda\rangle ~ \mathbf{J}^{\lambda}  \quad (\mu, \nu, \lambda = \pm1, 0)
\end{equation}
where $\mathbf{J}^{\mu} =  [ \mathbf{J}_{\mu} ]^{+} = (-)^{\mu} \mathbf{J}_{-\mu}$ are the contravariant spherical components of $\vec{\mathbf{J}}$. We shall take this opportunity to call to mind that $\hat{e}_{\mu} \cdot \hat{e}^{\nu} = \delta_{\mu, \nu}$ (= 1 if $\mu = \nu$ and = 0 if $\mu \neq \nu$) and that 
\begin{center}
if $\vec{A} = \sum_{\mu} A^{\mu}\hat{e}_{\mu} = \sum_{\mu} A_{\mu}\hat{e}^{\mu}$ and $\vec{B} = \sum_{\mu} B^{\mu}\hat{e}_{\mu} = \sum_{\mu} B_{\mu}\hat{e}^{\mu}$
\end{center}
\begin{center}
 then $\vec{A}\cdot\vec{B} = \sum_{\mu} A_{\mu}B^{\mu} = \sum_{\mu} (-)^{\mu} A_{\mu}B_{-\mu}$
\end{center}
\begin{center}
whereas $\{ \vec{A}\wedge\vec{B} \}_{\lambda} = -i ~ \sqrt{2} \sum_{\mu, \nu} \langle 1\mu1\nu\vert1\lambda\rangle A_{\mu}B_{\nu}$ and $\{ \vec{A}\wedge\vec{B} \}^{\lambda} = i ~ \sqrt{2} \sum_{\mu, \nu} \langle 1\mu1\nu\vert1\lambda\rangle A^{\mu}B^{\nu}$.
\end{center} 
Whatever the basis vectors $\hat{e}_{i}$ it is effortlessly inferred that $[~\vec{\mathbf{J}}^{2}, ~\hat{e}_{i}\cdot\vec{\mathbf{J}}~] = 0$, in particular
\begin{equation}
[~\vec{\mathbf{J}}^{2}, ~\hat{e}_{\mu}\cdot\vec{\mathbf{J}}~] = 0 \quad (\mu = -1, 0, 1) \qquad \vert \qquad [~\vec{\mathbf{J}}^{2}, ~\hat{e}_{m}\cdot\vec{\mathbf{J}}~] = 0 \quad (m = x, y, z)
\end{equation}
$\vec{\mathbf{J}}^{2}$ commuting with $\mathbf{J}_{0}$ share with it common eigenstates $\vert \tau JM\rangle$, which, because the operators are self-adjoint, are orthonormal : $\langle \tau JM \vert \tau^{\prime} J^{\prime}M^{\prime}\rangle = \delta_{\tau, \tau^{\prime}} \delta_{J, J^{\prime}} \delta_{M, M^{\prime}}$. Using the commutation relations $[~\mathbf{J}_{\mu}, ~\mathbf{J}_{\nu} ~] = -\sqrt{2} ~ \langle1\mu1\nu\vert1\lambda\rangle ~ \mathbf{J}_{\lambda}$ and $[~\vec{\mathbf{J}}^{2}, ~\mathbf{J}_{\mu}~] = 0 ~ (\mu, \nu, \lambda = \pm1, 0)$ together with the fact $\vec{\mathbf{J}}^{2} = \sum_{\mu}(-)^{\mu} \mathbf{J}_{\mu}\mathbf{J}_{-\mu}$, it is an easy matter to show that (cf. the case $\vec{\mathbf{J}} = \vec{\mathbf{L}}$ in Appendix~\ref{ME}).
\begin{equation}
\label{eq:AMR}
\begin{split}
\vec{\mathbf{J}}^{2} \vert \tau JM\rangle & = J(J+1) \vert \tau JM\rangle
\\
\mathbf{J}_{0}  \vert \tau JM\rangle & = M \vert \tau JM\rangle
\\
\mathbf{J}_{+1}  \vert \tau JM\rangle & = -X_{M} \vert \tau JM+1\rangle
\\
\mathbf{J}_{-1}  \vert \tau JM\rangle & = X_{M-1} \vert \tau JM-1\rangle
\end{split}
\end{equation}
where $X_{M} = \left[ \frac{1}{2} (J-M)(J+M+1) \right]^{\frac{1}{2}}$ with $2J\in\mathbb{N}$ and $2M\in\mathbb{Z} : -J \leq M \leq J$, that is $J = 0, \frac{1}{2}, 1, \frac{3}{2}, 2, \cdots$ and, $J$ being fixed, $M = -J, -J+1,  \cdots, J-1, J$. $\tau$ distinguishes between the orthogonal eigenstates of $\vec{\mathbf{J}}^{2}$ with the same eigenvalue $J(J+1)$. The set $\{ \vert \tau JM\rangle \}_{-J \leq M \leq J}$ engenders a state subspace $\mathcal{S}_{\tau, J}$ of dimension $2J+1$, which then admits the closure relation
\begin{equation}
\sum_{M = -J}^{J} \vert \tau JM\rangle \langle \tau JM \vert = \mathbf{1}_{\tau J}
\end{equation}
It finally may be shown using the same method as for the spherical harmonics (cf.~Appendix~\ref{ME}) that the set $\{\vert \tau JM\rangle\}_{\tau JM}$ is complete in the state space $\mathcal{S}$ of the quantum system under concern.

\subsection{\label{AMB}Wigner D-Matrix}

\noindent The matrix representatives of the operators $\{\mathbf{J}_{\mu}\}_{\mu = -1, 0, +1}$ over the basis $\{\vert \tau JM\rangle\}_{\tau JM}$, traceless and hermitian, materialize the irreducible matrix representations of the $\mathfrak{su}(2)$ Lie algebra. These ascend faithfully to the associated $SU(2)$ Lie group, because this is simply connected, namely the irreducible components $\mathcal{S}_{\tau, J}$ of the representation space $\mathcal{S}$ are also those of $SU(2)$ and the matrix representatives of the operators $\mathbf{U}(\omega, \hat{u})$ or $\mathbf{U}(\alpha, \beta, \gamma)$ over the  basis $\{\vert \tau JM\rangle\}_{\tau JM}$ provides with the irreducible matrix representations of $SU(2)$. Using the Euler angles 
\begin{gather}
\label{eq:WDF}
\langle\tau^{\prime} J^{\prime}M^{\prime}\vert ~\mathbf{U}(\alpha, \beta, \gamma)~ \vert \tau JM\rangle = {}
\\ \nonumber
= \langle\tau^{\prime} J^{\prime}M^{\prime}\vert ~\exp\{ - i ~ \alpha ~ ( \hat{z} \cdot \vec{\mathbf{J}} ) \} \exp\{ - i ~ \beta ~ ( \hat{y} \cdot \vec{\mathbf{J}} ) \} \exp\{ - i ~ \gamma ~ ( \hat{z} \cdot \vec{\mathbf{J}} ) \}~ \vert \tau JM\rangle = {}
\\ \nonumber
= \delta_{\tau^{\prime}, \tau} \delta_{J^{\prime}, J} ~ \exp\{ - i ~ \alpha M^{\prime} \} ~d_{M^{\prime}M}^{J}(\beta)~ \exp\{ - i ~ \gamma M \} = \delta_{\tau^{\prime}, \tau} \delta_{J^{\prime}, J} ~ \mathcal{D}_{M^{\prime}M}^{J} (\alpha, \beta, \gamma)
\end{gather}
$\mathcal{D}_{M^{\prime}M}^{J}(\alpha, \beta, \gamma)$ is sometimes dubbed a Wigner D-function and $\mathcal{D}^{J}(\alpha, \beta, \gamma)$ a Wigner D-matrix. 
\medskip 

\noindent $\vec{\mathbf{J}} \equiv \frac{1}{2}\vec{\boldsymbol \sigma}$ for $J=\frac{1}{2}$, where $\vec{\boldsymbol \sigma}$ is the Pauli operator. The matrix representatives of its Cartesian components, $\left(\hat{x} \cdot \vec{\boldsymbol \sigma}\right), \left(\hat{y} \cdot \vec{\boldsymbol \sigma}\right), \left(\hat{z} \cdot \vec{\boldsymbol \sigma}\right)$, over the basis $\{\vert \tau \frac{1}{2} +\frac{1}{2} \rangle, \vert \tau \frac{1}{2} -\frac{1}{2} \rangle\}$ are the Pauli matrices
\begin{equation}
\label{eq:pma}
\sigma_{x} = \left(\begin{array}{cc}0 & 1 \\1 & 0\end{array}\right), \quad \sigma_{y} = \left(\begin{array}{cc}0 & -i \\i & 0\end{array}\right), \quad \sigma_{z} = \left(\begin{array}{cc}1 & 0 \\0 & -1\end{array}\right)
\end{equation}
$\sigma_{x}^{2} = \sigma_{y}^{2} = \sigma_{z}^{2} = 1_{2\times2}$ (unit $2\times2$ matrix) and $\sigma_{m}\sigma_{n} = -\sigma_{n}\sigma_{m} = i\sigma_{l}$ for any cyclic permutation $(m, n, l)$ of $(x, y, z)$, which allows showing that 
\begin{equation}
\label{eq:PSI}
(\vec{u} \cdot \vec{\boldsymbol \sigma})(\vec{v} \cdot \vec{\boldsymbol \sigma}) = (\vec{u} \cdot \vec{v}) ~\mathbf{1} + i ~ \vec{\boldsymbol \sigma} \cdot (\vec{u} \wedge \vec{v})
\end{equation} 
for any pair $(\vec{u}, \vec{v})$ of ordinary vectors or of vector operators commuting with $\vec{\boldsymbol \sigma}$ but not necessarily with each other. In particular, $\forall \hat{u} ~ (\hat{u} \cdot \vec{\boldsymbol \sigma})^{2} = (\hat{u} \cdot \hat{u}) ~\mathbf{1} = \mathbf{1}$ ($\hat{u}$ :  unit vector). So 
\begin{equation}
\exp \{-i ~ \frac{\beta}{2} ~ (\hat{u} \cdot \vec{\boldsymbol \sigma}) \} = \cos (\frac{\beta}{2})~\mathbf{1} - i~ \sin (\frac{\beta}{2}) ~ ( \hat{u} \cdot \vec{\boldsymbol \sigma})
\end{equation} 
and
\begin{equation}
\mathcal{D}^{\frac{1}{2}}(\alpha, \beta, \gamma) = \left(\begin{array}{cc}\cos(\frac{\beta}{2}) \exp\{- i ~ \frac{\alpha+\gamma}{2}\} & -\sin(\frac{\beta}{2}) \exp\{- i ~ \frac{\alpha-\gamma}{2}\} \\ \\\sin(\frac{\beta}{2}) \exp\{ i ~ \frac{\alpha-\gamma}{2}\} & \cos(\frac{\beta}{2}) \exp\{ i ~ \frac{\alpha+\gamma}{2}\} \end{array}\right)
\end{equation}
\medskip 

\noindent $\mathcal{D}^{J}(\alpha, \beta, \gamma)$ or rather $d_{M^{\prime}M}^{J}(\beta)$ for any $J$ is elegantly evaluated from $\mathcal{D}^{\frac{1}{2}}(\alpha, \beta, \gamma)$ using the spinor formalism, which, for each fixed $\tau$, allows building up the angular momentum states $\vert \tau JM\rangle$ from tensor products of the spinor basis states $\chi_{\pm} = \exp\{i~\varphi_{\pm}\} \vert \tau \frac{1}{2} \pm\frac{1}{2} \rangle$. As to avoid writing heaviness the tensor products will be denoted multiplicatively, for instance $\chi_{1}\cdots\chi_{N} = \prod_{l = 1}^{N}\chi_{l}$ for $(\eta^{+}_{1}\chi_{+} + \eta^{-}_{1}\chi_{-}) \otimes \cdots \otimes(\eta^{+}_{N}\chi_{+} + \eta^{-}_{N} \chi_{-}) = \bigotimes_{l = 1}^{N} (\eta^{+}_{l}\chi_{+} + \eta^{-}_{l}\chi_{-}), ~ \eta^{\pm}_{l} \in \mathbb{C},$ and $\chi^{a}$ for $\chi \overset{\mathrm{a-times}}{\cdots} \chi$. One of the essential tools of the spinor calculus is provided by the mappings 
\begin{equation}
\partial_{\chi_{\pm}} : \chi_{1}\cdots\chi_{N} \mapsto \partial_{\chi_{\pm}} (\chi_{1}\cdots\chi_{N}) = \sum_{l = 1}^{N} ~ \chi_{1}\cdots\chi_{l-1}~\langle\chi_{\pm}~\vert~ \chi_{l} \rangle ~\chi_{l+1} \cdots\chi_{N}
\end{equation}
which might be interpreted as derivation operators. On examining the action of the components of the vector operator $\vec{\mathbf{J}}$ on the states $\chi_{\pm} = \exp\{i~\varphi_{\pm}\} \vert \tau \frac{1}{2} \pm\frac{1}{2} \rangle$ (cf.~eq.~\eqref{eq:AMR}) the following identities are inferred
\begin{gather}
\label{eq:AMS}
\mathbf{J}_{-1} = \frac{1}{\sqrt{2}} \exp\{i~[\varphi_{+} - \varphi_{-}]\} \chi_{-}\partial_{\chi_{+}}, \quad \mathbf{J}_{0} = \frac{1}{2}(\chi_{+}\partial_{\chi_{+}} - \chi_{-}\partial_{\chi_{-}}), 
\\ \nonumber
\mathbf{J}_{+1} = -\frac{1}{\sqrt{2}} \exp\{-i~[\varphi_{+} - \varphi_{-}]\}  \chi_{+}\partial_{\chi_{-}},
\\ \nonumber
\vec{\mathbf{J}}^{2} = \mathbf{K}(\mathbf{K}+ \mathbf{1}) \quad \text{with} \quad \mathbf{K} = \frac{1}{2}(\chi_{+}\partial_{\chi_{+}} + \chi_{-}\partial_{\chi_{-}})
\end{gather}
Using these it easily is shown that $\forall a \in \mathbb{N} ~\forall b \in \mathbb{N} ~\mathbf{J}_{0} (\chi_{+}^{a}\chi_{-}^{b}) =  \frac{1}{2}(a-b)(\chi_{+}^{a}\chi_{-}^{b})$ and $\vec{\mathbf{J}}^{2} (\chi_{+}^{a}\chi_{-}^{b}) =  (\frac{a+b}{2})(\frac{a+b}{2}+1)(\chi_{+}^{a}\chi_{-}^{b})$, that is any monomial $\chi_{+}^{a}\chi_{-}^{b}$ is an eigenstate of $\mathbf{J}_{0}$ and of $\vec{\mathbf{J}}^{2}$ simultaneously. It also is found out that $\mathbf{J}_{-1} (\chi_{+}^{a}\chi_{-}^{b}) =  \frac{1}{\sqrt{2}}a \exp\{i~[\varphi_{+} - \varphi_{-}]\} (\chi_{+}^{a-1}\chi_{-}^{b+1})$ and $\mathbf{J}_{+1} (\chi_{+}^{a}\chi_{-}^{b}) =  -\frac{1}{\sqrt{2}}b \exp\{-i~[\varphi_{+} - \varphi_{-}]\} (\chi_{+}^{a+1}\chi_{-}^{b-1})$, that is applying $\mathbf{J}_{-1}$ and $\mathbf{J}_{+1}$ on any monomial $\chi_{+}^{a}\chi_{-}^{b}$ allows generating the irreducible component $\mathcal{S}_{\tau, \frac{a+b}{2}}$ of the representation space $\mathcal{S}$. All this merely suggests that 
\begin{equation*}
\vert \tau JM\rangle \equiv C(J, M) ~ \chi_{+}^{J+M}\chi_{-}^{J-M}
\end{equation*}
If $C(J, J)$ is fixed to $\frac{1}{\sqrt{(2J)!}}$ then, since $(\mathbf{J}_{-1} )^{N} \vert \tau JJ\rangle = (\frac{1}{\sqrt{2}})^{N} \sqrt{\frac{(2J)!N!}{(2J-N)!}} \vert \tau JJ-N\rangle$, we find 
\begin{equation*}
C(J, M) = \frac{(\exp\{i~[\varphi_{+} - \varphi_{-}]\})^{J-M}}{\sqrt{(J+M)!(J-M)!}}
\end{equation*}
by applying $J-M$ times $\mathbf{J}_{-1}$ on $\vert \tau JJ\rangle$. We are free to also fix the phase $(\varphi_{+} - \varphi_{-})$ with some arbitrariness. A conventional choice is $(\varphi_{+} - \varphi_{-}) = (2m+1) \pi$, with $m \in \mathbb{Z}$, so that $\exp\{i~[\varphi_{+} - \varphi_{-}]\} = -1$ and
\begin{equation}
\label{eq:AMT}
\vert \tau JM\rangle \equiv (-)^{J-M} \frac{\chi_{+}^{J+M}\chi_{-}^{J-M}}{\sqrt{(J+M)!(J-M)!}}
\end{equation}
$\mathbf{U}(0, \beta, 0) \vert \tau JM\rangle$ now can be computed either in the form
\begin{equation}
(-)^{J-M}\frac{(\chi_{+} \cos\frac{\beta}{2} - \chi_{-} \sin\frac{\beta}{2})^{J+M} (\chi_{+} \sin\frac{\beta}{2} + \chi_{-} \cos\frac{\beta}{2} )^{J-M}}{\sqrt{(J+M)!(J-M)!}} 
\end{equation}
or
\begin{equation}
\sum_{M^{\prime} = -J}^{J} \vert \tau JM^{\prime}\rangle \langle \tau JM^{\prime} \vert \mathbf{U}(0, \beta, 0) \vert \tau JM\rangle = \sum_{M^{\prime} = -J}^{J} (-)^{J-M^{\prime}}\frac{\chi_{+}^{J+M^{\prime}}\chi_{-}^{J-M^{\prime}}}{\sqrt{(J+M^{\prime})!(J-M^{\prime})!}} ~d_{M^{\prime}M}^{J}(\beta)
\end{equation}
Using the identities
\begin{gather*}
\partial_{x}^{n} (1 \pm x)^{l} = (\pm)^{n} \frac{l!}{(n-l)!} ~(1\pm x)^{n-l}
\\
(x+y)^{n} = \sum_{k = 0}^{n} \frac{n!}{k!(n-k)!} ~ x^{k}~y^{n-k} \quad \text{and} \quad \partial_{x}^{n} (fg) = \sum_{k = 0}^{n} \frac{n!}{k!(n-k)!} ~\partial_{x}^{k} f ~\partial_{x}^{n-k} g
\end{gather*}
we deduce
\begin{multline}
\label{eq:AMV}
d_{M^{\prime}M}^{J}(\beta) = (-)^{J+M^{\prime}} \frac{1}{2^{J}} \left[ \frac{(J-M^{\prime})!}{(J+M^{\prime})! (J+M)! (J-M)!} \right]^{\frac{1}{2}} (1+\cos\beta)^{\frac{M^{\prime}+M}{2}} (1-\cos\beta)^{\frac{M^{\prime}-M}{2}} \times {} 
\\[8pt]
\shoveright{ \times \partial_{\cos\beta}^{J+M^{\prime}} \left[ (1+\cos\beta)^{J-M}(1-\cos\beta)^{J+M} \right] }
\\[8pt]
\shoveleft{\qquad \qquad = (-)^{J-M} \frac{1}{2^{J}} \left[ \frac{(J+M)!}{(J-M)! (J+M^{\prime})! (J-M^{\prime})!} \right]^{\frac{1}{2}} (1+\cos\beta)^{-\frac{M^{\prime}+M}{2}} (1-\cos\beta)^{\frac{M^{\prime}-M}{2}}  \times {} }
\\[8pt]
\times \partial_{\cos\beta}^{J-M} \left[ (1+\cos\beta)^{J+M^{\prime}}(1-\cos\beta)^{J-M^{\prime}} \right]
\end{multline}
then, from this, the expression of $\mathcal{D}^{J}(\alpha, \beta, \gamma)$ for any $J$. We find the symmetry properties
\begin{gather*}
d_{M^{\prime}M}^{J}(-\beta) = d_{MM^{\prime}}^{J}(\beta), \quad d_{-M^{\prime}-M}^{J}(\beta) = (-)^{M^{\prime}-M} d_{M^{\prime}M}^{J}(\beta), \\[4pt] d_{MM^{\prime}}^{J}(\beta) = (-)^{M^{\prime}-M} d_{M^{\prime}M}^{J}(\beta), \quad d_{-M-M^{\prime}}^{J}(\beta) = d_{M^{\prime}M}^{J}(\beta) \\[4pt] d_{M^{\prime}M}^{J}(\beta \pm 2n\pi) = (-)^{2nJ} d_{M^{\prime}M}^{J}(\beta) ~ (n\in\mathbb{N}), \\[4pt] d_{M^{\prime}M}^{J}(\beta \pm (2n+1)\pi) = (-)^{\pm (2n+1)J-M} d_{M^{\prime}-M}^{J}(\beta) ~ (n\in\mathbb{N})
\end{gather*}
\medskip

\noindent Setting $J = l \in \mathbb{N}, M^{\prime} = m \in \mathbb{N} ~(-l \leq m\leq l)\text{ and }M = 0$ or $J = l \in \mathbb{N}, M^{\prime} = 0$ and $M = m \in \mathbb{N} ~(-l \leq m\leq l)$ in the eq.~\eqref{eq:AMV} and comparing with the functional forms of the Legendre polynomials $P_{m}^{l}$ and of the spherical harmonics $Y_{m}^{l}$ (cf.~eq.~\eqref{eq:MEB}) it is found out that
\begin{equation}
d_{m 0}^{l}(\beta) = (-)^{m} \sqrt{\frac{(l-m)!}{(l+m)!}} ~ P_{m}^{l}(\cos\beta) \quad \vert \quad d_{0 m}^{l}(\beta) = (-)^{-m} \sqrt{\frac{(l+m)!}{(l-m)!}} ~ P_{-m}^{l}(\cos\beta)
\end{equation}
so that
\begin{equation}
\label{eq:AMU}
\mathcal{D}_{m 0}^{l}(\alpha, \beta, \gamma) = \sqrt{\frac{4\pi}{2l+1}} ~ Y_{m}^{l}(\beta, -\alpha) \quad \vert \quad \mathcal{D}_{0 m}^{l}(\alpha, \beta, \gamma) = \sqrt{\frac{4\pi}{2l+1}} ~ Y_{m}^{l}(-\beta, -\gamma)
\end{equation}
\medskip

\noindent If $R(\alpha_{3}, \beta_{3}, \gamma_{3}) = R(\alpha_{2}, \beta_{2}, \gamma_{2})~R(\alpha_{1}, \beta_{1}, \gamma_{1})$ describe two consecutive rotations then the Wigner function $\mathcal{D}_{M^{\prime}M}^{J}(\alpha_{3}, \beta_{3}, \gamma_{3}) = \sum_{M^{\prime \prime} = -J}^{J} ~ \mathcal{D}_{M^{\prime}M^{\prime \prime}}^{J}(\alpha_{2}, \beta_{2}, \gamma_{2}) ~ \mathcal{D}_{M^{\prime \prime}M^{\prime \prime}}^{J}(\alpha_{1}, \beta_{1}, \gamma_{1})$
whereas the Euler angles $\alpha_{3}, \beta_{3}, \gamma_{3}$ are given as functions of the Euler angles $\alpha_{1}, \beta_{1}, \gamma_{1}$ and $\alpha_{2}, \beta_{2}, \gamma_{2}$ by
$\cot(\alpha_{3}-\alpha_{2}) = f(\beta_{2}, \beta_{1})$, $ \cos\beta_{3} = \cos\beta_{1}\cos\beta_{2} - \sin\beta_{1}\sin\beta_{2} \cos(\alpha_{1}+\gamma_{2})$ and $\cot(\gamma_{3}-\gamma_{1}) = f(\beta_{1}, \beta_{2})$, where
$f(x,y) = \cos x \cot(\alpha_{1}+\gamma_{2}) + \cot y \frac{\sin x}{\sin(\alpha_{1}+ \gamma_{2})}$, whence by setting $J = l \in \mathbb{N}$ and $M^{\prime} = M = 0$  we recover the addition theorem for the spherical harmonics (cf.~eq.~\eqref{eq:MED}). 
\medskip

\noindent The rotation $R(\omega, \hat{u})$ can be described as the succession $R(\alpha_{3} = \varphi_{\hat{u}}, \beta_{3} = -\theta_{\hat{u}}, \gamma_{3} = -\varphi_{\hat{u}})$ $R(\alpha_{2}, \beta_{2} = 0, \gamma_{2} = 0)$ $R(\alpha_{1} = \varphi_{\hat{u}}, \beta_{1} = \theta_{\hat{u}}, \gamma_{1} = -\varphi_{\hat{u}})$ of a rotation (1) which align the $\hat{z}-$axis along $\hat{u} (\theta_{\hat{u}}, \varphi_{\hat{u}})$, then of a rotation (2) about $\hat{u} (\theta_{\hat{u}}, \varphi_{\hat{u}})$ through an angle $\omega$ and finally of a rotation (3) which send back the $\hat{z}-$axis to its initial direction. Using these rotation angle $\omega-$rotation axis $\hat{u} (\theta_{\hat{u}}, \varphi_{\hat{u}})$ variables $\langle\tau^{\prime} J^{\prime}M^{\prime}\vert ~\mathbf{U}(\omega, \hat{u})~ \vert \tau JM\rangle =  \delta_{\tau^{\prime}, \tau} \delta_{J^{\prime}, J} ~ \mathcal{U}_{M^{\prime}M}^{J}(\omega, \theta_{\hat{u}}, \varphi_{\hat{u}})$ with
\begin{equation}
\mathcal{U}_{M^{\prime}M}^{J}(\omega, \theta_{\hat{u}}, \varphi_{\hat{u}}) = \sum_{M^{\prime \prime} = -J}^{J} \mathcal{D}_{M^{\prime}M^{\prime \prime}}^{J}(\varphi_{\hat{u}}, \theta_{\hat{u}}, -\varphi_{\hat{u}}) ~ \exp\{ -iM^{\prime \prime}\omega \} ~  \mathcal{D}_{M^{\prime \prime}M}^{J}(\varphi_{\hat{u}}, -\theta_{\hat{u}}, -\varphi_{\hat{u}})
\end{equation}
Another way obtaining  $\mathcal{U}_{M^{\prime}M}^{J}(\omega, \theta_{\hat{u}}, \varphi_{\hat{u}})$ is by the direct $(\alpha, \beta, \gamma) \rightarrow (\omega, \theta_{\hat{u}}, \varphi_{\hat{u}})$ variable substitution in $\mathcal{D}_{M^{\prime}M}^{J}(\alpha, \beta, \gamma)$, with the aid of the relations 
\begin{center}
$\cos \left(\frac{\omega}{2} \right) = \cos \left(\frac{\beta}{2} \right)\cos \left(\frac{\alpha+\gamma}{2} \right), \quad \tan(\theta_{\hat{u}}) = \tan \left(\frac{\beta}{2} \right)\left[\sin \left(\frac{\alpha+\gamma}{2} \right)\right]^{-1}, \quad \varphi_{\hat{u}} = \frac{\pi}{2} + \frac{\alpha-\gamma}{2}$ :
\end{center}
\begin{equation}
\mathcal{U}_{M^{\prime}M}^{J}(\omega, \theta_{\hat{u}}, \varphi_{\hat{u}}) = i^{M^{\prime}-M} \exp\{-i(M^{\prime}-M)\varphi_{\hat{u}}\} \left( \frac{1-i\tan\frac{\omega}{2}\cos\theta_{\hat{u}}}{\sqrt{1+\tan^{2}\frac{\omega}{2}\cos^{2}\theta_{\hat{u}}}} \right)^{M^{\prime}+M} d_{M^{\prime}M}^{J}(\xi)
\end{equation}
where $\xi$ is determined by $\sin\frac{\xi}{2} =  \sin\frac{\omega}{2}\sin\theta_{\hat{u}}$. 
\medskip

\noindent The trace $Tr\left[\mathbf{U}(\omega, \hat{u})\right]$ of the operators $\mathbf{U}(\omega, \hat{u}) =  \exp\{ - i ~ \omega ~ ( \hat{u} \cdot \vec{\mathbf{J}} ) \}$ over the  basis $\{\vert \tau JM\rangle\}_{\tau JM}$ provides with the characteristic functions, or simply, the \emph{characters} of the $SU(2)$ irreducible representations. Since $Tr(P_{C}(A_{1} \cdots A_{N})) = Tr(A_{1} \cdots A_{N})$ for any cyclic permutation $P_{C}$ of the operators $A_{1} \cdots A_{N}$, the characters are invariant under any unitary transformation. As a result these are independent of the colatitude $\theta_{\hat{u}}$ and longitude $\varphi_{\hat{u}}$ of the rotation axis, to be precise $Tr\left[\mathbf{U}(\omega, \hat{u})\right] = \chi^{J}(\omega)$ with
\begin{equation}
\chi^{J}(\omega) = \sum_{M = -J}^{J}\exp\{-iM\omega\} = \frac{\sin\left[ (2J+1)\frac{\omega}{2} \right]}{\sin\frac{\omega}{2}} = \frac{1}{2J+1} \partial_{\cos\frac{\omega}{2}} \cos \left[ (2J+1) \frac{\omega}{2} \right]
\end{equation}
$\mathbf{U}$ being unitary $Tr\left[\mathbf{U}^{-1}(\omega, \hat{u})\right] = Tr\left[\mathbf{U}(\omega, \hat{u})\right]$. $\chi^{J}(\omega)$ is a real ($[\chi^{J}(\omega)]^{\ast} = \chi^{J}(\omega)$) and even ($\chi^{J}(-\omega) = \chi^{J}(\omega)$) function of $\omega$, which is $2\pi-$periodic if $2J$ is even and  $2\pi-$antiperiodic if $2J$ is odd ($\chi^{J}(\omega + 2\pi) = (-)^{2J} \chi^{J}(\omega)$). 
\medskip

\noindent The characters $\chi^{J}(\omega)$ of the irreducible representations of $SU(2)$ can be generalized in the form 
\begin{equation}
\chi_{\lambda}^{J}(\omega) = \sqrt{2J+1} \sqrt{\frac{(2J-\lambda)!}{(2J+ \lambda+1)!}} \left( \sin\frac{\omega}{2} \right)^{\lambda} \left( \partial_{\cos\frac{\omega}{2}} \right)^{\lambda} \chi^{J}(\omega)
\end{equation}
where $\lambda \in \mathbb{N}, ~ 0 \leq \lambda \leq 2J$. $\chi_{\lambda}^{J}(\omega)$ is dubbed the associated character of order $\lambda$ of the irreducible representation of rank $J$. $\chi_{0}^{J}(\omega) = \chi^{J}(\omega)$ and $[ \chi_{\lambda}^{J}(\omega) ]^{\ast} = \chi_{\lambda}^{J}(\omega) = (-)^{\lambda} \chi_{\lambda}^{J}(-\omega)$. The collection $\{\chi_{\lambda}^{J}(\omega)\}_{J \geq \frac{\lambda}{2}~\omega\in\left[0, 2\pi\right]}$ exhibits the orthonormality and completeness properties
\begin{equation}
\int_{0}^{2\pi} \sin^{2} \frac{\omega}{2} \chi_{\lambda}^{J_{1}}(\omega) \chi_{\lambda}^{J_{2}}(\omega)~d\omega = \pi \delta_{J_{1}, J_{2}} \quad \vert \quad \sum_{J = \frac{\lambda}{2}}^{\infty} \chi_{\lambda}^{J}(\omega_{1}) \chi_{\lambda}^{J}(\omega_{2}) = \frac{\pi \delta(\omega_{1} - \omega_{2})}{\sin^{2} \frac{\omega_{1}}{2}}
\end{equation}
Using the generalized characters the function $\mathcal{U}_{M^{\prime}M}^{J}(\omega, \theta_{\hat{u}}, \varphi_{\hat{u}})$ may be expanded in a series of spherical harmonics depending on the angles $(\theta_{\hat{u}}, \varphi_{\hat{u}})$ of the rotation axis :
\begin{equation}
\mathcal{U}_{M^{\prime}M}^{J}(\omega, \theta_{\hat{u}}, \varphi_{\hat{u}}) = \sum_{\lambda, \mu} (-i)^{\lambda} \frac{2 \lambda+1}{2J+1} ~\chi_{\lambda}^{J}(\omega)~ \langle JM^{\prime} \lambda\mu \vert JM \rangle~ \sqrt{\frac{4\pi}{2\lambda+1}} ~Y_{\mu}^{\lambda}(\theta_{\hat{u}}, \varphi_{\hat{u}})
\end{equation}
It then is seen that $\mathcal{U}_{M^{\prime}M}^{J}(\omega, \theta_{\hat{u}}, \varphi_{\hat{u}})$ depends on $M^{\prime}$ and $M$ only through the Clebsch-Gordan coefficients $\langle JM^{\prime} \lambda\mu \vert JM \rangle$ (cf.~Appendix~\ref{AMC}) \cite{Varshalovich}.

\subsection{\label{AMC}Clebsch-Gordan and 3jm Symbols}

\noindent Let $\mathcal{S}_{1}$ and $\mathcal{S}_{2}$ be the state spaces of two kinematically independent quantum systems or parts of a quantum system. The state space of their union is the tensor product $\mathcal{S} = \mathcal{S}_{1} \bigotimes \mathcal{S}_{2}$. Under the spatial rotation $R(\omega, \hat{u})$, the states $\vert \psi_{k} \rangle \in \mathcal{S}_{k} ~ (k = 1, 2)$ get transformed into $\vert \psi_{k}^{\prime} \rangle = \mathbf{U}_{k}(\omega, \hat{u}) \vert \psi_{k} \rangle = \exp\{ - i ~ \omega ~ ( \hat{u} \cdot \vec{\mathbf{J}}_{k} ) \} \vert \psi_{k} \rangle$, where $\vec{\mathbf{J}}_{k} ~ (k = 1, 2)$ defines the total angular momenta of each of the two quantum (sub-)systems, therefore the states $\vert \psi_{1} \rangle \otimes \vert \psi_{2} \rangle \in \mathcal{S}$ get transformed into $\vert \psi_{1}^{\prime} \rangle \otimes \vert \psi_{2}^{\prime} \rangle = \mathbf{U}_{1}(\omega, \hat{u}) \vert \psi_{1} \rangle \otimes \mathbf{U}_{2}(\omega, \hat{u}) \vert \psi_{2} \rangle = (\mathbf{U}_{1}(\omega, \hat{u})~\otimes~\mathbf{1}_{1})(\mathbf{1}_{2}~\otimes~\mathbf{U}_{1}(\omega, \hat{u}))  \vert \psi_{1} \rangle \otimes \vert \psi_{2} \rangle = (\mathbf{U}_{1}(\omega, \hat{u}) \otimes \mathbf{U}_{2}(\omega, \hat{u})) \vert \psi_{1} \rangle \otimes \vert \psi_{2} \rangle$, but the whole quantum system is featured by its own total angular momentum $\vec{\mathbf{J}}$ so that $\vert \psi_{1}^{\prime} \rangle \otimes \vert \psi_{2}^{\prime} \rangle = \mathbf{U}(\omega, \hat{u}) \vert \psi_{1} \rangle \otimes \vert \psi_{2} \rangle = \exp\{ - i ~ \omega ~ ( \hat{u} \cdot \vec{\mathbf{J}} ) \} \vert \psi_{1} \rangle \otimes \vert \psi_{2} \rangle$. As a result, $\mathbf{U}(\omega, \hat{u}) = \mathbf{U}_{1}(\omega, \hat{u}) \otimes \mathbf{U}_{2}(\omega, \hat{u})$, namely 
\begin{equation}
\exp\{ - i ~ \omega ~ ( \hat{u} \cdot \vec{\mathbf{J}} ) \} = \exp\{ - i ~ \omega ~ ( \hat{u} \cdot \vec{\mathbf{J}}_{1} ) \} \otimes \exp\{ - i ~ \omega ~ ( \hat{u} \cdot \vec{\mathbf{J}}_{2} ) \}
\end{equation}
and in the case of an infinitesimal rotation $R(\delta\omega, \hat{u})$
\begin{equation*}
\mathbf{1} - i ~ \delta\omega ~ \hat{u} \cdot \vec{\mathbf{J}} + O(\delta\omega)^{2} = \left[ \mathbf{1}_{1} - i ~ \delta\omega ~  \hat{u} \cdot \vec{\mathbf{J}}_{1} + O(\delta\omega)^{2} \right] ~\otimes~ \left[ \mathbf{1}_{2} - i ~ \delta\omega ~ \hat{u} \cdot \vec{\mathbf{J}}_{2} + O(\delta\omega)^{2} \right]
\end{equation*}
so that
\begin{equation}
\vec{\mathbf{J}} = \vec{\mathbf{J}}_{1}~\otimes~\mathbf{1}_{2} + \mathbf{1}_{1}~\otimes~\vec{\mathbf{J}}_{2}
\end{equation}
Usually denoted $\vec{\mathbf{J}} = \vec{\mathbf{J}}_{1} + \vec{\mathbf{J}}_{2}$ \emph{par abus d'\'ecriture}, this defines the addition of two angular momenta. Owing to the kinematic independence of the two quantum (sub-)systems, $[~ \vec{\mathbf{J}}_{1}, ~ \vec{\mathbf{J}}_{2}~] = 0$. As an outcome, $\vec{\mathbf{J}}$ shows the commutation relations of an angular momentum. Additionally, $[~ (\vec{\mathbf{J}}_{k})^{2}, ~\vec{\mathbf{J}}^{2}~] = 0$ and $[~ (\vec{\mathbf{J}}_{k})^{2}, ~\mathbf{J}_{0}~] = 0 ~(k = 1, 2)$, which implies that the eigenstates common to $\vec{\mathbf{J}}^{2}$ and $\mathbf{J}_{0}$ (cf.~eq.~\eqref{eq:AMR}), engendering the irreducible components $\mathcal{S}_{\tau, J}$ of the representation space $\mathcal{S}$, are eigenstates of $(\vec{\mathbf{J}}_{k})^{2}$ with the eigenvalues $J_{k}(J_{k}+1) ~(k = 1, 2)$, hence denoted $\vert \tau J_{1}J_{2}JM\rangle ~(\tau \equiv \tau_{1}\tau_{2} \eta)$. These form a complete set in $\mathcal{S}$. The eigentates $\vert \tau_{k} J_{k}M_{k}\rangle$ common to $(\vec{\mathbf{J}}_{k})^{2}$ and $(\mathbf{J}_{k})_{0}$, engendering the irreducible components $\mathcal{S}_{\tau_{k}, J_{k}}$ of the representation space $\mathcal{S}_{k}$, also form for each $k$ a complete set in $\mathcal{S}_{k}~(k = 1, 2)$, so the states $\vert \tau_{1} J_{1}M_{1}\rangle \otimes \vert \tau_{2} J_{2}M_{2}\rangle$, which we shall denote $\vert \tau_{1}\tau_{2} J_{1}M_{1}J_{2}M_{2} \rangle$, form another complete set in $\mathcal{S} = \mathcal{S}_{1} \bigotimes \mathcal{S}_{2}$. $\vec{\mathbf{J}}^{2}$ is computed equal to $(\vec{\mathbf{J}}_{1})^{2} + (\vec{\mathbf{J}}_{2})^{2} + 2 \sum_{\mu} (-)^{\mu} (\mathbf{J}_{1})_{\mu} (\mathbf{J}_{2})_{\mu}$, so that $[~\vec{\mathbf{J}}_{2}, ~ (\mathbf{J}_{k})_{0}~] \neq 0 ~(k = 1, 2)$, whereas $\mathbf{J}_{0} = (\mathbf{J}_{1})_{0} + (\mathbf{J}_{2})_{0}$ trivially, so that $\vert \tau_{1}\tau_{2} J_{1}M_{1}J_{2}M_{2} \rangle$ is an eigenstate of $\mathbf{J}_{0}$ with the eigenvalue 
\begin{equation}
M = M_{1}+M_{2}
\end{equation} 
Let $N_{\tau_{1}\tau_{2}}(J)$ be the number of sets $\{\vert \tau J_{1}J_{2}JM\rangle\}_{-J \leq M \leq J}$ associated to each eigenvalue $J$ of $\vec{\mathbf{J}}^{2}$ and let $n_{\tau_{1}\tau_{2}}(M)$ be the degeneracy attached to each eigenvalue $M$ of $\mathbf{J}_{0}$, both for fixed quantum numbers $\tau_{1}\tau_{2}$. $n_{\tau_{1}\tau_{2}}(M) = \sum_{J \geq \vert M \vert} N_{\tau_{1}\tau_{2}}(J)$ then $N_{\tau_{1}\tau_{2}}(J) = n_{\tau_{1}\tau_{2}}(J)-n_{\tau_{1}\tau_{2}}(J+1)$, but $n_{\tau_{1}\tau_{2}}(M)$ is merely the number of pairs $(M_{1}, M_{2})$ such that $M = M_{1}+M_{2}$. As to enumerate them it is helpful to take them as points with the abscissa $X = M_{1}$ and ordinate $Y = M_{2}$ in the plane. $n_{\tau_{1}\tau_{2}}(M)$ then is the number of points in the diagonal $X+Y = M$. We find $n_{\tau_{1}\tau_{2}}(M) = 0$ if $\vert M \vert > J_{1}+J_{2}$, $n_{\tau_{1}\tau_{2}}(M) = J_{1}+J_{2}+1-\vert M \vert$ if $J_{1}+J_{2} \geq \vert M \vert \geq \vert J_{1}-J_{2} \vert $ and $n_{\tau_{1}\tau_{2}}(M) = 2\min(J_{1}, J_{2})+1$ if $\vert J_{1}-J_{2} \vert \geq \vert M \vert \geq 0$, which implies that $N(J) = 1$ if $J = J_{1}+J_{2}, J_{1}+J_{2}-1, \cdots, \vert J_{1}-J_{2} \vert$ and $N(J) = 0$ otherwise: the quantum number $\eta$ is limited to a single value then useless ($\tau \equiv \tau_{1}\tau_{2}$) while $J$ must satisfy the \emph{triangular condition} 
\begin{equation}
\vert J_{1}-J_{2} \vert \leq J \leq J_{1}+J_{2}
\end{equation} 
Let $\mathcal{U}^{J_{k}}(\omega, \hat{u})$ be the matrix representatives of the transformation operators $\mathbf{U}_{k}(\omega, \hat{u})$ over the basis $\{\vert \tau_{k} J_{k}M_{k}\rangle\}_{-J_{k} \leq M_{k} \leq J_{k}}$ of $\mathcal{S}_{\tau_{k}, J_{k}}$, defining the irreducible matrix representations of $SU(2)$, for each quantum (sub-)systems $(k = 1, 2)$. The tensor product $\mathcal{U}^{J_{1}}(\omega, \hat{u}) \otimes \mathcal{U}^{J_{2}}(\omega, \hat{u})$ defines a matrix representation of $SU(2)$ for the whole quantum system. Its character is computed to be 
\begin{gather*}
\nonumber Tr\left[ \mathcal{U}^{J_{1}}(\omega, \hat{u}) \otimes \mathcal{U}^{J_{2}}(\omega, \hat{u}) \right] = Tr\left[ \mathcal{U}^{J_{1}}(\omega, \hat{u})\right] Tr\left[ \mathcal{U}^{J_{2}}(\omega, \hat{u}) \right] = {}
\\ \nonumber
= \sum_{M_{1} = -J_{1}}^{J_{1}} \exp \{-i~M_{1}~\omega\} \times \sum_{M_{2} = -J_{2}}^{J_{2}} \exp \{-i~M_{2}~\omega\} = \sum_{J = \vert J_{1}-J_{2} \vert}^{J_{1}+J_{2}} \sum_{M = -J}^{J} \exp \{-i~M~\omega\} :
\end{gather*}
that is
\begin{equation}
\chi^{J_{1}}(\omega)\chi^{J_{2}}(\omega) = \sum_{J} \{J_{1} ~ J_{2} ~ J\} ~ \chi^{J}(\omega)
\end{equation}
where $\{J_{1} ~ J_{2} ~ J\} = 1$ if $\vert J_{1} - J_{2} \vert \leq J \leq J_{1} + J_{2}$ and $\{J_{1} ~ J_{2} ~ J\} = 0$ otherwise. As a corollary, the sum of characters of ranks $J_{1}, J_{1} + 1, J_{1} + 2, \cdots, J_{2}$ may be factorized according to 
\begin{equation}
\sum_{J = J_{1}, J_{1} + 1, \cdots}^{J_{2}}  \chi^{J}(\omega)=  \chi^{\frac{J_{2} + J_{1}}{2}}(\omega)  \chi^{\frac{J_{2} - J_{1}}{2}}(\omega)
\end{equation}
$\mathcal{U}^{J_{1}}(\omega, \hat{u}) \otimes \mathcal{U}^{J_{2}}(\omega, \hat{u}) = \bigoplus_{J = \vert J_{1}-J_{2} \vert}^{J_{1}+J_{2}} \mathcal{U}^{J}(\omega, \hat{u})$, where $\mathcal{U}^{J}(\omega, \hat{u})$ designate the matrix representatives over the basis $\{\vert \tau J_{1}J_{2}JM\rangle\}_{-J \leq M \leq J}$ of the transformation operators $\mathbf{U}(\omega, \hat{u})$ defining the irreducible matrix representations of $SU(2)$ associated with the whole quantum system, and $\mathcal{S}_{\tau_{1}, J_{1}} \otimes \mathcal{S}_{\tau_{2}, J_{2}} = \bigoplus_{J = \vert J_{1}-J_{2} \vert}^{J_{1}+J_{2}}\mathcal{S}_{\tau, J}$. Incidentally, $(2J_{1}+1)(2J_{2}+1) = \sum_{J = \vert J_{1}-J_{2} \vert}^{J_{1}+J_{2}} (2J+1)$ which is easy to directly confirm. 
\medskip

\noindent The sets $\{\vert \tau J_{1}J_{2}JM\rangle\}$ and $\{\vert \tau J_{1}M_{1}J_{2}M_{2} \rangle\}$,  orthonormal and complete in $\mathcal{S}$, are related to each other by a unitary transformation:
\begin{equation}
\label{eq:CGD}
\begin{split}
\vert \tau J_{1}J_{2}JM\rangle = \sum_{M_{1} = -J_{1}}^{J_{1}}\sum_{M_{2} = -J_{2}}^{J_{2}} (\vert \tau J_{1}M_{1}J_{2}M_{2} \rangle \langle \tau J_{1}M_{1}J_{2}M_{2} \vert) ~\vert \tau J_{1}J_{2}JM\rangle
\\
\vert \tau J_{1}M_{1}J_{2}M_{2} \rangle = \sum_{J = \vert J_{1}-J_{2} \vert}^{J_{1}+J_{2}} \sum_{M = -J}^{J} (\vert \tau J_{1}J_{2}JM \rangle \langle \tau J_{1}J_{2}JM \vert) ~\vert \tau J_{1}M_{1}J_{2}M_{2} \rangle
\end{split}
\end{equation}
where the $\langle \tau J_{1}M_{1}J_{2}M_{2} \vert \tau J_{1}J_{2}JM \rangle$ coefficients are dubbed \emph{Clebsch-Gordan}. Complex conjugate to $\langle \tau J_{1}J_{2}JM \vert \tau J_{1}M_{1}J_{2}M_{2} \rangle$ they manifestly are $\tau-$independent, then rather denoted $\langle J_{1}M_{1}J_{2}M_{2} \vert JM \rangle$. These can be evaluated in a variety of ways exploiting unitary properties and recursion formulas or by operator techniques based on commutations or else by means of group theoretical methods. Use can also be made of the spinor formalism, taking advantage of its formal analogy with functional analysis to first determine $\langle J_{1}M_{1}J_{2}M_{2} \vert JJ \rangle$ then deduce $\langle J_{1}M_{1}J_{2}M_{2} \vert JM \rangle$ from the application of $\mathbf{J}_{-1}^{J-M}$ on both side of the equation $\vert \tau J_{1}J_{2}JJ\rangle = \sum_{M_{1}, M_{2}} \langle J_{1}M_{1}J_{2}M_{2} \vert JJ \rangle ~ \vert \tau J_{1}M_{1}J_{2}M_{2} \rangle$: let $(\chi_{k})_{\pm} = \pm \vert \tau \frac{1}{2} \pm\frac{1}{2} \rangle$ be the spinor basis states and $\partial_{(\chi_{k})\pm}$ the associated spinor derivative operators of each of the two quantum (sub-)systems $(k = 1, 2)$ for fixed quantum numbers $\tau_{1} \tau_{2}$. It is inferred from 
\begin{equation}
\mathbf{J}_{+1} \vert \tau J_{1}J_{2}JJ\rangle = \frac{1}{\sqrt{2}} \left[(\chi_{1})_{+}\partial_{(\chi_{1})_{-}} + (\chi_{2})_{+}\partial_{(\chi_{2})_{-}}\right] \vert \tau J_{1}J_{2}JJ\rangle  = 0
\end{equation}
(cf.~eq.~\eqref{eq:AMS} with $(\varphi_{k})_{+} - (\varphi_{k})_{-} = \pi$) and from our knowledge of homogeneous partial differential equations that
\begin{equation}
\vert \tau J_{1}J_{2}JJ\rangle = f \left( \left[ (\chi_{1})_{-}(\chi_{2})_{+})-(\chi_{2})_{-}(\chi_{1})_{+}\right], (\chi_{1})_{+}, (\chi_{2})_{+} \right)
\end{equation}
Additionally, $\vert \tau J_{1}J_{2}JJ\rangle$ is an eigenstate of $\vec{\mathbf{J}}_{k}^{2} = \mathbf{K}_{k}(\mathbf{K}_{k}+ \mathbf{1})$ with eigenvalues $J_{k}(J_{k} + 1)$ thus also of $\mathbf{K}_{k}$ with eigenvalues $J_{k}$. We then get an Euler partial differential equation 
\begin{equation}
(\chi_{k})_{+}\partial_{(\chi_{k})_{+}} + (\chi_{k})_{-}\partial_{(\chi_{k})_{-}} f = 2J_{k} f
\end{equation}
(cf.~eq.~\eqref{eq:AMS}). This implies that $f$ is an homogeneous function of order $2J_{k}$ in $(\chi_{k})_{\pm}$, that is of the form 
\begin{equation}
\vert \tau J_{1}J_{2}JJ\rangle = \sum_{L} C_{L}\left[(\chi_{1})_{-}(\chi_{2})_{+})-(\chi_{2})_{-}(\chi_{1})_{+}\right]^{L}(\chi_{1})_{+}^{2J_{1} - L}(\chi_{2})_{+}^{2J_{2} - L}
\end{equation}
It follows that $\sum_{L} C_{L}(L - J_{1}-J_{2}+J) [(\chi_{1})_{-}(\chi_{2})_{+})-(\chi_{2})_{-}(\chi_{1})_{+}]^{L}(\chi_{1})_{+}^{2J_{1} - L}(\chi_{2})_{+}^{2J_{2} - L} = 0$, since
\begin{equation}
\mathbf{J}_{0} \vert \tau J_{1}J_{2}JJ\rangle = \frac{1}{2}\left\{\sum_{k=1,2}\left[(\chi_{k})_{+}\partial_{(\chi_{k})_{+}} - (\chi_{k})_{-}\partial_{(\chi_{k})_{-}}\right] \right\} \vert \tau J_{1}J_{2}JJ\rangle = J \vert \tau J_{1}J_{2}JJ\rangle
\end{equation} 
(cf.~eq.~\eqref{eq:AMS}). $C_{L}$ then is null unless $L = J_{1}+J_{2}-J$. Accordingly, 
\begin{gather}
\vert \tau J_{1}J_{2}JJ\rangle = C~[(\chi_{1})_{-}(\chi_{2})_{+})-(\chi_{2})_{-}(\chi_{1})_{+}]^{ J_{1}+J_{2}-J}(\chi_{1})_{+}^{J + J_{1} - J_{2}}(\chi_{2})_{+}^{J + J_{2} - J_{1}} = {}
\\ \nonumber
= C~\sum_{X = 0}^{J_{1}+J_{2}-J} \binom{J_{1}+J_{2}-J}{X} \left[~ (\chi_{1})_{-}^{X} (\chi_{1})_{+}^{2J_{1}-X} (-)^{J_{1}+J_{2}-J-X}(\chi_{2})_{-}^{J_{2}-J+J_{1}-X} (\chi_{2})_{+}^{J_{2}+J-J_{1}-X} ~\right]
\end{gather}
Substituting $J_{1}-M_{1}$ for $X$ and $ M_{1} + M_{2}$ for $J$, the summation extends over all the pairs $(M_{1}, M_{2})$ with fixed $J = M_{1} + M_{2}$ (notice that we do have $n_{\tau_{1}\tau_{2}}(M = J) = J_{1}+J_{2}+1-J$). Then
\begin{gather}
\vert \tau J_{1}J_{2}JJ\rangle = C~(J_{1}+J_{2}-J)! \times {}
\\ \nonumber
\times \sum_{M_{1} = -J_{1}}^{J_{1}} \sum_{M_{2} = -J_{2}}^{J_{2}}  (-)^{J_{1}-M_{1}} ~\delta_{M_{1}+M_{2}, J} \sqrt{\frac{(J_{1}+M_{1})!(J_{2}+M_{2})!}{(J_{1}-M_{1})!(J_{2}-M_{2})!}} \vert \tau J_{1}M_{1}J_{2}M_{2}\rangle
\end{gather} 
since $(-)^{J_{k}-M_{k}} (\chi_{k})_{-}^{J_{k}-M_{k}} (\chi_{k})_{+}^{J_{k}+M_{k}} = \sqrt{(J_{k}-M_{k})!(J_{k}+M_{k})!} \vert \tau_{k} J_{k}M_{k}\rangle ~(k = 1, 2)$ (cf.~eq.~\eqref{eq:AMT}). Using the binomial identity 
\begin{gather*}
\sum_{\rho} \binom{n}{\rho} \binom{m}{r-\rho} = \binom{n+m}{r}
\\
\text{where} \quad \binom{p}{k}_{p \geq 0} = \frac{p!}{k!(p-k)!} \quad \text{and} \quad \binom{p}{k}_{p < 0} = (-)^{k} \frac{(-p-1+k)!}{k!(-p-1)!} = (-)^{k} \binom{k-p-1}{k} 
\end{gather*}
obtained by considering the binomial coefficients of $x^{r}y^{m+n-r}$ in the binomial expansion of both side of the equality $(x+y)^{n}(x+y)^{m} = (x+y)^{n+m}$, 
it is a matter of elementary algebra, putting $n = J_{1}-J_{2}-J-1 < 0$, $m = J_{2}-J_{1}-J-1 < 0$, $\rho = J_{1}-M_{1} \geq 0$ and $r = J_{1}+J_{2} - J \geq 0$, to prove that 
\begin{equation}
\sum_{\substack {M_{1} = -J_{1}, \cdots, J_{1} \\ M_{2} = -J_{2}, \cdots, J_{2} }} \delta_{M_{1}+M_{2}, J}\frac{(J_{1}+M_{1})!(J_{2}+M_{2})!}{(J_{1}-M_{1})!(J_{2}-M_{2})!} = \frac{(J+J_{1}-J_{2})!(J+J_{2}-J_{1})!(J+J_{1}+J_{2}+1)!}{(2J+1)!(J_{1}+J_{2}-J)!}
\end{equation} 
so with
\begin{equation*}
\vert C \vert = \left[\frac{(2J+1)!}{(J+J_{1}-J_{2})!(J+J_{2}-J_{1})!(J+J_{1}+J_{2}+1)!(J_{1}+J_{2}-J)!}\right]^{\frac{1}{2}}
\end{equation*} 
$\vert \tau J_{1}J_{2}JJ\rangle$ is normalized to unity. Adopting the conventional choice of phase \cite{Edmonds}, $C$ is fixed to $\vert C \vert$. Now to find the general Clebsch-Gordan coefficients it suffices to operate $J-M$ times on $\vert \tau J_{1}J_{2}JJ\rangle$ with either $\mathbf{J}_{-1} \equiv \frac{1}{\sqrt{2}} [(\chi_{1})_{-}\partial_{(\chi_{1})_{+}} + (\chi_{2})_{-}\partial_{(\chi_{2})_{+}}]$ (notice that $\mathbf{J}_{-1} [(\chi_{1})_{-}(\chi_{2})_{+})-(\chi_{2})_{-}(\chi_{1})_{+}] = 0$, which makes the computation easier) or directly with $\mathbf{J}_{-1} = (\mathbf{J}_{1})_{-1} + (\mathbf{J}_{2})_{-1}$, calling to mind that 
\begin{equation*}
(\mathbf{J}_{-1})^{N} \vert \tau JM \rangle = (\frac{1}{\sqrt{2}})^{N}  \left[ \frac{(J+M)!(J-M+N)!}{(J-M)!(J+M-N)!} \right]^{\frac{1}{2}}\vert \tau JM-N \rangle 
\end{equation*}
We find 
\begin{multline}
(\mathbf{J}_{-1})^{J-M} \vert \tau J_{1}J_{2}JJ\rangle = (\frac{1}{\sqrt{2}})^{J-M} \left[ \frac{(2J)!(J-M)!}{(0)!(J+M)!} \right]^{\frac{1}{2}} \vert \tau J_{1}J_{2}JM\rangle
\\
\shoveright{ = C~(J_{1}+J_{2}-J)! \sum_{M_{1} = -J_{1}}^{J_{1}} \sum_{M_{2} = -J_{2}}^{J_{2}}  (-)^{J_{1}-M_{1}} ~\delta_{M_{1}+M_{2}, J} \sqrt{\frac{(J_{1}+M_{1})!(J_{2}+M_{2})!}{(J_{1}-M_{1})!(J_{2}-M_{2})!}} \times {} }
\\
\shoveright{ \times \sum_{X = 0}^{J-M} \binom{J-M}{X} \left[(\mathbf{J}_{1})_{-1}\right]^{X} \left[(\mathbf{J}_{2})_{-1}\right]^{J-M-X} \vert \tau J_{1}M_{1}J_{2}M_{2}\rangle = {} }
\\
= (\frac{1}{\sqrt{2}})^{J-M}~C~(J_{1}+J_{2}-J)! \sum_{M_{1} = -J_{1}}^{J_{1}} \sum_{M_{2} = -J_{2}}^{J_{2}}  (-)^{J_{1}-M_{1}} ~\delta_{M_{1}+M_{2}, J} \frac{(J_{1}+M_{1})!(J_{2}+M_{2})!}{(J_{1}-M_{1})!(J_{2}-M_{2})!} \times {}
\\
\shoveright{ \times \sum_{ \substack{X = 0, \cdots, J-M \\ M_{1}-X \geq -J_{1} \\ M_{2}-J+M+X \geq -J_{2}} } \frac{(J-M)!}{X!(J-M-X)!} \left[ \frac{(J_{1}-M_{1}+X)!(J_{2}-M_{2}+J-M-X)!}{(J_{1}+M_{1}-X)!(J_{2}+M_{2}-J+M+X)!} \right]^{\frac{1}{2}}  \times {} }
\\
\times \vert \tau J_{1}(M_{1}-X)J_{2}(M_{2}-J+M+X)\rangle
\end{multline}
It finally is inferred, after the substitutions $M_{1} = X+\mu_{1}$ and $M_{2} = J-M-X+\mu_{2}$, that
\begin{multline}
\label{eq:CG0}
\langle J_{1} \mu_{1}J_{2} \mu_{2} \vert JM \rangle = \delta_{\mu_{1}+ \mu_{2}, M}~(-)^{J_{1}-\mu_{1}} \left[ \frac{(2J+1) (J_{1}+J_{2}-J)! (J+M)! (J-M)! }{ (J+J_{1}-J_{2})!(J+J_{2}-J_{1})!(J+J_{1}+J_{2}+1)! } \right]^{\frac{1}{2}}  \times {}
\\
\times \left[ \frac{(J_{1}-\mu_{1})! (J_{2}-\mu_{2})! }{ (J_{1}+\mu_{1})! (J_{2}+\mu_{2})! } \right]^{\frac{1}{2}}  \sum_{X \in \mathbb{N}} (-)^{-X} \frac{(J_{1}+\mu_{1}+X)! (J_{2}+J-\mu_{1}-X)!}{X! (J-M-X)! (J_{1}-\mu_{1}-X)! (J_{2}-J+\mu_{1}+X)!}
\end{multline}
with the summation over $X$ such that the arguments of the denominator are non negative. Interestingly this is limited to the single term $X = 0$ when $\mu_{1} =  J_{1}$ and to the single term $X = J_{1}-\mu_{1}$ when $J_{3} = J_{1}+J_{2}$ and $\mu_{3} = \mu_{1}+ \mu_{2}$, which provides with the simpler formulas :
\begin{multline}
\label{eq:CG1}
\langle J_{1}J_{1} J_{2}M-J_{1} \vert JM \rangle = {}
\\
= \left[ \frac{(2J+1) (2J_{1})! (-J_{1}+J_{2}+J)! (J_{1}+J_{2}-M)! (J+M)!}{(J_{1}+J_{2}-J)! (J_{1}-J_{2}+J)! (J_{1}+J_{2}+J+1)! (-J_{1}+J_{2}+M)! (J-M)!} \right]^{\frac{1}{2}}
\end{multline}
\begin{multline}
\label{eq:CG2}
\langle J_{1} \mu_{1} J_{2} \mu_{2} \vert J_{1}+J_{2} \mu_{1}+ \mu_{2} \rangle = {}
\\
= \left[ \frac{(2J_{1})!(2J_{2})!(J_{1}+J_{2}+ \mu_{1}+ \mu_{2})!(J_{1}+J_{2}- \mu_{1}- \mu_{2})!}{(2J_{1}+2J_{2})!(J_{1}-\mu_{1})!(J_{1}+\mu_{1})!(J_{2}-\mu_{2})!(J_{2}+\mu_{2})!} \right]^{\frac{1}{2}}
\end{multline}
\medskip

\noindent 
As evident from its algebraic expression, \emph{the Clebsch-Gordan coefficient $\langle J_{1} \mu_{1}J_{2} \mu_{2} \vert JM \rangle$ is real}. It also \emph{exhibit a high degree of symmetry}. With the substitution $X = J-M-Y$, for instance, we get the result 
\begin{equation}
\label{eq:CG3}
\langle J_{1} \mu_{1}J_{2} \mu_{2} \vert JM \rangle = (-)^{J_{1}+J_{2}-J} \langle J_{2} \mu_{2} J_{1} \mu_{1} \vert JM \rangle
\end{equation}
A number of other symmetry relations can be found out. These are better displayed by defining the \emph{3jm symbol}
\begin{equation}
\label{eq:3jm}
\left(\begin{array}{ccc}j_{1} & j_{2} & j_{3} \\m_{1} & m_{2} & m_{3}\end{array}\right) = (-)^{j_{3}+m_{3}+2j_{1}} \left[ j_{3}^{-1} \right] \langle j_{1}-m_{1}j_{2}-m_{2} \vert j_{3}m_{3} \rangle
\end{equation}
where $\left[ j_{1}^{\epsilon_{1}},  j_{2}^{\epsilon_{2}}, \cdots \right]$ is a standard abbreviation for $\left \{ (2 j_{1}+1)^{\epsilon_{1}}(2 j_{2}+1)^{\epsilon_{2}}\cdots \right \}^{\frac{1}{2}}$. The inverse relation is
\begin{equation*}
\langle j_{1}m_{1}j_{2}m_{2} \vert j_{3}m_{3} \rangle = (-)^{j_{1}-j_{2}+m_{3}} \left[ j_{3} \right] \left(\begin{array}{ccc}j_{1} & j_{2} & j_{3} \\m_{1} & m_{2} & -m_{3}\end{array}\right)
\end{equation*}
The 3jm symbol are null unless the \emph{triangular condition} $\vert j_{1} - j_{2}\vert \leq j_{3} \leq j_{1} + j_{2}$ and the \emph{zero-sum condition} $m_{1} + m_{2} + m_{3} = 0$ are satisfied. 
\medskip

\noindent The 3jm symbols are invariant by cyclic permutations of the columns and multiplied by the phase factor $(-)^{j_{1}+j_{2}+j_{3}}$ by exchange of two columns or by sign change of all the projection quantum numbers :
\begin{equation}
\label{eq:3s1}
\left(\begin{array}{ccc}j_{1} & j_{2} & j_{3} \\m_{1} & m_{2} & m_{3}\end{array}\right) = \left(\begin{array}{ccc}j_{3} & j_{1} & j_{2} \\m_{3} & m_{1} & m_{2}\end{array}\right) = \left(\begin{array}{ccc}j_{2} & j_{3} & j_{1} \\m_{2} & m_{3} & m_{1}\end{array}\right)
\end{equation}
\begin{equation}
\label{eq:3s2}
\left(\begin{array}{ccc}j_{1} & j_{2} & j_{3} \\m_{1} & m_{2} & m_{3}\end{array}\right) = (-)^{j_{1}+j_{2}+j_{3}} \left(\begin{array}{ccc}j_{2} & j_{1} & j_{3} \\m_{2} & m_{1} & m_{3}\end{array}\right)
\end{equation}
\begin{equation}
\label{eq:3s3}
\left(\begin{array}{ccc}j_{1} & j_{2} & j_{3} \\m_{1} & m_{2} & m_{3}\end{array}\right) = (-)^{j_{1}+j_{2}+j_{3}} \left(\begin{array}{ccc}j_{1} & j_{2} & j_{3} \\-m_{1} & -m_{2} & -m_{3}\end{array}\right)
\end{equation}
\medskip

\noindent The  orthonormal properties of the quantum states
\begin{center}
$\langle j_{1}m_{1}j_{2}m_{2} \vert j_{1}m^{\prime}_{1}j_{2}m^{\prime}_{2} \rangle = \langle j_{1}m_{1}j_{2}m_{2} \vert (\sum_{j_{3}, m_{3}} \vert j_{3}m_{3}\rangle \langle j_{3}m_{3} \vert) \vert j_{1}m^{\prime}_{1}j_{2}m^{\prime}_{2} \rangle = \delta_{m_{1}, m^{\prime}_{1}}~\delta_{m_{2}, m^{\prime}_{2}}$ and $\langle j_{1}j_{2}j_{3}m_{3} \vert  j_{1}j_{2}j^{\prime}_{3}m^{\prime}_{3} \rangle = \langle j_{1}j_{2}j_{3}m_{3} \vert (\sum_{m_{1}, m_{2}} \vert  j_{1}m_{1}j_{2}m_{2} \rangle \langle  j_{1}m_{1}j_{2}m_{2} \vert) \vert j_{1}j_{2}j^{\prime}_{3}m^{\prime}_{3} \rangle = \delta_{j_{3}, j^{\prime}_{3}}~\delta_{m_{3}, m^{\prime}_{3}}$
\end{center}
impose on the 3jm symbols the orthonormal conditions
\begin{gather}
\sum_{j_{3}, m_{3}} \left[ j_{3}^{2} \right] \left(\begin{array}{ccc}j_{1} & j_{2} & j_{3} \\m_{1} & m_{2} & m_{3}\end{array}\right) \left(\begin{array}{ccc}j_{1} & j_{2} & j_{3} \\m^{\prime}_{1} & m^{\prime}_{2} & m_{3}\end{array}\right) = \delta_{m_{1}, m^{\prime}_{1}}~\delta_{m_{2}, m^{\prime}_{2}}
\\
\sum_{m_{1}, m_{2}} \left(\begin{array}{ccc}j_{1} & j_{2} & j_{3} \\m_{1} & m_{2} & m_{3}\end{array}\right) \left(\begin{array}{ccc}j_{1} & j_{2} & j^{\prime}_{3} \\m_{1} & m_{2} & m^{\prime}_{3}\end{array}\right) = \left[ j_{3}^{-2} \right] \delta_{j_{3}, j^{\prime}_{3}}~\delta_{m_{3}, m^{\prime}_{3}} ~ \{ j_{1} ~ j_{2} ~ j_{3} \} \qquad
\end{gather}
where $\{ j_{1} ~ j_{2} ~ j_{3} \} = 1$ if $j_{1}, j_{2}, j_{3}$ satisfy the triangular condition and $\{ j_{1} ~ j_{2} ~ j_{3} \} = 0$ otherwise. Since $(\mathbf{j}_{1})_{0} + (\mathbf{j}_{2})_{0} = (\mathbf{j}_{3})_{0}$ is diagonal in both the $\vert \tau j_{1}m_{1}j_{2}m_{2}\rangle$ and $\vert \tau j_{1}j_{2}j_{3}m_{3}\rangle$ representations the orthonormal properties of the quantum states materialize themselves for fixed values of $m_{3} = - m_{1} - m_{2} = m^{\prime}_{3} =  - m^{\prime}_{1} - m^{\prime}_{2}$ as well, which provides with the additional formulas 
\begin{gather}
\sum_{j_{3}} \left[ j_{3}^{2} \right] \left(\begin{array}{ccc}j_{1} & j_{2} & j_{3} \\m_{1} & -m_{1}-m_{3} & m_{3}\end{array}\right) \left(\begin{array}{ccc}j_{1} & j_{2} & j_{3} \\m^{\prime}_{1} & -m^{\prime}_{1}-m_{3} & m_{3}\end{array}\right) = \delta_{m_{1}, m^{\prime}_{1}}
\\
\sum_{m_{1}} \left(\begin{array}{ccc}j_{1} & j_{2} & j_{3} \\m_{1} & -m_{1}-m_{3} & m_{3}\end{array}\right) \left(\begin{array}{ccc}j_{1} & j_{2} & j^{\prime}_{3} \\m_{1} & -m_{1}-m_{3} & m_{3}\end{array}\right) = \left[ j_{3}^{-2} \right] \delta_{j_{3}, j^{\prime}_{3}} ~ \{ j_{1} ~ j_{2} ~ j_{3} \} \qquad
\end{gather}
The Clebsch-Gordan coefficients or the 3jm symbols are the basic quantities of angular momentum coupling and get ubiquitously into the formulation of a number of physical quantities. Let, for instance, $\mathcal{D}_{M^{\prime}M}^{J}(\alpha, \beta, \gamma) = \langle JM^{\prime}\vert ~\mathbf{U}(\alpha, \beta, \gamma)~ \vert JM\rangle$ be the Wigner D-function of a quantum system associated with the rotation $R(\alpha, \beta, \gamma)$ (cf.~eq.~\eqref{eq:WDF}) and let $\mathcal{D}_{M^{\prime}_{k}M_{k}}^{J_{k}}(\alpha, \beta, \gamma) = \langle JM^{\prime}\vert ~\mathbf{U}_{k}(\alpha, \beta, \gamma)~ \vert JM\rangle~ (k = 1, 2)$ be the Wigner D-functions of two kinematically independent components of this system. Calling to mind that $\mathbf{U}(\alpha, \beta, \gamma) = \mathbf{U}_{1}(\alpha, \beta, \gamma) \otimes \mathbf{U}_{2}(\alpha, \beta, \gamma)$ and making use of the appropriate closure relations, it effortlessly is shown that 
\begin{gather*}
\mathcal{D}_{M^{\prime}M}^{J}(\alpha, \beta, \gamma) = \sum_{ \substack{ M^{\prime}_{1}, M^{\prime}_{2} \\ M_{1}, M_{2} } } \langle JM^{\prime} \vert  J_{1}M^{\prime}_{1}J_{2}M^{\prime}_{2} \rangle ~\mathcal{D}_{M^{\prime}_{1}M_{1}}^{J_{1}}(\alpha, \beta, \gamma)~ \mathcal{D}_{M^{\prime}_{2}M_{2}}^{J_{2}}(\alpha, \beta, \gamma) ~\langle J_{1}M_{1}J_{2}M_{2} \vert JM \rangle
\\
\mathcal{D}_{M^{\prime}_{1}M_{1}}^{J_{1}}(\alpha, \beta, \gamma)~ \mathcal{D}_{M^{\prime}_{2}M_{2}}^{J_{2}}(\alpha, \beta, \gamma) = \sum_{J, M^{\prime}, M}  \langle  J_{1}M^{\prime}_{1}J_{2}M^{\prime}_{2} \vert JM^{\prime} \rangle ~\mathcal{D}_{M^{\prime}M}^{J}(\alpha, \beta, \gamma)~ \langle JM \vert  J_{1}M_{1}J_{2}M_{2} \rangle
\end{gather*} 
Setting $J = L \in \mathbb{N}, J_{1} = L_{1} \in \mathbb{N}, J_{2} = L_{2} \in \mathbb{N}$ and $M^{\prime}_{1} = M^{\prime}_{2} = 0$ in the second equation interestingly provides with the frequently used multipole expansion of the product of two spherical harmonics. We indeed get
\begin{equation*}
\mathcal{D}_{0M_{1}}^{L_{1}}(\alpha, \beta, \gamma)~ \mathcal{D}_{0M_{2}}^{L_{2}}(\alpha, \beta, \gamma) = \sum_{L, M}  \langle  L_{1}0L_{2}0 \vert L0 \rangle ~\mathcal{D}_{0M}^{L}(\alpha, \beta, \gamma)~  \langle LM \vert  L_{1}M_{1}L_{2}M_{2} \rangle
\end{equation*}
that is
\begin{equation}
\label{eq:AM1}
\begin{split}
Y_{M_{1}}^{L_{1}}(\Omega) Y_{M_{2}}^{L_{2}}(\Omega) & = \sum_{L, M} \left \{ \frac{(2L_{1}+1)(2L_{2}+1)(2L+1)}{4\pi} \right \}^{\frac{1}{2}} \times {}
\\
& \qquad \times \left(\begin{array}{ccc}L_{1} & L_{2} & L \\0 & 0 & 0\end{array}\right) [Y_{M}^{L}(\Omega)]^{\ast} \left(\begin{array}{ccc}L_{1} & L_{2} & L \\M_{1} & M_{2} & M\end{array}\right) 
\end{split}
\end{equation}
(cf.~eq.~\eqref{eq:AMU}) where $\Omega \equiv (-\beta, -\alpha)$ and $\left[ Y_{M}^{L} (-\beta, -\alpha) \right]^{\ast} = (-)^{M} Y_{-M}^{L} (-\beta, -\alpha)$. Using the orthonormality property of the spherical harmonics (cf.~eq.~\eqref{eq:MEF}) it also is found out that
\begin{equation}
\label{eq:AM2}
\begin{split}
\int d\Omega~ Y_{M_{1}}^{L_{1}}(\Omega) Y_{M_{2}}^{L_{2}}(\Omega) Y_{M}^{L}(\Omega) & = \left \{ \frac{(2L_{1}+1)(2L_{2}+1)(2L+1)}{4\pi} \right \}^{\frac{1}{2}} \times {}
\\
& \qquad \times \left(\begin{array}{ccc}L_{1} & L_{2} & L \\0 & 0 & 0\end{array}\right) \left(\begin{array}{ccc}L_{1} & L_{2} & L \\M_{1} & M_{2} & M\end{array}\right)
\end{split}
\end{equation}
\medskip

\noindent Specialized formulas for the 3jm symbols can be derived from those obtained for the Clebsh-Gordan coefficients. Using the eq. \eqref{eq:CG0} we compute, for instance,
\begin{equation}
\langle 10jm \vert jm \rangle \underset{X = 0, 1}{=} -\frac{m}{\sqrt{j(j+1)}}
\end{equation}
whence
\begin{equation}
\label{eq:3j6}
\left(\begin{array}{ccc}1 & j & j \\0 & -m & m\end{array}\right) = (-)^{j+m-1} \left[ j^{-1} \right] \frac{m}{\sqrt{j(j+1)}} 
\end{equation}
As from the eq.~\eqref{eq:CG1} we deduce 
\begin{multline}
\label{eq:3j1}
\left(\begin{array}{ccc}j_{1} & j_{2} & j_{3} \\j_{1} & -j_{1}+m_{3} & -m_{3}\end{array}\right) = (-)^{-j_{1} + j_{2} - m_{3}} \left[  j_{3}^{-1} \right] \langle j_{1}j_{1} j_{2}-j_{1}+m_{3} \vert j_{3}m_{3} \rangle = {}
\\
= (-)^{-j_{1} + j_{2} - m_{3}} \left[ \frac{ (2j_{1})! (-j_{1}+j_{2}+j_{3})! (j_{1}+j_{2}-m_{3})! (j_{3}+m_{3})!} {(j_{1}+j_{2}- j_{3})! (j_{1}-j_{2}+ j_{3})! (j_{1}+j_{2}+ j_{3}+1)! (-j_{1}+j_{2}+m_{3})! ( j_{3}-m_{3})!} \right]^{\frac{1}{2}}
\\[-12pt]
\end{multline}
and from the eq.~\eqref{eq:CG2}
\begin{multline}
\label{eq:3j2}
\left(\begin{array}{ccc}j_{1} & j_{2} & j_{1}+j_{2} \\m_{1} & m_{2} & -m_{1}-m_{2}\end{array}\right) = (-)^{ -j_{1} + j_{2} - m_{1} - m_{2}} \left[ (j_{1}+j_{2})^{-1} \right]\langle j_{1} m_{1} j_{2} m_{2} \vert j_{1}+j_{2} m_{1}+ m_{2} \rangle = {}
\\
= (-)^{- j_{1} + j_{2} - m_{1} - m_{2}} \left[ \frac{(2j_{1})!(2j_{2})!(j_{1}+j_{2}+ m_{1}+ m_{2})!(j_{1}+j_{2}- m_{1}- m_{2})!}{(2j_{1}+2j_{2}+1)!(j_{1}-m_{1})!(j_{1}+m_{1})!(j_{2}-m_{2})!(j_{2}+m_{2})!} \right]^{\frac{1}{2}}
\end{multline}
\medskip

\noindent A few particular instances of these formulas are
\begin{equation}
\label{eq:3j0}
\left(\begin{array}{ccc}0 & j & j\\0 & -m & m\end{array}\right) = (-)^{j+m} \left[ j^{-1} \right]
\end{equation}
\begin{gather}
\label{eq:3j3}
\left(\begin{array}{ccc}\frac{1}{2} & j-\frac{1}{2} & j \\[4pt] \frac{1}{2} & -m-\frac{1}{2} & m\end{array}\right)  = (-)^{j+m-1} \left[ \frac{j-m}{2j(2j+1)} \right]^{\frac{1}{2}}
\\[8pt]
\left(\begin{array}{ccc}1 & j-1 & j \\1 & -m-1 & m\end{array}\right)  = (-)^{j+m} \left[ \frac{(j-m)(j-m-1)}{(2j-1) 2j (2j+1)} \right]^{\frac{1}{2}}
\end{gather}
as from which, using the symmetry properties of the 3jm symbols (cf.~eq.~\eqref{eq:3s1}, eq.~\eqref{eq:3s2} and eq.~\eqref{eq:3s3}), several other may easily be inferred, for instance
\begin{gather}
\left(\begin{array}{ccc}\frac{1}{2} & j-\frac{1}{2} & j \\[4pt] -\frac{1}{2} & -m+\frac{1}{2} & m\end{array}\right)  = (-)^{-j-m-1} \left[ \frac{j+m}{2j(2j+1)} \right]^{\frac{1}{2}}
\\ \nonumber
\left(\begin{array}{ccc}\frac{1}{2} & j & j-\frac{1}{2} \\[4pt] \frac{1}{2} & m & -m-\frac{1}{2}\end{array}\right)  = (-)^{-j+m-1} \left[ \frac{j-m}{2j(2j+1)} \right]^{\frac{1}{2}}
\\ \nonumber
\vdots
\\ \nonumber
\left(\begin{array}{ccc}1 & j-1 & j \\-1 & -m+1 & m\end{array}\right)  = (-)^{-j-m} \left[ \frac{(j+m)(j+m-1)}{(2j-1) 2j (2j+1)} \right]^{\frac{1}{2}}
\\ \nonumber
\vdots
\end{gather}
A number of recursion relations further easing the computation of the 3jm symbols may be obtained from the expression displayed in the eq.~\eqref{eq:AM3} of the expansion over 3jm symbols of the product of a 3jm symbol with a 6j symbol by giving special values to the arguments $l_{1}$, $l_{2}$ and $l_{3}$ of the 6j symbol and taking help, for instance, from the eq.~\eqref{eq:6j1} to compute the involved particular 6j symbol. An illustrating case is the one with the values $l_{1} = \frac{1}{2}$, $l_{2} = j_{3}-\frac{1}{2}$ and $l_{3} = j_{2}-\frac{1}{2}$. $\mu_{1}$ can take only the two values $\pm  \frac{1}{2}$ and, to comply with the zero-sum condition for the 3jm symbols, $\mu_{2}$ must be fixed to $\mu_{1} + m_{3}$ and $\mu_{3}$ to $\mu_{1} - m_{2}$. Thus
\begin{multline}
\left\{\begin{array}{ccc}j_{1} & j_{2} & j_{3} \\\frac{1}{2} & j_{3}-\frac{1}{2} & j_{2}-\frac{1}{2}\end{array}\right\} \left(\begin{array}{ccc}j_{1} & j_{2} & j_{3} \\m_{1} & m_{2} & m_{3}\end{array}\right) = (-)^{j_{2}+j_{3}-m_{2}+m_{3}} \sum_{\mu_{1} = \pm \frac{1}{2}} (-)^{\mu_{1}-\frac{1}{2}}  \times {}
\\
\times \left(\begin{array}{ccc}j_{1} & j_{3}-\frac{1}{2} & j_{2}-\frac{1}{2} \\m_{1} & \mu_{1} + m_{3} & -\mu_{1} + m_{2}\end{array}\right) \left(\begin{array}{ccc}\frac{1}{2} & j_{2} & j_{2}-\frac{1}{2} \\-\mu_{1} & m_{2} & \mu_{1} - m_{2} \end{array}\right) \left(\begin{array}{ccc}\frac{1}{2} & j_{3}-\frac{1}{2} & j_{3} \\\mu_{1} & -\mu_{1} - m_{3} & m_{3}\end{array}\right)
\\[-12pt]
\end{multline}
The 6j symbol to compute is the last among those displayed in the eq.~\eqref{eq:6j2} while four among the six involved 3jm symbols may be derived, using the symmetry properties (cf.~eq.~\eqref{eq:3s1}, eq.~\eqref{eq:3s2} and eq.~\eqref{eq:3s3}), from the one displayed in the eq.~\eqref{eq:3j3}. After substitution, we get
\begin{multline}
\label{eq:3j4}
\left[ (j+1)(j-2j_{1}) \right]^{\frac{1}{2}} \left(\begin{array}{ccc}j_{1} & j_{2} & j_{3} \\m_{1} & m_{2} & m_{3}\end{array}\right) =
\\
\shoveright{ \left[ (j_{2}+m_{2})(j_{3}-m_{3}) \right]^{\frac{1}{2}} \left(\begin{array}{ccc}j_{1} & j_{2}-\frac{1}{2} & j_{3}-\frac{1}{2} \\m_{1} & m_{2} - \frac{1}{2} & m_{3} + \frac{1}{2}\end{array}\right) }
\\
-\left[ (j_{2}+m_{2})(j_{3}-m_{3}) \right]^{\frac{1}{2}} \left(\begin{array}{ccc}j_{1} & j_{2}-\frac{1}{2} & j_{3}-\frac{1}{2} \\m_{1} & m_{2} + \frac{1}{2} & m_{3} - \frac{1}{2}\end{array}\right)
\end{multline}
where $j = j_{1} + j_{2} + j_{3}$. Similarly, by setting $l_{1} = 1$, $l_{2} = j_{3}-1$ and $l_{3} = j_{2}$,  it may be shown that 
\begin{multline}
\label{eq:3j5}
\left[ (j+1)(j-2j_{1})(j-2j_{2})(j-2j_{3}+1) \right]^{\frac{1}{2}} \left(\begin{array}{ccc}j_{1} & j_{2} & j_{3} \\m_{1} & m_{2} & m_{3}\end{array}\right) =
\\
\shoveright{ \left[ (j_{2}-m_{2})(j_{2}+m_{2}+1)(j_{3}+m_{3})(j_{3}+m_{3}-1) \right]^{\frac{1}{2}} \left(\begin{array}{ccc}j_{1} & j_{2} & j_{3}-1 \\m_{1} & m_{2} + 1 & m_{3} - 1\end{array}\right) }
\\
\shoveright{ -2m_{2}\left[ (j_{3}+m_{3})(j_{3}-m_{3}) \right]^{\frac{1}{2}} \left(\begin{array}{ccc}j_{1} & j_{2} & j_{3}-1 \\m_{1} & m_{2} & m_{3}\end{array}\right) }
\\
-\left[ (j_{2}+m_{2})(j_{2}-m_{2}+1)(j_{3}-m_{3})(j_{3}-m_{3}-1) \right]^{\frac{1}{2}} \left(\begin{array}{ccc}j_{1} & j_{2} & j_{3}-1 \\m_{1} & m_{2} - 1 & m_{3} + 1\end{array}\right)
\end{multline}
Use can be made of this recurrence relation to compute, for instance, the frequently occurring 3jm symbol with all the projection quantum numbers $m$ null. If $j = j_{1} + j_{2} + j_{3}$ is odd then by the symmetry described in the eq.~\eqref{eq:3s3}) this 3jm symbol is null.  If $j = j_{1} + j_{2} + j_{3}$ is even then from the recurrence relation displayed in the eq.~\eqref{eq:3j5}, using the symmetry specified in the eq.~\eqref{eq:3s2} and the eq.~\eqref{eq:3s3}, we may write
\begin{multline}
\left(\begin{array}{ccc}j_{1} & j_{2} & j_{3} \\0 & 0 & 0\end{array}\right) = 2 \left[ \frac{j_{2} (j_{2} +1) j_{3} (j_{3} -1)}{(j+1) (j-2j_{1}) (j-2j_{2}) (j-2j_{3}+1)} \right]^{\frac{1}{2}}  \left(\begin{array}{ccc}j_{1} & j_{2} & j_{3}-1 \\0 & 1 & -1\end{array}\right) = {}
\\
 = \left[ \frac{(j-2j_{2}-1)(j-2j_{3}+2)}{(j-2j_{2}) (j-2j_{3}+1)} \right]^{\frac{1}{2}} \left(\begin{array}{ccc}j_{1} & j_{2} +1& j_{3}-1 \\0 & 0 & 0\end{array}\right)
\end{multline}
and, iterating this $k$ times, 
\begin{multline*}
\left(\begin{array}{ccc}j_{1} & j_{2} & j_{3} \\0 & 0 & 0\end{array}\right) = \left[ \frac{ \left\{ \frac{(j-2j_{2})!}{(j-2j_{2}-2k)! \frac{2^{k}(\frac{j}{2}-j_{2})!}{(\frac{j}{2}-j_{2}-k)!}}\right\} \left\{ \frac{2^{k}(\frac{j}{2}-j_{3} +k)!}{(\frac{j}{2}-j_{3}-k)!} \right\} }{ \left\{\frac{2^{k}(\frac{j}{2}-j_{2})!}{(\frac{j}{2}-j_{2}-k)!}\right\} \left\{ \frac{(j-2j_{3}+2k)!}{(j-2j_{3})!\frac{2^{k}(\frac{j}{2}-j_{3} +k)!}{(\frac{j}{2}-j_{3}-k)!}}\right\} } \right]^{\frac{1}{2}} \left(\begin{array}{ccc}j_{1} & j_{2} +k & j_{3}-k \\0 & 0 & 0\end{array}\right) = {}
\\
= \left[ \frac{(j-2j_{2})!(j-2j_{3})!}{(j-2j_{2}-2k)!(j-2j_{3}+2k)!} \right]^{\frac{1}{2}} \frac{(\frac{j}{2}-j_{2}-k)!(\frac{j}{2}-j_{3}+k)!}{(\frac{j}{2}-j_{2})!(\frac{j}{2}-j_{3})!}\left(\begin{array}{ccc}j_{1} & j_{2} +k & j_{3}-k \\0 & 0 & 0\end{array}\right)
\end{multline*}
whence, setting $k = j_{3}-\frac{j}{2}$ and making use of the eq.~\eqref{eq:3j2},
\begin{equation}
\label{eq:AM0}
\begin{split}
\left(\begin{array}{ccc} j_{1} &  j_{2} &  j_{3} \\0 & 0 & 0\end{array}\right) = (-)^{j/2} \left[ \frac{(j-2j_{1})! (j-2j_{2})! (j-2j_{3})!}{(j+1)!} \right]^{\frac{1}{2}} \times {} 
\\
\times \frac{(\frac{j}{2})!}{(\frac{j}{2}-j_{1})! (\frac{j}{2}-j_{2})! (\frac{j}{2}-j_{3})!}
\end{split}
\end{equation}

\subsection{\label{AMD}6j Symbols and 9j Symbols}

The addition of $n\geq3$ angular momenta $\vec{\mathbf{J}}_{1} + \vec{\mathbf{J}}_{2} + \cdots + \vec{\mathbf{J}}_{n} = \vec{\mathbf{J}}$ is conceptually not complex, but now there is not a single way to build up the coupled states. Let us consider the case $n=3$. We can either couple the states $\vert J_{1} M_{1}\rangle$ and $\vert J_{2} M_{2}\rangle$ to give the resultant $\vert J_{12} M_{12}\rangle$ then couple this with $\vert J_{3} M_{3}\rangle$ to finally give $\vert JM\rangle$ (we can drop the $\tau_{k}$ without loss of generality) : 
\begin{multline}
\vert (J_{1} J_{2})J_{12} J_{3} JM \rangle = \sum_{M_{12}, M_{3}} \vert (J_{1} J_{2})J_{12}M_{12} J_{3}M_{3} \rangle ~ \langle J_{12}M_{12} J_{3}M_{3} \vert J_{12} J_{3} JM \rangle = {}
\\
= \sum_{\substack{ M_{12}, M_{3} \\ M_{1}, M_{2} }} \vert J_{1}M_{1}J_{2}M_{2}J_{3}M_{3} \rangle ~ \langle J_{12}M_{12} J_{3}M_{3} \vert J_{12} J_{3} JM \rangle ~ \langle J_{1}M_{1}J_{2}M_{2} \vert J_{1} J_{2}J_{12}M_{12} \rangle
\end{multline}
or alternatively couple the states $\vert J_{2} M_{2}\rangle$ and $\vert J_{3} M_{3}\rangle$ to give the resultant $\vert J_{23} M_{23}\rangle$ then couple this with $\vert J_{1} M_{1}\rangle$ to again give $\vert JM\rangle$ : 
\begin{multline}
\vert J_{1} (J_{2}J_{3}) J_{23}  JM \rangle = \sum_{M_{1}, M_{23}} \vert J_{1}M_{1} (J_{2}J_{3}) J_{23} M_{23} \rangle ~ \langle J_{1}M_{1}J_{23} M_{23} \vert J_{1} J_{23} JM \rangle = {}
\\
= \sum_{\substack{ M_{1}, M_{23} \\ M_{2}, M_{3} }} \vert J_{1}M_{1}J_{2}M_{2}J_{3}M_{3} \langle J_{1}M_{1}J_{23} M_{23} \vert J_{1} J_{23} JM \rangle ~ \langle J_{2}M_{2}J_{3}M_{3} \vert J_{2} J_{3}J_{23}M_{23} \rangle
\end{multline}
The transformation which connects the states in these two coupling schemes can be written as
\begin{equation}
\vert J_{1} (J_{2}J_{3}) J_{23}  JM \rangle = \sum_{J_{12}} \vert (J_{1} J_{2})J_{12} J_{3} JM \rangle ~ \langle (J_{1} J_{2})J_{12} J_{3} JM \vert J_{1} (J_{2}J_{3}) J_{23} JM \rangle
\end{equation}
Operating on both side of this equality with $\vec{\mathbf{J}}_{+1}$ it is found out that the transformation coefficients $\langle (J_{1} J_{2})J_{12} J_{3} JM \vert J_{1} (J_{2}J_{3}) J_{23}  JM \rangle$ are independent of $M$. It then is customary to omit it and write these in the form
\begin{equation}
\label{eq:6jS}
\langle (J_{1} J_{2})J_{12} J_{3} J \vert J_{1} (J_{2}J_{3}) J_{23}  J \rangle = \left[ J_{12}\right]\left[ J_{23}\right] (-)^{J_{1}+J_{2}+J_{3}+J} \left\{\begin{array}{ccc}J_{3} & J & J_{12} \\J_{1} & J_{2} & J_{23}\end{array}\right\}
\end{equation}
which defines the \emph{6j symbol}. Inserting the equations of the two coupling schemes into the transformation equation which connects them, equating the coefficients of $\vert J_{1}M_{1}J_{2}M_{2}J_{3}M_{3} \rangle$ and using the 3jm symbols in place of the Clebsch-Gordan coefficients, it follows that
\begin{multline*}
\sum_{M_{23}} (-)^{J_{2}-J_{3}+M_{23}+J_{1}-J_{23}+M} \left[ J_{23} \right] \left[ J \right] \left(\begin{array}{ccc}J_{2} & J_{3} & J_{23} \\M_{2} & M_{3} & -M_{23}\end{array}\right) \left(\begin{array}{ccc}J_{1} & J_{23} & J \\M_{1} & M_{23} & -M\end{array}\right)
\\
= \sum_{J_{12}, M_{12}} \left[ J_{12}\right]\left[ J_{23}\right] (-)^{J_{1}+J_{2}+J_{3}+J} \left\{\begin{array}{ccc}J_{3} & J & J_{12} \\J_{1} & J_{2} & J_{23}\end{array}\right\} \times {}
\\
\times  (-)^{J_{1}-J_{2}+M_{12}+J_{12}-J_{3}+M} \left[ J_{12} \right] \left[ J \right] \left(\begin{array}{ccc}J_{1} & J_{2} & J_{12} \\M_{1} & M_{2} & -M_{12}\end{array}\right) \left(\begin{array}{ccc}J_{12} & J_{3} & J \\M_{12} & M_{3} & -M\end{array}\right)
\end{multline*}
Using the symmetry properties of the 3jm symbols (cf.~eqs.~\eqref{eq:3s1}-\eqref{eq:3s3}) this, after appropriate notation substitutions, is more conveniently rewritten
\begin{multline}
\label{eq:AM4}
\sum_{m_{3}} (-)^{l_{1}+l_{2}+\mu_{1}+\mu_{2}} \left(\begin{array}{ccc}j_{1} & j_{2} & j_{3} \\m_{1} & m_{2} & m_{3}\end{array}\right) \left(\begin{array}{ccc}l_{1} & l_{2} & j_{3} \\\mu_{1} & -\mu_{2} & m_{3}\end{array}\right) = {}
\\
\sum_{l_{3}, \mu_{3}}  (-)^{l_{3}+\mu_{3}} \left[ l_{3}^{2} \right] \left\{\begin{array}{ccc}j_{1} & j_{2} & j_{3} \\l_{1} & l_{2} & l_{3}\end{array}\right\} \left(\begin{array}{ccc}l_{1} & j_{2} & l_{3} \\-\mu_{1} & m_{2} & \mu_{3}\end{array}\right) \left(\begin{array}{ccc}j_{1} & l_{2} & l_{3} \\m_{1} & -\mu_{2} & -\mu_{3}\end{array}\right)
\end{multline}
Multiplying both sides by $(-)^{-M^{\prime}_{12}}\left(\begin{array}{ccc}J_{1} & J_{2} & J^{\prime}_{12} \\M_{1} & M_{2} & -M^{\prime}_{12}\end{array}\right)$ and summing over $M_{1}$ and $M_{2}$, 
\begin{multline*}
\left\{\begin{array}{ccc}J_{3} & J & J^{\prime}_{12} \\J_{1} & J_{2} & J_{23}\end{array}\right\} \left(\begin{array}{ccc}J^{\prime}_{12} & J_{3} & J \\M^{\prime}_{12} & M_{3} & -M\end{array}\right)= \sum_{M_{23}, M_{1}, M_{2}}(-)^{J_{1}+J_{2}+J_{23}-M_{1}+M_{2}+M_{23}} \times {}
\\
\times \left(\begin{array}{ccc}J_{3} & J_{2} & J_{23} \\M_{3} & M_{2} & -M_{23}\end{array}\right) \left(\begin{array}{ccc}J_{1} & J & J_{23} \\M_{1} & -M & M_{23}\end{array}\right) \left(\begin{array}{ccc}J_{1} & J_{2} & J^{\prime}_{12} \\-M_{1} & -M_{2} & M^{\prime}_{12}\end{array}\right)
\end{multline*}
by making use of the orthonormal conditions and symmetry properties of the 3jm symbols. After appropriate notation substitutions, this rewrites 
\begin{multline}
\label{eq:AM3}
\left\{\begin{array}{ccc}j_{1} & j_{2} & j_{3} \\l_{1} & l_{2} & l_{3}\end{array}\right\} \left(\begin{array}{ccc}j_{1} & j_{2} & j_{3} \\m_{1} & m_{2} & m_{3}\end{array}\right) = \sum_{\mu_{1},\mu_{2},\mu_{3}} (-)^{l_{1}+l_{2}+l_{3}+\mu_{1}+\mu_{2}+\mu_{3}} \times {}
\\
\times \left(\begin{array}{ccc}j_{1} & l_{2} & l_{3} \\m_{1} & \mu_{2} & -\mu_{3}\end{array}\right) \left(\begin{array}{ccc}l_{1} & j_{2} & l_{3} \\-\mu_{1} & m_{2} & \mu_{3}\end{array}\right) \left(\begin{array}{ccc}l_{1} & l_{2} & j_{3} \\\mu_{1} & -\mu_{2} & m_{3}\end{array}\right)
\end{multline}
It immediately is inferred that the 6j symbol is null unless the four triple of angular momenta $(j_{1}, j_{2}, j_{3})$, $(j_{1}, l_{2}, l_{3})$, $(l_{1}, j_{2}, l_{3})$ and $(l_{1}, l_{2}, j_{3})$ all satisfy the triangular condition. Inserting in the eq.~\eqref{eq:AM3} the algebraic form of the Clebsch-Gordan coefficients gives a complicated expression even though to determine the 6j symbol we can set $m_{1} = j_{1}$ and $m_{2} = -j_{2}$. According to \cite{Racah} this should give
\begin{multline*}
\left\{\begin{array}{ccc}j_{1} & j_{2} & j_{3} \\l_{1} & l_{2} & l_{3}\end{array}\right\}  = \triangle(j_{1}, j_{2}, j_{3})\triangle(j_{1}, l_{2}, l_{3})\triangle(l_{1}, j_{2}, l_{3})\triangle(l_{1}, l_{2}, j_{3}) \times {}
\\
\times \sum_{z}\frac{(-)^{z} (z+1)!}{\substack{(z-j_{1}-j_{2}-j_{3})! (z-j_{1}-l_{2}-l_{3})! (z-l_{1}-j_{2}-l_{3})! (z-l_{1}-l_{2}-j_{3})! (j_{1}+j_{2}+l_{1}+l_{2}-z)! (j_{2}+j_{3}+l_{2}+l_{3}-z)! \times {} \\ \times (j_{3}+j_{1}+l_{3}+l_{1}-z)!}}
\end{multline*}
where
\begin{equation*}
\triangle(a, b, c) = \left[\frac{(a+b-c)! (a-b+c)! (b+c-a)!}{(a+b+c+1)!}\right]^{\frac{1}{2}} \qquad
\end{equation*}
An advantage of this general expression is to put into evidence the invariance of the 6j symbol by any permutation of its columns:
\begin{equation}
\label{eq:6s1}
\left\{\begin{array}{ccc}j_{1} & j_{2} & j_{3} \\l_{1} & l_{2} & l_{3}\end{array}\right\} = 
\left\{\begin{array}{ccc}j_{3} & j_{1} & j_{2} \\l_{3} & l_{1} & l_{2}\end{array}\right\} = \left\{\begin{array}{ccc}j_{2} & j_{1} & j_{3} \\l_{2} & l_{1} & l_{3}\end{array}\right\} = \cdots
\end{equation}
and by interchange of any of the upper and lower arguments in each of any two of its columns
\begin{equation}
\label{eq:6s2}
\left\{\begin{array}{ccc}j_{1} & j_{2} & j_{3} \\l_{1} & l_{2} & l_{3}\end{array}\right\} = 
\left\{\begin{array}{ccc}l_{1} & l_{2} & j_{3} \\j_{1} & j_{2} & l_{3}\end{array}\right\} =  \cdots
\end{equation}
It is not be a difficult matter to show that the coupled states of the angular momenta form an orthonormal and complete set in the state space of the quantum system under concern and that the transformation connecting the states of the different coupling schemes are unitary. It then is inferred, using the eq.~\eqref{eq:6jS}), that
\begin{gather}
\langle (j_{1} j_{2}) j^{\prime} j_{3} j_{4} \vert (j_{1} j_{2}) j^{\prime \prime} j_{3} j_{4} \rangle = \langle (j_{1} j_{2}) j^{\prime} j_{3} j_{4} \vert \left[ \sum_{j} \vert j_{1} (j_{2}j_{3}) j j_{4} \rangle \langle  j_{1} (j_{2}j_{3}) j j_{4} \vert \right] \vert (j_{1} j_{2}) j^{\prime \prime} j_{3} j_{4} \rangle = {}
\nonumber \\
= \sum_{j} \left[ j^{2} \right] \left[ j^{\prime} \right] \left[ j^{\prime \prime} \right] \left\{\begin{array}{ccc}j_{1} & j_{2} & j^{\prime} \\j_{3} & j_{4} & j\end{array}\right\} \left\{\begin{array}{ccc}j_{1} & j_{2} & j^{\prime \prime} \\j_{3} & j_{4} & j \end{array}\right\} = \delta_{j^{\prime}, j^{\prime \prime}}
\end{gather}
or else from
\begin{equation*}
\langle (j_{1} j_{2}) j_{12} j_{3} j \vert j_{2} (j_{3} j_{1}) j_{31} j \rangle = \langle (j_{1} j_{2}) j_{12} j_{3} j \vert \left[ \sum_{j_{23}} \vert j_{1} (j_{2}j_{3}) j_{23} j \rangle \langle  j_{1} (j_{2}j_{3}) j_{23} j \vert \right] \vert j_{2} (j_{3} j_{1}) j_{31} j \rangle
\end{equation*}
and from
\begin{gather*}
\langle (j_{1} j_{2}) j_{12} j_{3} j \vert j_{2} (j_{3} j_{1}) j_{31} j \rangle = (-)^{j_{1} + j_{2} - j_{12} + j_{1} + j_{3} - j_{13}} \langle (j_{2} j_{1}) j_{12} j_{3} j \vert j_{2} (j_{1} j_{3}) j_{31} j \rangle 
\\
\langle  j_{1} (j_{2}j_{3}) j_{23} j \vert j_{2} (j_{3} j_{1}) j_{31} j \rangle = (-)^{j_{1} + j_{23} -j}\langle  (j_{2}j_{3}) j_{23} j_{1} j \vert j_{2} (j_{3} j_{1}) j_{31} j \rangle
\end{gather*}
(cf.~eq.~\eqref{eq:CG3}) that
\begin{equation}
\sum_{j_{23}} (-)^{j_{12}+j_{23}+j_{31}} \left[ j_{23}^{2} \right] \left\{\begin{array}{ccc}j_{1} & j_{2} & j_{12} \\j_{3} & j & j_{23}\end{array}\right\} \left\{\begin{array}{ccc}j_{2} & j_{3} & j_{23} \\j_{1} & j & j_{31}\end{array}\right\} = \left\{\begin{array}{ccc}j_{3} & j_{1} & j_{31} \\j_{2} & j & j_{12}\end{array}\right\}
\end{equation}
Although the general algebraic formula might be used to numerically compute any 6j symbol it reveals more convenient to work out simpler expression for special values of the arguments. A nice example is provided by setting $j_{3} = l_{1} + l_{2} = m_{3}$ in the eq.~\eqref{eq:AM3}, which then writes
\begin{multline}
\left\{\begin{array}{ccc}j_{1} & j_{2} & l_{1} + l_{2} \\l_{1} & l_{2} & l_{3}\end{array}\right\} \left(\begin{array}{ccc}j_{1} & j_{2} & l_{1} + l_{2} \\m_{1} & m_{2} & l_{1} + l_{2}\end{array}\right) = \sum_{\mu_{1},\mu_{2},\mu_{3}} (-)^{l_{1}+l_{2}+l_{3}+\mu_{1}+\mu_{2}+\mu_{3}} \times {}
\\
\times \left(\begin{array}{ccc}j_{1} & l_{2} & l_{3} \\m_{1} & \mu_{2} & -\mu_{3}\end{array}\right) \left(\begin{array}{ccc}l_{1} & j_{2} & l_{3} \\-\mu_{1} & m_{2} & \mu_{3}\end{array}\right) \left(\begin{array}{ccc}l_{1} & l_{2} & l_{1} + l_{2} \\\mu_{1} & -\mu_{2} & l_{1} + l_{2}\end{array}\right)
\end{multline}
The last 3jm symbol is null unless the zero-sum condition $\mu_{1}-\mu_{2} = -l_{1} - l_{2}$ is satisfied, but $\vert \mu_{1} \vert \leq l_{1}$ and $\vert \mu_{2} \vert \leq l_{2}$ so that necessarily $\mu_{1} = -l_{1}$ and $\mu_{2} = l_{2}$. On further setting  $m_{1} = -j_{1}$ the zero-sum condition on the other 3jm symbols impose that $m_{2} = j_{1}-l_{1} - l_{2}$ and $\mu_{3} = -j_{1} + l_{2}$. The summation in the right hand side then is reduced to a single term and, using the symmetry properties of the 3jm symbols, we may write
\begin{multline*}
\left\{\begin{array}{ccc}j_{1} & j_{2} & l_{1} + l_{2} \\l_{1} & l_{2} & l_{3}\end{array}\right\} (-)^{j_{1}+j_{2}+l_{1}+l_{2}} \left(\begin{array}{ccc}j_{1} & j_{2} & l_{1} + l_{2} \\j_{1} & -j_{1} + l_{1} + l_{2} & -l_{1} - l_{2}\end{array}\right) = (-)^{-j_{1}-l_{2}+l_{3}} \times{}
\\
\times \left(\begin{array}{ccc}l_{2} & l_{3} & j_{1} \\l_{2} & -l_{2} + j_{1} & -j_{1}\end{array}\right) \left(\begin{array}{ccc}l_{1} & j_{2} & l_{3} \\l_{1} & - l_{1} -j_{1} - l_{2} & j_{1}+l_{2}\end{array}\right) \left(\begin{array}{ccc}l_{1} & l_{2} & l_{1} + l_{2} \\-l_{1} & -l_{2} & l_{1} + l_{2}\end{array}\right)
\end{multline*}
The 3jm symbol in the $l.h.s$ and the two first  3jm symbols in the $r.h.s.$ can be computed using the eq.~\eqref{eq:3j1} while the third 3jm symbol in the $r.h.s$ can be computed using the eq.~\eqref{eq:3j2} to finally give
\begin{multline}
\label{eq:6j1}
\left\{\begin{array}{ccc}j_{1} & j_{2} & l_{1} + l_{2} \\l_{1} & l_{2} & l_{3}\end{array}\right\} = (-)^{j_{1}+j_{2}+l_{1}+l_{2}} 
\\
\left[ \frac{ \substack{ (2l_{1})! (2l_{2})! (j_{1} + j_{2} + l_{1} + l_{2} +1)! (j_{1} - j_{2} + l_{1} + l_{2})! (-j_{1} + j_{2} + l_{1} + l_{2})! \\ (j_{1} - l_{2} + l_{3})! (j_{2} - l_{1} + l_{3})!} }{ \substack{ (2l_{1} + 2l_{2} +1)! (j_{1} + j_{2} - l_{1} - l_{2})! (j_{1} + l_{2} - l_{3})! (-j_{1} + l_{2} + l_{3})! (l_{1} + j_{2} - l_{3})! (l_{1} - j_{2} + l_{3})! \\ (j_{1} + l_{2} + l_{3} +1)! (l_{1} + j_{2} + l_{3} +1)! } } \right]^{\frac{1}{2}}
\end{multline}
A few particular instances of this formula are
\begin{multline}
\label{eq:6j2}
\left\{\begin{array}{ccc}j_{1} & j_{2} & j_{3} \\0 & j_{3} & j_{2}\end{array}\right\}  = (-)^{j} \left[ j_{2}^{-1} j_{3}^{-1} \right] \qquad (j = j_{1}+j_{2}+j_{3})
\\ 
\left\{\begin{array}{ccc}j_{1} & j_{2} & j_{3} \\1 & j_{3}-1 & j_{2}\end{array}\right\}  = (-)^{j} \left[ \frac{2(j+1) (j-2j_{1}) (j-2j_{2}) (j-2j_{3}+1)}{(2j_{2}) (2j_{2}+1) (2j_{2}+2) ( 2j_{3}-1) 2j_{3} (2j_{3}+1)} \right]^{\frac{1}{2}}
\\ 
\left\{\begin{array}{ccc}j_{1} & j_{2} & j_{3} \\\frac{1}{2} & j_{3}-\frac{1}{2} & j_{2}+\frac{1}{2}\end{array}\right\}  = (-)^{j} \left[ \frac{(j-2j_{2}) (j-2j_{3}+1)}{(2j_{2}+1) (2j_{2}+2) 2j_{3} (2j_{3}+1)} \right]^{\frac{1}{2}}
\\ 
\left\{\begin{array}{ccc}j_{1} & j_{2} & j_{3} \\\frac{1}{2} & j_{3}-\frac{1}{2} & j_{2}-\frac{1}{2}\end{array}\right\}  = (-)^{j} \left[ \frac{(j+1) (j-2j_{1})}{(2j_{2}) (2j_{2}+1) 2j_{3} (2j_{3}+1)} \right]^{\frac{1}{2}}
\end{multline}
\medskip

\noindent Let us now consider the coupling of $n= 4$ angular momenta. We can either separately couple on one side the states $\vert J_{1} M_{1}\rangle$ and $\vert J_{2} M_{2}\rangle$ to give the resultant $\vert J_{12} M_{12}\rangle$ and on the other side the states $\vert J_{3} M_{3}\rangle$ and $\vert J_{4} M_{4}\rangle$ to give the resultant $\vert J_{34} M_{34}\rangle$ then couple the states $\vert J_{12} M_{12}\rangle$ and $\vert J_{34} M_{34}\rangle$ to finally give $\vert JM\rangle$ or else separately couple on one side the states $\vert J_{1} M_{1}\rangle$ and $\vert J_{3} M_{3}\rangle$ to give the resultant $\vert J_{13} M_{13}\rangle$ and on the other side the states $\vert J_{2} M_{2}\rangle$ and $\vert J_{4} M_{4}\rangle$ to give the resultant $\vert J_{24} M_{24}\rangle$ then couple the states $\vert J_{13} M_{13}\rangle$ and $\vert J_{24} M_{24}\rangle$ to finally give $\vert JM\rangle$. The transformation coefficient which connect these two coupling schemes may be written in the form
\begin{multline}
\label{eq:9jS}
\left \langle (J_{1}J_{2})J_{12}~ (J_{3}J_{4})J_{34}~J ~\vert~  (J_{1}J_{3})J_{13}~ (J_{2}J_{4})J_{24}~J  \right \rangle = {}
\\
= \left[ J_{12}~J_{34}~J_{13}~J_{24} \right] \left\{\begin{array}{ccc}J_{1} & J_{2} & J_{12} \\J_{3} & J_{4} & J_{34} \\J_{13} & J_{24} & J\end{array}\right\}
\end{multline}
which defines the \emph{9j symbol}. It is clear that other coupling schemes can be anticipated giving rise to different transformations coefficients and the coupling by pairs of angular momenta we have described can also be viewed in terms of three angular momenta couplings, which provides with relations between transformation coefficients, for instance
\begin{multline}
\left \langle (J_{1}J_{2})J_{12}~ (J_{3}J_{4})J_{34}~J ~ \vert~  (J_{1}J_{3})J_{13}~ (J_{2}J_{4})J_{24}~J  \right \rangle = {}
\\[8pt]
\shoveleft{ = \sum_{J_{234}} \left \langle (J_{1}J_{2})J_{12}~ (J_{3}J_{4})J_{34}~J ~\vert~ J_{1} \left\{ J_{2}(J_{3}J_{4})J_{34} \right\} J_{234}~J \right \rangle \times {}}
\\
\times \left \langle J_{1} \left\{ J_{2}(J_{3}J_{4})J_{34} \right\} J_{234}~J ~\vert~ J_{1} \left\{ J_{3}(J_{2}J_{4})J_{24} \right\} J_{234}~J \right \rangle \times {}
\\[12pt]
\shoveright{ \times \left \langle J_{1} \left\{ J_{3}(J_{2}J_{4})J_{24} \right\} J_{234}~J ~\vert~  (J_{1}J_{3})J_{13}~ (J_{2}J_{4})J_{24}~J  \right \rangle = {} }
\\
\shoveleft{ = \left[ J_{12}~J_{34}~J_{13}~J_{24} \right] }
\\
\sum_{J_{234}} (-)^{2J_{234}} \left[ J_{234}^{2} \right] \left\{\begin{array}{ccc}J_{1} & J_{2} & J_{12} \\J_{34} & J & J_{234}\end{array}\right\} \left\{\begin{array}{ccc}J_{3} & J_{4} & J_{34} \\J_{2} & J_{234} & J_{24}\end{array}\right\}
\left\{\begin{array}{ccc}J_{13} & J_{24} & J \\J_{234} & J_{1} & J_{3}\end{array}\right\}
\end{multline}
(cf.~eq.~\eqref{eq:6jS}). Accordingly, after appropriate notation substitutions,
\begin{equation}
\left\{\begin{array}{ccc}j_{11} & j_{12} & j_{13} \\j_{21} & j_{22} & j_{23} \\j_{31} & j_{32} & j_{33}\end{array}\right\} = \sum_{z} (-)^{2z} \left[ z^{2} \right] \left\{\begin{array}{ccc}j_{11} & j_{21} & j_{31} \\j_{32} & j_{33} & z\end{array}\right\}
\left\{\begin{array}{ccc}j_{12} & j_{22} & j_{32} \\j_{21} & z & j_{23}\end{array}\right\}
\left\{\begin{array}{ccc}j_{13} & j_{23} & j_{33} \\z & j_{11} & j_{12}\end{array}\right\}
\end{equation}
Using the appropriate closure relations (cf.~eq.~\eqref{eq:CGD}), it may be shown that
\begin{multline}
\left \vert~ (J_{1}J_{2})J_{12}~ (J_{3}J_{4})J_{34}~JM \right \rangle = \sum_{M_{12}, M_{34}} \left \vert~ (J_{1}J_{2})J_{12}M_{12}~ (J_{3}J_{4})J_{34}M_{34} \right \rangle \times 
\\
\shoveright{ \times \left \langle (J_{1}J_{2})J_{12}M_{12}~ (J_{3}J_{4})J_{34}M_{34} ~\vert~ (J_{1}J_{2})J_{12}~ (J_{3}J_{4})J_{34}~JM \right \rangle = }
\\[12pt]
\shoveleft{ = \sum_{M_{12}, M_{34}, M_{1}, M_{2}, M_{3}, M_{4}}  \left \vert J_{1}M_{1}J_{2}M_{2} \right \rangle \left \langle J_{1}M_{1}J_{2}M_{2} \vert (J_{1}J_{2})J_{12}M_{12}\right \rangle \times {} }
\\
\left \vert J_{3}M_{3}J_{4}M_{4} \right \rangle \left \langle J_{3}M_{3}J_{4}M_{4} \vert (J_{3}J_{4})J_{34}M_{34}\right \rangle \times {}
\\[10pt]
\times \left \langle (J_{1}J_{2})J_{12}M_{12}~ (J_{3}J_{4})J_{34}M_{34} ~\vert~ (J_{1}J_{2})J_{12}~(J_{3}J_{4})J_{34}~JM \right \rangle
\end{multline}
and similarly for $\vert~ (J_{1}J_{3})J_{13}~ (J_{2}J_{4})J_{24}~JM \rangle$ then, computing the scalar product of the two vectors, using the eq.~\eqref{eq:3jm} to replace the Clecbsch-gordan coefficients by 3jm symbols and performing the appropriate notation substitutions,
\begin{multline}
\left\{\begin{array}{ccc}j_{11} & j_{12} & j_{13} \\j_{21} & j_{22} & j_{23} \\j_{31} & j_{32} & j_{33}\end{array}\right\} = 
\\
\sum_{\text{All m's}} \left(\begin{array}{ccc}j_{11} & j_{12} & j_{13} \\m_{11} & m_{12} & m_{13}\end{array}\right) \left(\begin{array}{ccc}j_{21} & j_{22} & j_{23} \\m_{21} & m_{22} & m_{23}\end{array}\right) \left(\begin{array}{ccc}j_{31} & j_{32} & j_{33} \\m_{31} & m_{32} & m_{33}\end{array}\right) \times {}
\\ \quad \times \left(\begin{array}{ccc}j_{11} & j_{21} & j_{31} \\m_{11} & m_{21} & m_{31}\end{array}\right) \left(\begin{array}{ccc}j_{12} & j_{22} & j_{32} \\m_{12} & m_{22} & m_{32}\end{array}\right) \left(\begin{array}{ccc}j_{13} & j_{23} & j_{33} \\m_{13} & m_{23} & m_{33}\end{array}\right)
\end{multline}
A merit of this formula is to immediately reveal the symmetry properties of the 9j symbol. This is invariant under even permutations of the columns or of the rows and by swap of the lines into columns and the columns into rows. It is multiplied by the phase factor $(-)^{\sum_{\alpha,\beta} j_{\alpha,\beta}}$ under odd permutations of the columns or of the rows. Similarly as for the 6j symbols the unitary nature of the coupling schemes lead to the orthogonality relation
\begin{equation}
\sum_{j_{12}, j_{34}} \left[ J_{12}~J_{34}~J_{13}~J_{24} \right] \left\{\begin{array}{ccc}J_{1} & J_{2} & J_{12} \\J_{3} & J_{4} & J_{34} \\J_{13} & J_{24} & J\end{array}\right\} \left\{\begin{array}{ccc}J_{1} & J_{2} & J_{12} \\J_{3} & J_{4} & J_{34} \\J^{\prime}_{13} & J^{\prime}_{24} & J\end{array}\right\} = \delta_{J_{13},J^{\prime}_{13}} \delta_{J_{24},J^{\prime}_{24}}
\end{equation}
A number of sum rules and recursion relations derived from these can be obtained for the 6j symbols and 9j symbols by considering various coupling schemes. Unfortunately, the expressions often get rather complicated, involving so many parameters that it is difficult to find their interrelations and to extract the symmetry of the formulas through the appropriate notation substitutions. It gets of course a further puzzling task to analyze the coupling of $n = 5, \cdots$ angular momenta, which leads to 3(n-1)j symbols. Graphical methods were proposed that allows dealing with more ease with the inherent difficulties of this poorly intuitive algebra \cite{Varshalovich}.

\section{\label{STO}Spherical Irreducible Tensor Operators}

\subsection{\label{STD}Definition and Examples}

A tensor operator $\mathbf{T}$ with respect to a symmetry group $G$ is a set of operators ${\mathbf T}_{i}$ that satisfy the equivariance property
\begin{equation}
\forall g \in G \quad U(g) \mathbf{T}_{i}U(g)^{-1} = \sum_{j} \mathbf{T}_{j} \mathcal{D}_{j, i} (g)
\end{equation} 
where $U(g) \in \mathcal{G}$ is an image of $g$ in the group $\mathcal{G}$ of the state transformations associated with $G$ and $\mathcal{D} (g)$ its matrix representative with respect to a given basis in the state space of the quantum system under concern. A spherical irreducible tensor operator is a tensor operator that shows the equivariance property for the group of rotations $g = R(\hat{u}, \omega)$ with respect to matrix representatives $\mathcal{D}^{J}(g)$ of an irreducible representation $\mathcal{D}^{J}$ of the Lie group $\mathcal{G} = SU(2)$. It then consists in a set of $2J+1$ operators and is naturally labeled by the index $J$. Since any finite rotation can be built from a succession of infinitesimal rotations it is equivalent to consider the equivariance property in terms of infinitesimal generators and matrix representations of the $\mathfrak{su}(2)$ Lie algebra. It then may be stated that a set of $2k+1$ operators $\mathbf{T}_{q}^{(k)}~(q = -k, -k+1, \cdots, k)$ form the components of a spherical irreducible tensor operator $\mathbf{T}^{(k)}$ of rank $k$ if and only if these operators satisfy the commutation relations 
\begin{equation}
\label{eq:STA}
[~ \mathbf{J}_{\mu}, \mathbf{T}_{q}^{(k)} ] = \sum_{q^{\prime} = -k}^{k} \mathbf{T}_{q^{\prime}}^{(k)} \left \langle kq^{\prime} \vert \mathbf{J}_{\mu} \vert kq \right \rangle \quad (\mu = 0, \pm 1)
\end{equation}
more precisely if an only if $[~ \mathbf{J}_{0}, \mathbf{T}_{q}^{(k)} ] = q \mathbf{T}_{q}^{(k)}$ and $[~\mathbf{J}_{\pm 1}, \mathbf{T}_{q}^{(k)} ] = \mp\left[ \frac{1}{2} (k \mp q)(k \pm q+1) \right]^{\frac{1}{2}} \mathbf{T}_{q \pm 1}^{(k)}$ (cf.~eq.~\eqref{eq:AMR}). It immediately is checked that scalar $(k = 0)$ and vector $(k = 1)$ operators are effectively those defined in the section \ref{AMA}. An example of vector operator $(k = 1)$ is the differential operator $\vec{\boldsymbol {\nabla}}$ the spherical components of which, in Cartesian coordinates $(x, y, z)$ and spherical coordinates $(r, \theta, \varphi)$, are given by
\begin{gather*}
\boldsymbol {\nabla}_{0} = \partial_{z} = \cos\theta~\partial_{r} - \frac{\sin\theta}{r}~\partial_{\theta}
\\
\boldsymbol {\nabla}_{\pm 1} = \mp \frac{1}{\sqrt{2}} (\partial_{x} \pm i \partial_{y}) = \frac{e^{\pm i\varphi}}{\sqrt{2}} \left\{ \sin\theta~\partial_{r} + \frac{\cos\theta}{r}~\partial_{\theta} \pm \frac{i}{r\sin\theta}~\partial_{\varphi} \right\}
\end{gather*}
Taking the adjoint of the defining relation for  $\mathbf{T}^{(k)}$, calling to mind that $\mathbf{U}^{+}(\alpha, \beta, \gamma) = \mathbf{U}^{-1}(\alpha, \beta, \gamma)$ and that $(ABC)^{+} = C^{+}B^{+}A^{+}$, and using the symmetry properties of the Wigner D-matrices given in the section \ref{AMB} it is found that $\mathbf{U}(\alpha, \beta, \gamma) (\mathbf{T}_{q}^{(k)})^{+} \mathbf{U}^{-1}(\alpha, \beta, \gamma) = \sum_{q^{\prime}} (\mathbf{T}_{q^{\prime}}^{(k)})^{+} \mathcal{D}^{J~\ast}_{q^{\prime}q}(\alpha, \beta, \gamma) = \sum_{q^{\prime}} (\mathbf{T}_{q^{\prime}}^{(k)})^{+} (-)^{q^{\prime}-q}\mathcal{D}^{J}_{-q^{\prime}-q}(\alpha, \beta, \gamma)$. The set $(-)^{q}~(\mathbf{T}_{-q}^{(k)})^{+}$ form the components of a spherical irreducible tensor operator $(\mathbf{T}^{+})^{(k)}$ called the adjoint to $\mathbf{T}^{(k)}$. \medskip

\noindent Let $\mathbf{T}^{(k_{1})}$ and $\mathbf{U}^{(k_{2})}$ be two spherical irreducible tensor operators wth components $\mathbf{T}_{q_{1}}^{(k_{1})}$ and $\mathbf{U}_{q_{2}}^{(k_{2})}$. The quantities
\begin{equation}
\label{eq:STB}
\mathbf{X}_{Q}^{(K)} = \sum_{q_{1}, q_{2}} \mathbf{T}_{q_{1}}^{(k_{1})} \mathbf{U}_{q_{2}}^{(k_{2})} \left \langle k_{1}q_{1} k_{2}q_{2} \vert k_{1} k_{2} K Q \right \rangle
\end{equation}
transform under rotation similarly as $\sum_{q_{1}, q_{2}}  \vert  k_{1}q_{1} k_{2}q_{2} \rangle \langle k_{1}q_{1} k_{2}q_{2} \vert k_{1} k_{2} K Q \rangle = \vert k_{1} k_{2} K Q \rangle$, that is according to the irreducible matrix representation $\mathcal{D}^{K}$ of $SU(2)$. They thus form the components of a spherical irreducible tensor operator $\mathbf{X}^{(K)}$ of rank $K$. It also may be shown more directly that the quantities $\mathbf{X}_{Q}^{(K)}$ do satisfy the required commutation relations. The values of $K$ are restricted to $K = \vert k_{1} - k_{2} \vert, \vert k_{1} - k_{2} \vert + 1, \cdots,  k_{1} + k_{2}$ and those of $Q$ for a fixed $K$ to $Q = q_{1} + q_{2} = -K, -K+1, \cdots, K$. $\mathbf{X}^{(K)}$ is often denoted $\{ \mathbf{T}^{(k_{1})} \otimes \mathbf{U}^{(k_{2})}\}^{(K)}$. \medskip

\noindent The set of $(2k_{1}+1)(2k_{2}+1)$ operators $\mathbf{T}_{q_{1}}^{(k_{1})} \mathbf{U}_{q_{2}}^{(k_{2})}$ define the direct product of  $\mathbf{T}^{(k_{1})}$ and $\mathbf{U}^{(k_{2})}$ and can be interpreted as the components of a spherical irreducible tensor operator $\mathbf{X}^{(K)}$
\begin{equation}
\label{eq:STC}
\mathbf{T}_{q_{1}}^{(k_{1})} \mathbf{U}_{q_{2}}^{(k_{2})} = \sum_{K} \left \langle k_{1}q_{1} k_{2}q_{2} \vert k_{1} k_{2} K Q \right \rangle \mathbf{X}_{Q}^{(K)}
\end{equation}
If $k_{1} = k_{2} = k$ then setting $K = 0$ it is found that
\begin{multline}
\label{eq:STD}
\left\{ \mathbf{T}^{(k)} \otimes \mathbf{U}^{(k)} \right\}_{0}^{(0)} = \sum_{q}  \mathbf{T}_{q}^{(k)} \mathbf{U}_{-q}^{(k)} \left \langle kq k-q \vert 00 \right \rangle = \sum_{q}  \mathbf{T}_{q}^{(k)} \mathbf{U}_{-q}^{(k)} \left(\begin{array}{ccc}k & k & 0 \\q & -q & 0\end{array}\right) = {}
\\
= (-)^{k} \left[ k^{-1} \right] \sum_{q} (-)^{q} \mathbf{T}_{q}^{(k)} \mathbf{U}_{-q}^{(k)} = (-)^{k} \left[ k^{-1} \right] (\mathbf{T}^{(k)} \cdot \mathbf{U}^{(k)})
\end{multline}
(cf.~eq.~\eqref{eq:3j0}) where $(\mathbf{T}^{(k)} \cdot \mathbf{U}^{(k)})$ defines the scalar product of $\mathbf{T}^{(k)}$ and $\mathbf{U}^{(k)}$. \medskip

\noindent If $k_{1} = k_{2} = 1$ then setting $K = 1$ it is found that 
\begin{equation}
\left\{ \mathbf{T}^{(1)} \otimes \mathbf{U}^{(1)} \right\}_{Q}^{(1)} = \sum_{q_{1}q_{2}} \mathbf{T}_{q_{1}}^{(1)} \mathbf{U}_{q_{2}}^{(1)} \left \langle 1q_{1} 1q_{2} \vert 1Q \right \rangle = \sum_{q_{1}q_{2}}  \mathbf{T}_{q_{1}}^{(1)} \mathbf{U}_{q_{2}}^{(1)}(-)^{Q} \sqrt{3}\left(\begin{array}{ccc}1 & 1 & 1 \\q_{1} & q_{2} & -Q\end{array}\right)
\end{equation}
(cf.~eq.~\eqref{eq:3jm}). Using the eq.~\eqref{eq:3j6} we compute 
\begin{equation*}
\left\{ \mathbf{T}^{(1)} \otimes \mathbf{U}^{(1)} \right\}_{0}^{(1)} = \frac{i}{\sqrt{2}} (\mathbf{T}_{x}\mathbf{U}_{y} - \mathbf{T}_{y}\mathbf{U}_{x}) = \frac{i}{\sqrt{2}} (\vec{\mathbf{T}} \wedge \vec{\mathbf{U}})_{z} = \frac{i}{\sqrt{2}} (\vec{\mathbf{T}} \wedge \vec{\mathbf{U}})_{0}
\end{equation*}
where $\mathbf{T}_{n}$ and $\mathbf{U}_{m}~(n, m =x, y, z)$ are the Cartesian components of the vector operators $\vec{\mathbf{T}}$ and $\vec{\mathbf{U}}$ :  $\mathbf{T}_{x} = -\frac{\mathbf{T}_{+1}^{(1)}-\mathbf{T}_{-1}^{(1)}}{\sqrt{2}}$, $\mathbf{T}_{y} = -\frac{\mathbf{T}_{+1}^{(1)}+\mathbf{T}_{-1}^{(1)}}{i\sqrt{2}}$, $\mathbf{T}_{z} = \mathbf{T}_{0}^{(1)}$, $\mathbf{U}_{x} = -\frac{\mathbf{U}_{+1}^{(1)}-\mathbf{U}_{-1}^{(1)}}{\sqrt{2}}$, $\mathbf{U}_{y} = -\frac{\mathbf{U}_{+1}^{(1)}+\mathbf{U}_{-1}^{(1)}}{i\sqrt{2}}$, $\mathbf{U}_{z} = \mathbf{U}_{0}^{(1)}$. It may similarly be computed that 
\begin{equation*}
\left\{ \mathbf{T}^{(1)} \otimes \mathbf{U}^{(1)} \right\}_{\pm 1}^{(1)} = \frac{i}{\sqrt{2}} (\vec{\mathbf{T}} \wedge \vec{\mathbf{U}})_{\mp 1}
\end{equation*}
Accordingly,
\begin{equation}
\label{eq:ST0}
(\vec{\mathbf{T}} \wedge \vec{\mathbf{U}})_{q} = i (-)^{1+q}\sqrt{6} \sum_{q_{1}, q_{2}} \mathbf{T}_{q_{1}} \mathbf{U}_{q_{2}} \left(\begin{array}{ccc}1 & 1 & 1 \\q_{1} & q_{2} & q\end{array}\right)
\end{equation}

\subsection{\label{WET}Wigner-Eckart Theorem}

Let $\mathbf{T}^{(k)}$ be a spherical irreducible tensor operator with components $\mathbf{T}_{q}^{(k)}$. Applying these on the angular momentum basis states $\vert \tau J M \rangle$, we may build, for each $\tau$ and $J$, $(2k+1)(2J+1)$ states $\mathbf{T}_{q}^{(k)} \vert \tau J M \rangle ~ (q = -k, -k+1, \cdots, k; ~ M = -J, -J+1, \cdots, J)$ and from these the states 
\begin{equation}
\psi_{\tau, J, k}(J^{\prime} M^{\prime}) = \sum_{q, M} \mathbf{T}_{q}^{(k)} \vert \tau J M \rangle \left \langle kqJM \vert J^{\prime}M^{\prime}\right \rangle
\end{equation}
It is an easy matter to invert this, thanks to the orthonormal properties of the Clebsch-Gordan coefficients : $\sum_{J^{\prime}M^{\prime}} \langle kqJM \vert J^{\prime}M^{\prime} \rangle \langle k\tilde{q}J\tilde{M} \vert J^{\prime}M^{\prime} \rangle = \delta_{q\tilde{q}}~\delta_{M\tilde{M}}$. We get
\begin{equation}
\mathbf{T}_{q}^{(k)} \vert \tau J M \rangle = \sum_{J^{\prime}M^{\prime}} \psi_{\tau, J, k}(J^{\prime} M^{\prime}) \left \langle kqJM \vert J^{\prime}M^{\prime}\right \rangle
\end{equation}
Using the eq.~\eqref{eq:AMR} and the commutation relations given in the eq.~\eqref{eq:STA}, it may be shown that
\begin{gather*}
\mathbf{J}_{0} \mathbf{T}_{q}^{(k)} \vert \tau J M \rangle = [~\mathbf{J}_{0}, \mathbf{T}_{q}^{(k)} ] \vert \tau J M \rangle + \mathbf{T}_{q}^{(k)} \mathbf{J}_{0} \vert \tau J M \rangle = q \mathbf{T}_{q}^{(k)} \vert \tau J M \rangle + M\mathbf{T}_{q}^{(k)} \vert \tau J M \rangle
\\[8pt]
\shoveleft{ \mathbf{J}_{\pm 1} \mathbf{T}_{q}^{(k)} \vert \tau J M \rangle = [~\mathbf{J}_{\pm 1}, \mathbf{T}_{q}^{(k)} ] \vert \tau J M \rangle + \mathbf{T}_{q}^{(k)} \mathbf{J}_{\pm 1} \vert \tau J M \rangle = {} }
\\
= \mp\left[ \frac{1}{2} (k \mp q)(k \pm q+1) \right]^{\frac{1}{2}}  \mathbf{T}_{q \pm 1}^{(k)} \vert \tau J M \rangle \mp\left[ \frac{1}{2} (J \mp M)(J \pm M+1) \right]^{\frac{1}{2}}  \mathbf{T}_{q}^{(k)} \vert \tau J M \pm 1 \rangle  
\end{gather*}
and, applying $\mathbf{J}_{\mu} ~ (\mu = 0, \pm 1)$ on $\vert J_{1}J_{2}JM \rangle = \sum_{M_{1}M_{2}} \vert J_{1}M_{1}J_{2}M_{2} \rangle \langle J_{1}M_{1}J_{2}M_{2} \vert J_{1}J_{2}JM \rangle$, that
\begin{gather*}
(q+M)  \left \langle kqJM \vert J^{\prime}M^{\prime}\right \rangle = M^{\prime} \left \langle kqJM \vert J^{\prime}M^{\prime}\right \rangle
\\
\mp\left[ \frac{1}{2} (k \mp q)(k \pm q+1) \right]^{\frac{1}{2}} \left \langle kq \pm 1 JM \vert J^{\prime}M^{\prime}\right \rangle \mp\left[ \frac{1}{2} (J \mp M)(J \pm M+1) \right]^{\frac{1}{2}} \left \langle kqJM \pm 1 \vert J^{\prime}M^{\prime}\right \rangle = {}
\\
=  \pm\left[ \frac{1}{2} (J \pm M^{\prime})(J \mp M^{\prime}+1) \right]^{\frac{1}{2}} \left \langle kqJM \vert J^{\prime}M^{\prime} \mp 1\right \rangle
\end{gather*}
Accordingly,
\begin{gather}
J_{0}\psi_{\tau, J, k}(J^{\prime} M^{\prime}) = M^{\prime} \psi_{\tau, J, k}(J^{\prime} M^{\prime}) \\ \nonumber J_{\pm 1} \psi_{\tau, J, k}(J^{\prime} M^{\prime}) =  \mp\left[ \frac{1}{2} (J^{\prime} \mp M^{\prime})(J^{\prime} \pm M^{\prime}+1) \right]^{\frac{1}{2}}  \psi_{\tau, J, k}({J^{\prime} M^{\prime} \pm 1})
\end{gather}
which tells that the states $\psi_{\tau, J, k}(J^{\prime} M^{\prime})$ are either null or proportional to basis states  $\vert \tau^{\prime} J^{\prime} M^{\prime} \rangle$ :  $\psi_{\tau, J, k}(J^{\prime} M^{\prime}) = \alpha_{\tau, J, k} \vert \tau^{\prime} J^{\prime} M^{\prime} \rangle$ where the $\alpha_{\tau, J, k}$ are some unknown coefficients that depends solely on $(\tau, J, k)$. It then is inferred, since $\langle \tau^{\prime\prime} J^{\prime\prime} M^{\prime\prime} ~\vert~\tau^{\prime} J^{\prime} M^{\prime} \rangle = \delta_{\tau^{\prime\prime}, \tau^{\prime}} \delta_{J^{\prime\prime}, J^{\prime}}\delta_{M^{\prime\prime}, M^{\prime}}$ is $M^{\prime\prime}-$independent, that
\begin{equation}
\langle \tau^{\prime\prime} J^{\prime\prime} M^{\prime\prime} ~\vert~ \mathbf{T}_{q}^{(k)} ~\vert~ \tau J M \rangle = f(\tau, J, k, \tau^{\prime \prime}, J^{\prime \prime}) \left \langle kqJM \vert J^{\prime \prime}M^{\prime \prime}\right \rangle
\end{equation}
This is the essence of the \emph{Wigner-Eckart theorem}. It allows factorizing the matrix elements of the spherical irreducible tensor operators into a geometric component, the Clebsch-Gordan coefficient $\left \langle kqJM \vert J^{\prime \prime}M^{\prime \prime}\right \rangle$, and a physical component, the quantity $f(\tau, J, k, \tau^{\prime \prime}, J^{\prime \prime})$. It actually may be derived more elegantly from group theory arguments and is a particular instance of the so-called theorem of the three representations. According to this, as applied to the $SU(2)$ Lie group, the matrix elements $\langle \tau^{\prime\prime} J^{\prime\prime} M^{\prime\prime} ~\vert~ \mathbf{T}_{q}^{(k)} ~\vert~ \tau J M \rangle$ under a rotation get transformed according to the $\mathcal{D}^{J^{\prime \prime}\ast} \otimes \mathcal{D}^{k} \otimes \mathcal{D}^{J}$ matrix representation. The reduction of this either contains the trivial representation once, in which case the matrix element necessarily must not depend on the azimuthal quantum numbers, or does not contain the trivial representation, in which case the matrix element is null. No other case exists and this is why the geometric component factors out. It is customary to express the theorem with a 3jm symbol in the form 
\begin{equation}
\label{eq:wet}
\left \langle (\upsilon jm)_{f} \vert~ \mathbf{T}_{q}^{(k)} ~\vert (\upsilon jm)_{i}  \right \rangle = (-)^{j_{f}-m_{f}} \left(\begin{array}{ccc}j_{f} & k & j_{i} \\-m_{f} & q & m_{i}\end{array}\right) \left ( (\upsilon j)_{f} ~\Vert~ \mathbf{T}^{(k)} ~\Vert~(\upsilon j)_{i} \right )
\end{equation}
where $( (\upsilon j)_{f} ~\Vert~ \mathbf{T}^{(k)} ~\Vert~(\upsilon j)_{i} )$ is called a \emph{reduced matrix element}. This is often computed in an obvious way, for instance from 
\begin{equation*}
\langle s=\frac{1}{2} m_{s} = \frac{1}{2} \vert ~\mathbf{s}_{0}~ \vert s=\frac{1}{2} m_{s}=\frac{1}{2} \rangle = \frac{1}{2} = \left(\begin{array}{ccc}\frac{1}{2} & 1 & \frac{1}{2} \\[4pt]-\frac{1}{2} & 0 & \frac{1}{2}\end{array}\right) \left ( s_{f} ~\Vert~ \mathrm{{\bf s}} ~\Vert~s_{i} \right )
\end{equation*}
we find
\begin{equation}
\label{eq:rms}
\left ( s_{f} ~\Vert~ \mathrm{{\bf s}} ~\Vert~s_{i} \right ) = \sqrt{\frac{3}{2}}
\end{equation}
and, more generally, from 
\begin{equation*}
\langle l m = l \vert ~\mathbf{l}_{0}~ \vert l m = l\rangle = l = \left(\begin{array}{ccc}l & 1 & l \\-l & 0 & l \end{array}\right) \left ( l_{f} ~\Vert~ \mathrm{{\bf l}} ~\Vert~l_{i} \right )
\end{equation*}
we get
\begin{equation}
\label{eq:rml}
\left ( l_{f} ~\Vert~ \mathrm{{\bf l}} ~\Vert~l_{i} \right ) = \sqrt{l(l+1)(2l+1)}
\end{equation}
while from the eq.~\eqref{eq:AM2} we deduce
\begin{equation}
\label{eq:rmy}
\left ( l_{f} ~\Vert~ Y_{L} ~\Vert~l_{i} \right ) = \sqrt{\frac{1}{4 \pi}} (-)^{ l_{f}} \{ (2l_{f}+1)(2L+1)(l_{i}+1)\}^{\frac{1}{2}} \left ( \begin{array}{ccc}l_{f} & L & l_{i} \\0 & 0 & 0\end{array} \right )
\end{equation}
The Wigner-Eckart shows its full strength in the computing of the matrix elements of spherical irreducible tensor operators. Let $\mathbf{T}^{(k_{1})}$ and $\mathbf{U}^{(k_{2})}$ be two spherical irreducible tensor operators that act on part 1 and 2 of a system with angular momenta $\vec{\mathbf{j}}_{1}$ and $\vec{\mathbf{j}}_{2}$. The matrix element of $\{ \mathbf{T}^{(k_{1})} \otimes \mathbf{U}^{(k_{2})} \}_{Q}^{(K)} = \mathbf{X}_{Q}^{(K)}$ between the states $\vert \upsilon j_{1}j_{2} JM)_{f, i} \rangle$ may be written
\begin{multline}
\langle (\upsilon j_{1}j_{2} JM)_{f} \vert \mathbf{X}_{Q}^{(K)} \vert (\upsilon j_{1}j_{2} JM)_{i} \rangle = {} \\ = (-)^{J_{f}-M_{f}} \left(\begin{array}{ccc}J_{f} & K & J_{i} \\-M_{f} & Q & M_{i}\end{array}\right) \left(  (\upsilon j_{1}j_{2} J)_{f} \Vert \mathbf{X}^{(K)} \Vert (\upsilon j_{1}j_{2} J)_{i}  \right)
\end{multline}
but since $\mathbf{T}^{(k_{1})}$ and $\mathbf{U}^{(k_{2})}$ are supposed acting on different state spaces, it follows that
\begin{multline}
\langle (\upsilon j_{1}j_{2} JM)_{f} \vert \mathbf{X}_{Q}^{(K)} \vert (\upsilon j_{1}j_{2} JM)_{i} \rangle = \sum_{m_{1f}, m_{2f},m_{1i}, m_{2i}} \left[ J_{f}J_{i} \right] (-)^{j_{1f}-j_{2f}+M_{f} + j_{1i}-j_{2i}+M_{i}} \times {}
\\ 
\times \left(\begin{array}{ccc}j_{1f} & j_{2f} & J_{f} \\m_{1f} & m_{2f}& -M_{f}\end{array}\right)\left(\begin{array}{ccc}j_{1i} & j_{2i} & J_{i} \\m_{1i} & m_{2i}& -M_{i}\end{array}\right)  \sum_{q_{1}, q_{2}} \left(\begin{array}{ccc}k_{1} & k_{2} & K \\q_{1} & q_{2} & -Q\end{array}\right) (-)^{k_{1} - k_{2} + Q} \left[ K \right] \times {}
\\
\times \left\langle  (\upsilon j_{1}m_{1}j_{2}m_{2})_{f} \vert~\mathbf{T}_{q_{1}}^{(k_{1})} \mathbf{U}_{q_{2}}^{(k_{2})} \vert (\upsilon j_{1}m_{1}j_{2}m_{2})_{f} \right \rangle
\\[12pt]
= \sum_{m_{1f}, m_{2f},m_{1i}, m_{2i},q_{1}, q_{2}, \rho} \left[ J_{f}J_{i}K \right] (-)^{j_{1f}-j_{2f}+M_{f} + j_{1i}-j_{2i}+M_{i} + k_{1} - k_{2} + Q } \times {}
\\
\shoveright{ \times \left(\begin{array}{ccc}j_{1f} & j_{2f} & J_{f} \\m_{1f} & m_{2f}& -M_{f}\end{array}\right)\left(\begin{array}{ccc}j_{1i} & j_{2i} & J_{i} \\m_{1i} & m_{2i}& -M_{i}\end{array}\right) \left(\begin{array}{ccc}k_{1} & k_{2} & K \\q_{1} & q_{2} & -Q\end{array}\right) \times {} }
\\
\shoveright{ \times (-)^{ j_{1f}-m_{1f}} \left(\begin{array}{ccc}j_{1f} & k_{1} & j_{1i} \\-m_{1f} & q_{1} & m_{1i}\end{array}\right) \left(  (\upsilon j_{1})_{f} \Vert \mathbf{T}^{(k_{1})} \Vert \rho (j_{1})_{i}  \right) \times {} }
\\
\times (-)^{ j_{2f}-m_{2f}} \left(\begin{array}{ccc}j_{2f} & k_{2} & j_{2i} \\-m_{2f} & q_{2} & m_{2i}\end{array}\right) \left( \rho(j_{2})_{f} \Vert \mathbf{U}^{(k_{2})} \Vert (\upsilon j_{2})_{i}  \right) \quad
\end{multline}
On multiplying by $(-)^{J_{f}-M_{f}} \left(\begin{array}{ccc}J_{f} & K & J_{i} \\-M_{f} & Q & M_{i}\end{array}\right)$ and summing over $M_{f}$, $M_{i}$ and $Q$ it may be shown that
\begin{multline}
\left(  (\upsilon j_{1}j_{2} J)_{f} \Vert \mathbf{X}^{(K)} \Vert (\upsilon j_{1}j_{2} J)_{i}  \right) = {} \\ =\sum_{\rho} \left(  (\upsilon j_{1})_{f} \Vert \mathbf{T}^{(k_{1})} \Vert \rho (j_{1})_{i}  \right) \left( \rho(j_{2})_{f} \Vert \mathbf{U}^{(k_{2})} \Vert (\upsilon j_{2})_{i}  \right) \left[ J_{f}J_{i}K \right] \left\{\begin{array}{ccc}j_{1f} & j_{1i} & k_{1} \\j_{2f} & j_{2i} & k_{2} \\J_{f} & J_{i} & K\end{array}\right\}
\end{multline}
Setting $k_{2} = 0$ the reduced matrix element for the operator $\mathbf{T}^{(k_{1})}$ acting only on part 1 of the system is given by
\begin{multline}
\label{eq:RME}
\left(  (\upsilon j_{1}j_{2} J)_{f} \Vert \mathbf{T}^{(k_{1})} \Vert (\upsilon j_{1}j_{2} J)_{i}  \right) = {} \\ = \delta_{j_{2f}, j_{2i}} (-)^{j_{1f}+j_{2f}+J_{i}+k_{1}} \left[ J_{f}J_{i} \right]  \left\{\begin{array}{ccc}J_{f} & k_{1} & J_{i} \\j_{1i} & j_{2f} & j_{1f}\end{array}\right\} \left(  (\upsilon j_{1})_{f} \Vert \mathbf{T}^{(k_{1})} \Vert (\upsilon j_{1})_{i}  \right)
\end{multline}
Setting $k_{1} = 0$ the reduced matrix element for the operator $\mathbf{U}^{(k_{2})}$ acting only on part 2 of the system is given by
\begin{multline}
\left(  (\upsilon j_{1}j_{2} J)_{f} \Vert \mathbf{U}^{(k_{2})} \Vert (\upsilon j_{1}j_{2} J)_{i}  \right) = {} \\ = \delta_{j_{1f}, j_{1i}} (-)^{j_{1i}+j_{2i}+J_{f}+k_{2}} \left[ J_{f}J_{i} \right]  \left\{\begin{array}{ccc}J_{f} & k_{2} & J_{i} \\j_{2i} & j_{1f} & j_{2f}\end{array}\right\} \left(  (\upsilon j_{2})_{f} \Vert \mathbf{U}^{(k_{2})} \Vert (\upsilon j_{2})_{i}  \right)
\end{multline}
If there is no possibility to regard $\mathbf{T}^{(k_{1})}$ and $\mathbf{U}^{(k_{2})}$ as acting on separate parts of a system then there is no point in indicating that $\vec{\mathbf{J}}$ is the resultant of $\vec{\mathbf{j}}_{1}$ and $\vec{\mathbf{j}}_{2}$. In this case
\begin{multline}
\langle (\upsilon JM)_{f} \vert \mathbf{X}_{Q}^{(K)} \vert (\upsilon JM)_{i} \rangle = {} \\ = \sum_{q_{1}, q_{2}}(-)^{k_{1}-k_{2} + Q} \left[ K \right] \left(\begin{array}{ccc}k_{1} & k_{2} & K \\q_{1} & q_{2} & -Q\end{array}\right) \langle (\upsilon JM)_{f} \vert \mathbf{T}_{q_{1}}^{(k_{1})} \mathbf{U}_{q_{2}}^{(k_{2})} \vert (\upsilon JM)_{i} \rangle = {}
\\ 
= \sum_{q_{1}, q_{2}, \overline{J}, \rho, \overline{M}}(-)^{k_{1}-k_{2} + Q} \left[ K \right] \left(\begin{array}{ccc}k_{1} & k_{2} & K \\q_{1} & q_{2} & -Q\end{array}\right) \langle (\upsilon JM)_{f} \vert \mathbf{T}_{q_{1}}^{(k_{1})} \vert \overline{J} \rho \overline{M} \rangle \langle \overline{J} \rho \overline{M} \vert \mathbf{U}_{q_{2}}^{(k_{2})} \vert (\upsilon JM)_{i} \rangle = {}
\\
= \sum_{q_{1}, q_{2}, \overline{J}, \rho, \overline{M}} (-)^{k_{1}-k_{2} + Q + J_{f} -M_{f} + \overline{J} - \overline{M}} \left[ K \right] \left(\begin{array}{ccc}k_{1} & k_{2} & K \\q_{1} & q_{2} & -Q\end{array}\right) \times {} \\ \times \left(\begin{array}{ccc}J_{f} & k_{1} & \overline{J} \\-M_{f} & q_{1} & \overline{M}\end{array}\right) 
\left(\begin{array}{ccc}\overline{J} & k_{2} & J_{i} \\-\overline{M} & q_{2} & M_{f}\end{array}\right) \left(  (\upsilon J)_{f} \Vert \mathbf{T}^{(k_{1})} \Vert \rho \overline{J}  \right)
\left(  \rho \overline{J} \Vert \mathbf{U}^{(k_{2})} \Vert (\upsilon J)_{i}  \right) = {}
\\
= \sum_{\overline{J}, \rho} (-)^{J_{f} + J_{i} + K + J_{f} - M_{f}} \left[ K \right] \left(\begin{array}{ccc}J_{f} & K & J_{i} \\-M_{f} & Q & M_{i}\end{array}\right) \left\{\begin{array}{ccc}J_{f} & K & J_{i} \\k_{2} & \overline{J} & k_{1}\end{array}\right\} \times {} \\ \times  \left(  (\upsilon J)_{f} \Vert \mathbf{T}^{(k_{1})} \Vert \rho \overline{J}  \right)
\left(  \rho \overline{J} \Vert \mathbf{U}^{(k_{2})} \Vert (\upsilon J)_{i}  \right) \left\{\begin{array}{ccc}J_{f} & K & J_{i} \\k_{2} & \overline{J} & k_{1}\end{array}\right\} 
\end{multline}
so that
\begin{multline}
\left(  (\upsilon J)_{f} \Vert \mathbf{X}^{(K)} \Vert \rho \overline{J}  \right) = (-)^{J_{f}  + K+ J_{i}} \left[ K \right] \sum_{\overline{J}, \rho} \left\{\begin{array}{ccc}J_{f} & K & J_{i} \\k_{2} & \overline{J} & k_{1}\end{array}\right\} \times {} \\ \times
\left(  (\upsilon J)_{f} \Vert \mathbf{T}^{(k_{1})} \Vert \rho \overline{J}  \right)
\left(  \rho \overline{J} \Vert \mathbf{U}^{(k_{2})} \Vert (\upsilon J)_{i}  \right)
\end{multline}

\section{\label{MEE}Many equivalent electron states}

It may be shown that an $\mathcal{N}-$ electron states $\vert \varsigma_{f, i} \rangle = \vert (\upsilon LM_{L}SM_{S})_{f, i} \rangle$ of the configuration $(n l)_{\mathcal{N}}$ can be built up from $(\mathcal{N}-1)-$ electron states $\vert \overline{\varsigma}_{f, i} \rangle = \vert (\overline{\upsilon} \overline{L}\overline{M_{L}}\overline{S}\overline{M_{S}})_{f, i} \rangle$ of the configuration $(n l)_{\mathcal{N}-1}$ and $1-$ electron state $\vert \eta_{f, i} \rangle = \vert (n lm_{l}sm_{s})_{f, i} \rangle$ of the configuration $(n l)$ as
\begin{gather*}
\vert (\upsilon LM_{L}SM_{S})_{f, i} \rangle = \sum_{(\overline{\upsilon} \overline{L} \overline{S})_{f, i}} \left ( (\overline{\upsilon} \overline{L} \overline{S}_{f, i}) \left \vert \right \} (\upsilon LS)_{f, i} \right ) \left \vert (\overline{\upsilon} \overline{L} \overline{S} ~n ls)_{f, i} \right \rangle
\end{gather*}
where
\begin{gather*}
\left \vert (\overline{\upsilon} \overline{L} \overline{S} ~n ls)_{f, i} \right \rangle = \sum_{(\overline{M_{L}}, \overline{M_{S}})_{f, i}} \sum_{(m_{l}m_{s})_{f, i}} \left \langle (\overline{L} \overline{M_{L}} lm_{l})_{f, i} \vert (LM_{L})_{f, i} \right \rangle \left \langle (\overline{S} \overline{M_{S}} sm_{s})_{f, i} \vert (SM_{S})_{f, i} \right \rangle \times {}
\\
\times \left \vert (\overline{\upsilon} \overline{L} \overline{M_{L}} \overline{S} \overline{M_{S}})_{f, i} \right \rangle \left \vert (n lm_{l}sm_{s})_{f, i} \right \rangle
\end{gather*}
are the coupled states of $(\mathcal{N}-1)$ electrons with the $\mathcal{N}^{th}$ electron and where $( (\overline{\upsilon} \overline{L} \overline{S})_{f, i} \vert \} (\upsilon LS)_{f, i} )$ are called \emph{coefficients of fractional parentage}. This concept is extremely useful when dealing with matrix elements of sums $\mathbf{O} = \sum_{u} \mathbf{O}(u)$ of one-electron operators $\mathbf{O}(u)$. Indeed, owing to the equivalence of the electrons, the matrix element of $\mathbf{O}(u)$ between the states $\vert (\upsilon LM_{L}SM_{S})_{f, i}\rangle$ does not depend on which electron $u$ has been chosen to compute it, so that
\begin{equation}
\langle  (\upsilon LM_{L}SM_{S})_{f} \vert ~\mathbf{O}~ \vert (\upsilon LM_{L}SM_{S})_{i}\rangle = \mathcal{N} \langle  (\upsilon LM_{L}SM_{S})_{f} \vert ~\mathbf{O}(\mathcal{N})~ \vert (\upsilon LM_{L}SM_{S})_{i}\rangle
\end{equation}
that is 
\begin{multline}
\langle  (\upsilon LM_{L}SM_{S})_{f} \vert ~\mathbf{O}~ \vert (\upsilon LM_{L}SM_{S})_{i}\rangle = \mathcal{N} \sum_{\overline{\upsilon} \overline{L} \overline{S}} ((\upsilon LS)_{f} \{ \vert \overline{\upsilon} \overline{L} \overline{S}) (\overline{\upsilon} \overline{L} \overline{S} \vert \} (\upsilon LS)_{i} ) \times {} \\ \times \sum_{M_{Lf}\overline{M}_{L}M_{Li}} \langle (\overline{L} \overline{M_{L}} lm_{l})_{f} \vert (LM_{L})_{f} \rangle \langle (\overline{L} \overline{M_{L}} lm_{l})_{i} \vert (LM_{L})_{i} \rangle \times {} \\ \times \sum_{M_{Sf}\overline{M}_{S}M_{Si}} \langle (\overline{S} \overline{M_{S}} sm_{s})_{f} \vert (SM_{S})_{f} \rangle \langle (\overline{L} \overline{M_{S}} sm_{s})_{i} \vert (SM_{S})_{i}  \rangle \times {} \\ \times \langle  (lm_{l}sm_{s})_{i} \vert ~\mathbf{O}(\mathcal{N})~  (lm_{l}sm_{s})_{i} \rangle = {}
\end{multline}
or else
\begin{multline}
\langle  (\upsilon LM_{L}SM_{S})_{f} \vert ~\mathbf{O}~ \vert (\upsilon LM_{L}SM_{S})_{i}\rangle = {} \\ = \mathcal{N} \sum_{\overline{\theta}} (\theta_{f} \{ \vert \overline{\theta}) (\overline{\theta} \vert \} \theta_{i} ) \langle (\overline{\theta} LM_{L}SM_{S})_{f} \vert ~\mathbf{O}(\mathcal{N})~ \vert (\overline{\theta} LM_{L}SM_{S})_{i} \rangle
\end{multline}
by introducing self-explicit shorthand notations to simplify the equations. \medskip

\noindent If the operator $\mathbf{O}$ is a tensor operator of rank $k$ independent of the spin
\begin{equation}
\mathbf{T}^{(k)} = \sum_{u} \mathbf{t}^{(k)}(u)
\end{equation}
then applying the Wigner-Eckart theorem
\begin{multline}
\langle  (\upsilon LM_{L}SM_{S})_{f} \vert ~\mathbf{T}_{q}^{(k)}~ \vert (\upsilon LM_{L}SM_{S})_{i}\rangle = {}
\\
= (-)^{L_{f}-M_{Lf}} \left(\begin{array}{ccc}L_{f} & k & L_{i} \\-M_{Lf} & q & M_{Li}\end{array}\right) \left( (\upsilon LSM)_{f} \Vert ~\mathbf{T}^{(k)}~\Vert (\upsilon LSM)_{i} \right) = {}
\\
= (-)^{L_{f}-M_{Lf}} \left(\begin{array}{ccc}L_{f} & k & L_{i} \\-M_{Lf} & q & M_{Li}\end{array}\right) \mathcal{N} \sum_{\overline{\theta}} (\theta_{f} \{ \vert \overline{\theta}) (\overline{\theta} \vert \} \theta_{i} )  \times {} 
\\ 
\times \left( (\overline{\theta} LSM)_{f} \Vert ~\mathbf{t}^{(k)}(\mathcal{N})~\Vert (\overline{\theta} LSM)_{i} \right) 
\end{multline}
Use can be made of the eq.~\eqref{eq:RME} to compute the reduced matrix element assuming that the part 1 of the system is played by the $\mathcal{N}^{th}$ electron. We get
\begin{multline}
\label{eq:emo}
\langle  (\upsilon LM_{L}SM_{S})_{f} \vert ~\mathbf{T}_{q}^{(k)}~ \vert (\upsilon LM_{L}SM_{S})_{i}\rangle = {}
\\
= (-)^{L_{f}-M_{Lf}} \left(\begin{array}{ccc}L_{f} & k & L_{i} \\-M_{Lf} & q & M_{Li}\end{array}\right) \mathcal{N} \sum_{\overline{\theta}} (\theta_{f} \{ \vert \overline{\theta}) (\overline{\theta} \vert \} \theta_{i} )  \times {} 
\\ 
\times \delta_{S_{f}, S_{i}} (-)^{\overline{L}+l+L+k} \left[ L_{f}L_{i} \right] \left\{\begin{array}{ccc}L_{f} & k & L_{i} \\l & \overline{L} & l\end{array}\right\}  \left( l \Vert ~\mathbf{t}^{(k)}(\mathcal{N})~\Vert l \right) 
\end{multline}
The operator $\mathbf{T}^{(k_{1}, k_{2})}$ comprising the $(2k_{1}+1)(2k_{2}+1)$ components $\mathbf{T}_{q_{1}, q_{2}}^{(k_{1}, k_{2})}$ is a double tensor if it behaves as a tensor of rank $k_{1}$ with respect to the spin angular momentum states and as a tensor of rank $k_{2}$ with respect to the orbital angular momentum states. If this is the sum of one-electron double tensors, spin (x) - orbit (y) direct products, then 
\begin{equation}
\mathbf{T}^{(k_{1}, k_{2})} = \sum_{u} \mathbf{t}^{(k_{1}, k_{2})}(u) = \sum_{u} \mathbf{x}^{(k_{1})}(u)\mathbf{y}^{(k_{2})}(u)
\end{equation}
Its matrix elements are given by
\begin{multline}
\langle  (\upsilon LM_{L}SM_{S})_{f} \vert ~\mathbf{T}^{(k_{1}, k_{2})} ~ \vert (\upsilon LM_{L}SM_{S})_{i}\rangle = {}
\\
= (-)^{S_{f}-M_{Sf} + L_{f}-M_{Lf}} \left(\begin{array}{ccc}S_{f} & k_{1} & S_{i} \\-M_{Sf} & q_{1} & M_{Si}\end{array}\right) \left(\begin{array}{ccc}L_{f} & k_{2} & L_{i} \\-M_{Lf} & q_{2} & M_{Li}\end{array}\right) \times {} 
\\
\times \mathcal{N} \sum_{\overline{\theta}} (\theta_{f} \{ \vert \overline{\theta}) (\overline{\theta} \vert \} \theta_{i} ) \left( (\overline{\theta} LSM)_{f} \Vert ~\mathbf{t}^{(k_{1}, k_{2})}(\mathcal{N})~\Vert (\overline{\theta} LSM)_{i} \right) 
\end{multline}
Use again can be made of the eq.~\eqref{eq:RME} to compute the reduced matrix element applying it separately to the spin and to the orbital part of the double tensor with the part 1 of the system played by the $\mathcal{N}^{th}$ electron. We get
\begin{multline}
\label{eq:ems}
\langle  (\upsilon LM_{L}SM_{S})_{f} \vert ~\mathbf{T}^{(k_{1}, k_{2})} ~ \vert (\upsilon LM_{L}SM_{S})_{i}\rangle = {}
\\
= (-)^{S_{f}-M_{Sf} + L_{f}-M_{Lf}} \left(\begin{array}{ccc}S_{f} & k_{1} & S_{i} \\-M_{Sf} & q_{1} & M_{Si}\end{array}\right) \left(\begin{array}{ccc}L_{f} & k_{2} & L_{i} \\-M_{Lf} & q_{2} & M_{Li}\end{array}\right) \times {} 
\\
\times \mathcal{N} \sum_{\overline{\theta}} (\theta_{f} \{ \vert \overline{\theta}) (\overline{\theta} \vert \} \theta_{i} ) (-)^{\overline{S}+\overline{L}+s+l+S+L+k_{1}+k_{2}}  \left[ S_{f}S_{i} L_{f}L_{i} \right]  \times {} 
\\ 
\times \left\{\begin{array}{ccc}S_{f} & k_{1} & S_{i} \\s & \overline{S} & s\end{array}\right\}  \left\{\begin{array}{ccc}L_{f} & k_{2} & L_{i} \\l & \overline{L} & l\end{array}\right\}  \left( sl \Vert ~\mathbf{t}^{(k_{1}, k_{2})}(\mathcal{N})~\Vert sl \right) 
\end{multline}
where
\begin{equation}
\left( sl \Vert ~\mathbf{t}^{(k_{1}, k_{2})}(\mathcal{N})~\Vert sl \right) = \left( s \Vert ~\mathbf{x}^{(k_{1})}(\mathcal{N})~\Vert s \right) \left( l \Vert ~\mathbf{y}^{(k_{2})}(\mathcal{N})~\Vert l \right)
\end{equation}


\end{document}